\documentclass[11pt]{article}
\usepackage{epsfig}
\usepackage{graphics}
\usepackage{longtable}
\usepackage{multicol}
\usepackage{lscape}
\usepackage{array}
\usepackage{longtable}
\usepackage{url}
\usepackage[sort,numbers,square]{natbib}
\usepackage[colorlinks=true,
            linkcolor=blue,
            urlcolor=blue,
            citecolor=blue]{hyperref}
\usepackage[latin1]{inputenc}

\begin{document}
\centerline{\large{\bf The Observer's Guide to the Gamma-Ray Burst$-$Supernova Connection}}

\vspace*{0.5cm}

\centerline{{\bf Zach Cano$^{1}$, Shan-Qin Wang$^{2,3}$, Zi-Gao Dai$^{2,3}$ and Xue-Feng Wu$^{4,5}$}}
\vspace*{0.5cm}
{\small \noindent $^1$Centre for Astrophysics and Cosmology, Science Institute, University of Iceland, Dunhagi 5, 107 Reykjavik, Iceland. (zewcano@gmail.com)\\
$^2$School of Astronomy and Space Science, Nanjing University, Nanjing 210093, China. (dzg@nju.edu.cn)\\
$^3$Key Laboratory of Modern Astronomy and Astrophysics (Nanjing University), Ministry of Education, China.\\
$^4$Purple Mountain Observatory, Chinese Academy of Sciences, Nanjing 210008, China. (xfwu@pmo.ac.cn)\\
$^5$Joint Center for Particle, Nuclear Physics and Cosmology, Nanjing University-Purple Mountain Observatory, Nanjing 210008, China.}

\vskip 0.5cm

{\bf Abstract.} 

In this review article we present an up-to-date progress report of the connection between long-duration (and their various sub-classes) gamma-ray bursts (GRBs) and their accompanying supernovae (SNe).  The analysis presented here is from the point of view of an observer, with much of the emphasis placed on how observations, and the modelling of observations, have constrained what we known about GRB-SNe.  We discuss their photometric and spectroscopic properties, their role as cosmological probes, including their measured luminosity$-$decline relationships, and how they can be used to measure the Hubble constant.  We present a statistical analysis of their bolometric properties, and use this to determine the properties of the ``average'' GRB-SN: which has a kinetic energy of $E_{\rm K} \approx 2.5\times10^{52}$~erg ($\sigma_{E_{\rm K}} = 1.8\times10^{52}$~erg), an ejecta mass of $M_{\rm ej} \approx 6$~M$_{\odot}$ ($\sigma_{M_{\rm ej}} = 4$~M$_{\odot}$), a nickel mass of $M_{\rm Ni} \approx 0.4$~M$_{\odot}$ ($\sigma_{M_{\rm Ni}} = 0.2$~M$_{\odot}$), an ejecta velocity at peak light of $v \approx 20,000$~km~s$^{-1}$ ($\sigma_{v_{\rm ph}} = 8,000$~km~s$^{-1}$), a peak bolometric luminosity of $L_{\rm p} \approx 1\times10^{43}$~erg~s$^{-1}$ ($\sigma_{L_{\rm p}} = 0.4\times10^{43}$~erg~s$^{-1}$), and it reaches peak bolometric light in $t_{\rm p} \approx 13$~days ($\sigma_{t_{\rm p}} = 2.7$~days).  We discuss their geometry, as constrained from observations, and consider the various physical processes that are thought to power the luminosity of GRB-SNe, and whether differences exist between GRB-SNe and the SNe associated with ultra-long duration GRBs such as GRB~111209A/SN~2011kl.  We discuss how observations of the environments of GRB-SNe further constrain the physical properties of their pre-explosion progenitor stars, and give a brief overview of the current theoretical paradigms of the central engines that produce the various types of GRB-SNe.  Furthermore, we present an overview of the $r$-process, radioactively powered transients that have been photometrically associated with short-duration GRBs, and we conclude the review by discussing what additional research is needed to further our understanding of GRB-SNe, in particular the role of binary-formation channels and the connection of GRB-SNe with superluminous SNe.


\section{Introduction}
\label{sec:Intro}
\vspace*{0.5cm}

\begin{figure*}[t!]
 \centering
 \includegraphics[bb=0 0 576 432,scale=0.8,keepaspectratio=true]{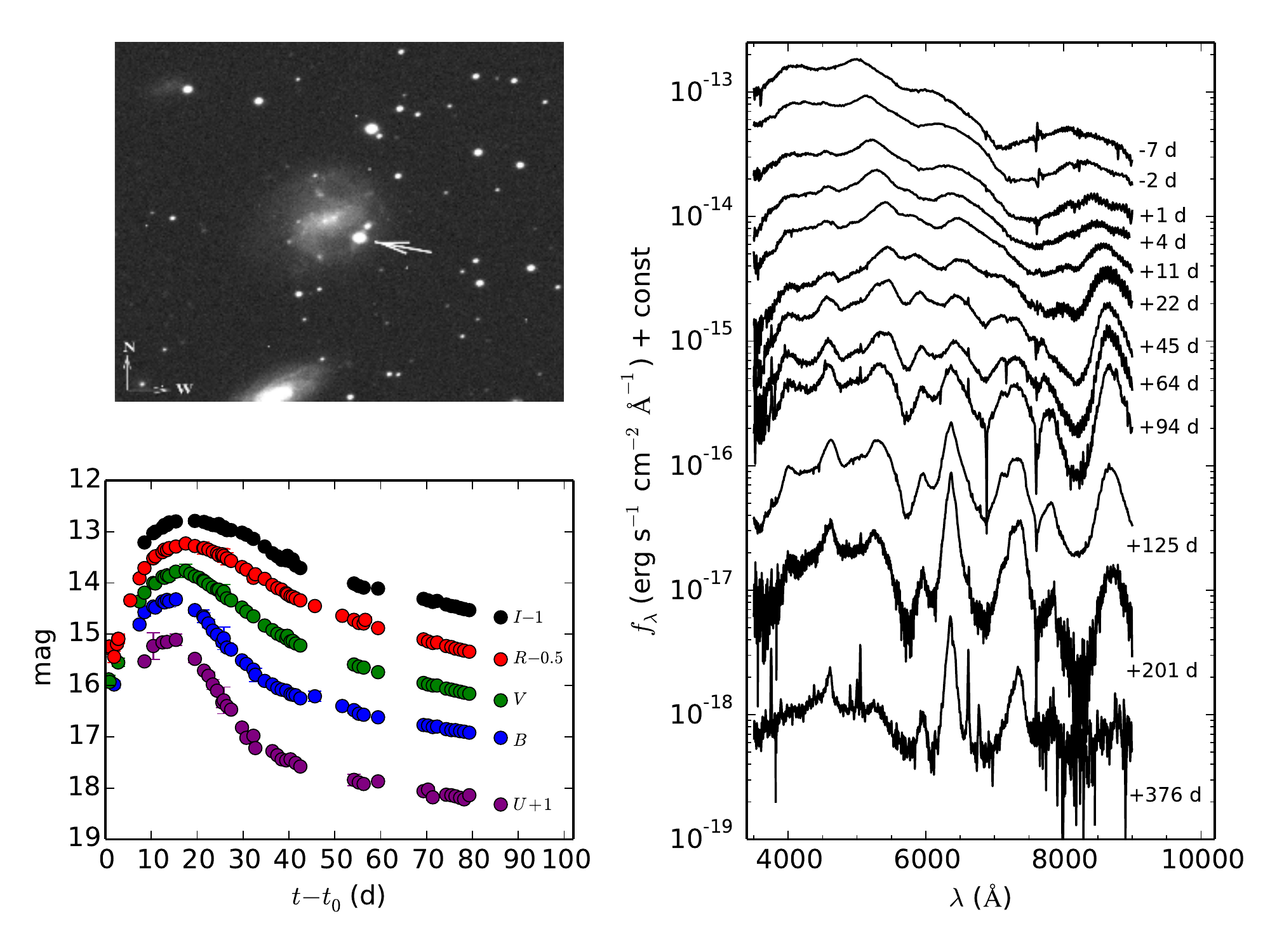}
 \caption{GRB~980425 / SN~1998bw: the archetype GRB-SN.  Host image (ESO 184-G82) is from \cite{Galama1998}, where the position of the optical transient is clearly visible. Optical light curves are from \cite{Clocchiatti2011}, and spectra from \cite{Patat2001}.}
 \label{fig:1998bw_mosiac}
\end{figure*}

\begin{figure}
 \centering
 \includegraphics[bb=0 0 1035 573,scale=0.45,keepaspectratio=true]{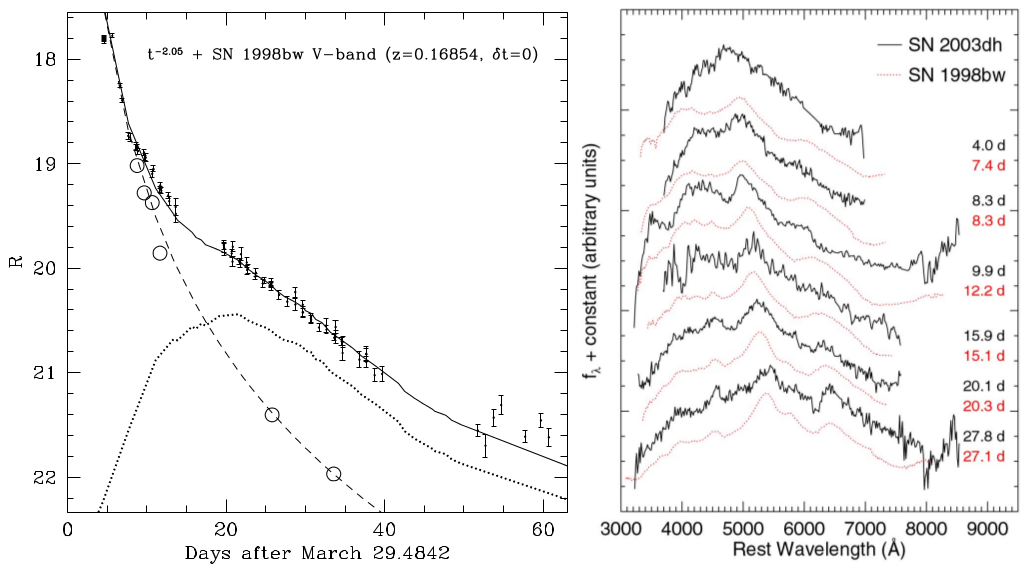}
 \caption{\textbf{Left}: The photometric ($R$-band) evolution of GRB~030329/SN~2003dh, from \cite{Matheson2003};  \textbf{Right}: The spectral evolution of GRB~030329/SN~2003dh, as compared with that of SN~1998bw, from \cite{Hjorth2003}.}
 \label{fig:GRB030329}
\end{figure}

Observations have proved the massive-star origins of long-duration GRBs (LGRBs) beyond any reasonable doubt.  The temporal and spatial connection between GRB~980425 and broad-lined type Ic (IcBL) SN~1998bw offered the first clues to their nature \cite{Galama1998,Patat2001} (Fig. \ref{fig:1998bw_mosiac}).  The close proximity of this event ($z=0.00866$; $\sim40$~Mpc), which is still the closest GRB to date, resulted in it becoming one of the most, if not \textit{the} most, scrutinized GRB-SN in history.  It was shown that SN~1998bw had a very large kinetic energy (see Section \ref{sec:Physical_Properties} and Table \ref{table:master_table_2_SN}) of $\sim 2-5\times10^{52}$~erg, which led it to being referred to as a hypernova \cite{Iwamoto1998}.  However,  given several peculiarities of its $\gamma$-ray properties, including its underluminous $\gamma$-ray luminosity ($L_{\gamma,\rm iso} \sim 5 \times 10^{46}$ erg~s$^{-1}$), it was doubted whether this event was truly representative of the general LGRB population.  This uncertainty persisted for almost five years until the spectroscopic association between cosmological/high-luminosity GRB~030329 ($L_{\gamma,\rm iso} \sim 8 \times 10^{50}$ erg~s$^{-1}$) and SN~2003dh \cite{Hjorth2003,Stanek2003,Matheson2003}.  GRB~030329 had an exceptionally bright optical afterglow (AG; see Figs. \ref{fig:GRB030329} and \ref{fig:grb_bumps}), and a careful  decomposition of the photometric and spectroscopic observations was required in order to isolate the SN features from the dominant AG light \cite{Deng2005} (see Section \ref{sec:Obs_photo} and Fig. \ref{fig:LC_decomposition}).  As was seen for SN~1998bw, SN~2003dh was a type IcBL SN, and its kinetic energy was in excess of $10^{52}$~erg, showing that it too was a hypernova. 

The launch of the \emph{Swift} satellite \cite{Gehrels2004} dramatically changed the way we studied GRBs and the GRB-SN association, and the number of events detected by this mission has helped increase the GRB-SN sample size by a factor of three since the pre-\emph{Swift} era.  This includes, among many others, the well-studied events GRB~060218/SN~2006aj, GRB~100316D/SN~2010bh, GRB~111209A/SN~2011kl, GRB~120422A/SN~2012bz, GRB~130427A/SN~2013cq and GRB~130702A /SN~2013dx.  A full list of the references to these well-studied spectroscopic GRB-SN associations are found in Table \ref{table:references}.

This review paper represents a continuation of other review articles presented to date, including the seminal work by Woosley \& Bloom (2006) \cite{WoosleyBloom2006}.  As such, we have focused the majority of the content to achievements made in the 10 years since \cite{WoosleyBloom2006} was published.  In this review and many others \cite{WW1998,Nomoto2007,Bissaldi2007,DV2011,Modjaz2011,HjorthBloom2012,Cano2013,Hjorth2013}, thorough historical accounts of the development of the gamma-ray burst supernova (GRB-SN) connection are presented, and we encourage the reader to consult the detailed presentation given in section one of \cite{WoosleyBloom2006} for further details.  In Tables \ref{table:master_table_I_highE} and \ref{table:master_table_2_SN} we present the most comprehensive database yet compiled of the observational and physical properties of the GRB prompt emission and GRB-SNe, respectively, which consists of 46 GRB-SNe.  It is the interpretation of these data which forms a substantial contribution to this review.  We have adopted the grading scheme devised by \cite{HjorthBloom2012} to assign a significance\footnote{\textbf{A}: Strong spectroscopic evidence. \textbf{B}: A clear light curve bump as well as some spectroscopic evidence resembling a GRB-SN. \textbf{C}: A clear bump consistent with other GRB-SNe at the spectroscopic redshift of the GRB. \textbf{D}: A bump, but the inferred SN properties are not fully consistent with other GRB-SNe or the bump was not well sampled or there is no spectroscopic redshift of the GRB. \textbf{E}: A bump, either of low significance or inconsistent with other GRB-SNe.} of the GRB-SN association to each event.  These are found in Table \ref{table:master_table_2_SN}.

Throughout this article we use a $\Lambda$CDM cosmology constrained by Planck \cite{Planck2014} of $H_{0} = 67.3$~km~s$^{-1}$ Mpc$^{-1}$, $\Omega_{\rm M} = 0.315$, $\Omega_{\Lambda} = 0.685$.  All published data, where applicable, have been renormalized to this cosmological model.  Foreground extinctions were calculated using the dust extinction maps of \cite{Schlegel1998} and \cite{SF2011}.  Unless stated otherwise, errors are statistical only.  \textit{Nomenclature}: $\sigma$ denotes the standard deviation of a sample, whereas the root-mean square of a sample is expressed as RMS.  A symbol with an overplotted bar denote an average value.  LGRB and SGRB are long- and short-duration GRBs, respectively, while a GRB-SN is implicitly understood to be associated with an LGRB.  The term $t_{0}$ refers to the time that a given GRB is detected by a GRB satellite.


\section{Observational Properties}
\label{sec:Obs_properties}
\vspace*{0.5cm}

\subsection{Photometric Properties}
\label{sec:Obs_photo}

\begin{figure*}
 \centering
 \includegraphics[bb=0 0 316 175,scale=1.46,keepaspectratio=true]{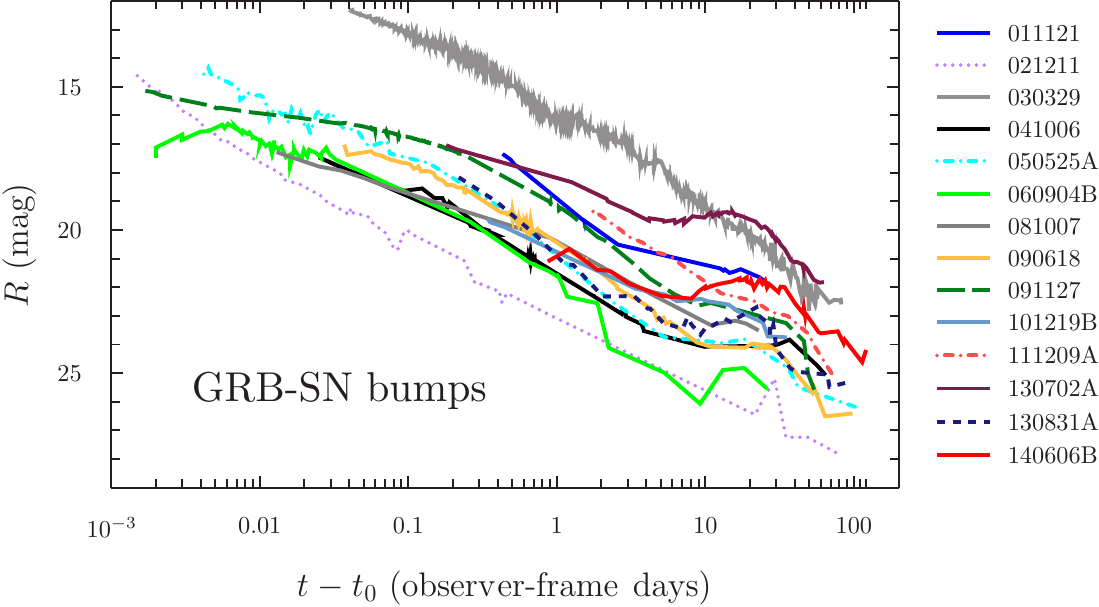}
 \caption{A mosaic of GRB-SNe (AG + SN).  Clear SN bumps are observed for all events except SN~2003dh, for which the SN's properties had to be carefully decomposed from photometric and spectroscopic observations \cite{Deng2005}.  The lack of an unambiguous SN bump in this case is not surprising given the brightness of its AG relative to the other GRB-SN in the plot -- SN~2013dx was at a comparable redshift ($z=0.145$, compared with $z=0.1685$ for 2003dh), but its AG was much fainter ($2-5$~mag) at a given moment in time.  The redshift range probed in this mosaic spans almost an order of magnitude ($0.145 < z < 1.006$), and shows the variation in peak observed magnitude for GRB-SNe.  It is important to remember that given the large span of distances probed here, observer-frame $R$-band samples a wide range of rest-frame SEDs (from $U$-band to $V$-band). }
 \label{fig:grb_bumps}
\end{figure*}

The observer-frame, optical light curves (LCs) of GRBs span more than $8-10$ magnitudes at a given observer-frame post-explosion epoch (see, e.g. fig. 1 in \cite{Kann2010}).  Similarly, if we inspect the observer-frame $R$-band LCs of GRB-SNe (redshift range $0.145 < z < 1.006$) shown in Fig. \ref{fig:grb_bumps}, they too span a similar range at a given epoch.  Indeed, the peak SN brightness during the SN ``bump'' phase ranges from $R=19.5$ for GRB~130702A (the brightest GRB-SN bump observed to date) to $R=25$ for GRB~021211.  

\begin{figure*}
 \centering
 \includegraphics[bb=0 0 1009 216,scale=0.39,keepaspectratio=true, scale=1.15]{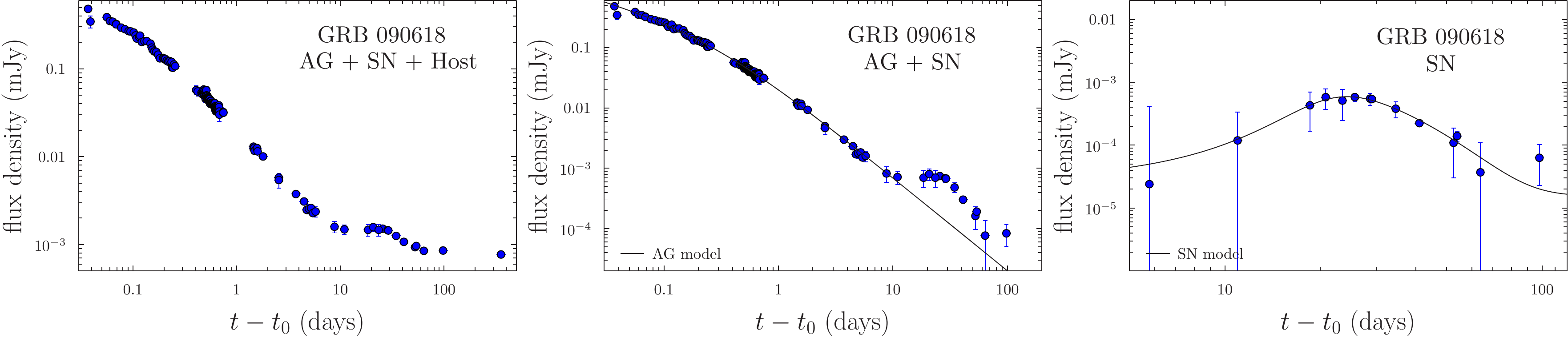}
 \caption{An example decomposition of the optical ($R$-band) light curve of GRB~090618 \cite{Cano2011b}.  \textbf{Left:} For a given GRB-SN event, the single-filter monochromatic flux is attributed as arising from three sources: the AG, the SN, and a constant source of flux from the host galaxy.  \textbf{Middle:} Once the observations have been dereddened, the host flux is removed, either via the image-subtraction technique or mathematically subtracted away.  At this point a mathematical model composed of one or more power-laws punctuated by break-times are fit to the early light curve to determine the temporal behaviour of the AG.  \textbf{Right:} One the AG model has been determined, it is subtracted from the observations leaving just light from the SN.}
 \label{fig:LC_decomposition}
\end{figure*}

For a typical GRB-SN, there are three components of flux being measured: (1) the AG, which is associated with the GRB event, (2) the SN, and (3) the constant source of flux coming from the host galaxy.  A great deal of information can be obtained from modelling each component, but for the SN component to be analysed, it needs to be \textit{decomposed} from the optical/NIR LCs (Fig. \ref{fig:LC_decomposition}).  To achieve this task, the temporal behaviour of the AG, the constant source of flux from the host galaxy,  and the line of sight extinction, including foreground extinction arising from different sight-lines through the Milky Way (MW) \cite{Schlegel1998,SF2011}, and extinction local to the event itself \cite{Kann2006,Kann2010,Kann2011,Cano2011b,Japelj2015}, in a given filter needs to be modelled and quantified.  The host contribution can be considered either by removing it via the image-subtraction technique \cite{Alard1998,Alard2000,Strolger2004}, subtracting the host flux mathematically \cite{Cano2011a,Cano_etal_2014,Cano2015}, or by including it as an additional component in the fitting routine \cite{Ferrero2006,Thone2011,Sollerman2007,Greiner2015}.  The AG component is modelled using either a single or a set of broken power-laws (SPL/BPL; \cite{Beuermann1999}).  This phenomenological approach is rooted in theory however, as standard GRB theory states that the light powering the AG is synchrotron in origin, and therefore follows a power-law behaviour in both time and frequency ($f_{\nu} \propto (t - t_{0})^{-\alpha}\nu^{-\beta}$, where the respective decay and energy spectral indices are $\alpha$ and $\beta$).

Once the SN LC has been obtained, traditionally it is compared to a template supernova, i.e. SN~1998bw, where the relative brightness ($k$) and width (also known as a stretch factor, $s$) are determined.  Such an approach has been used extensively over the years \cite{Zeh2004,Stanek2005,Ferrero2006,Cano2011b,Thone2011,Cano2013,Xu2013,Schulze2014,Cano2014,Cano_etal_2014,Cano2015,Toy2016}.  Another approach to determining the SN's properties is to fit a phenomenological model to the resultant SN LC \cite{Cano2011b,Cano2014,CJG14,Toy2016}, such as the Bazin function \cite{Bazin2011}, in order to determine the magnitude/flux at peak SN light, the time it takes to rise and fade from peak, and the width of the LC, such as the $\Delta m_{15}$  parameter (in a given filter, the amount a SN fades in magnitudes from peak light to 15 days later).  All published values of these observables are presented in Table \ref{table:master_table_2_SN}.

\begin{figure}
 \centering
 \includegraphics[bb=0 0 252 335,scale=1.7,keepaspectratio=true]{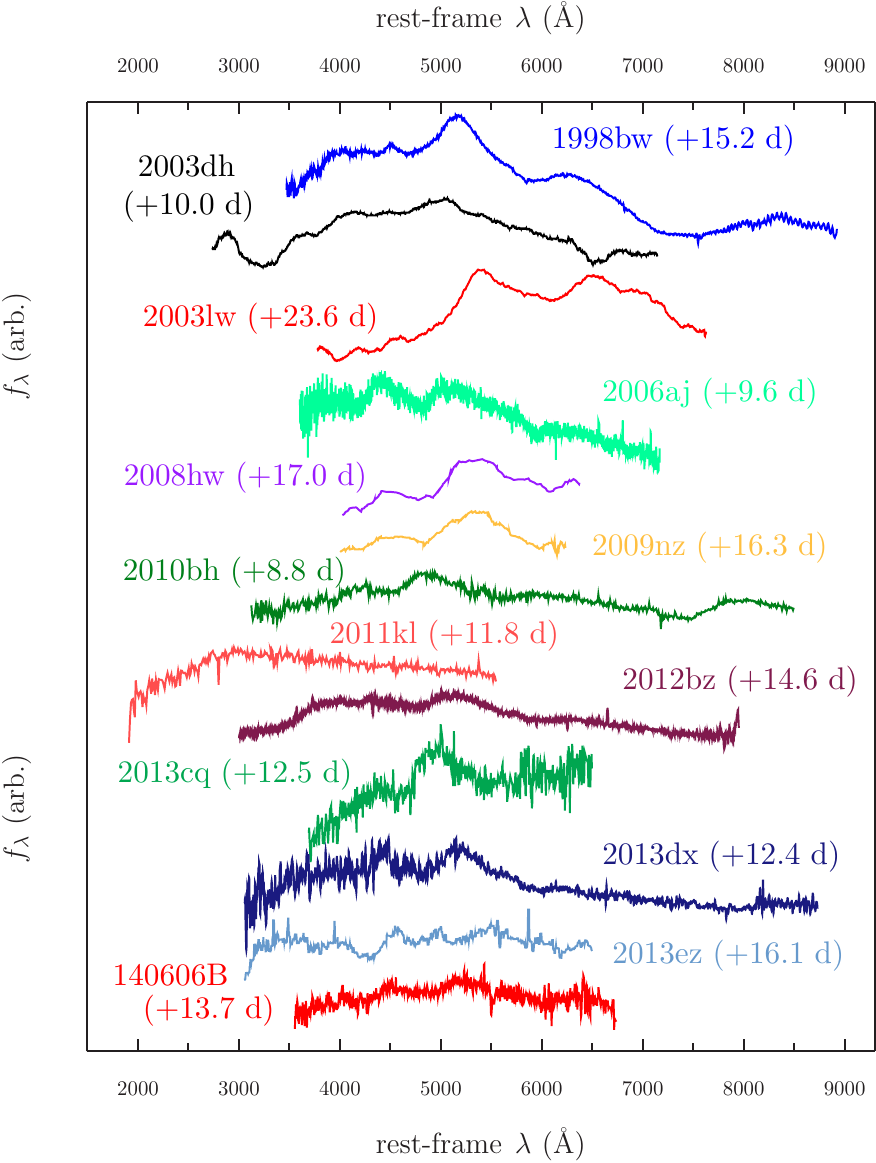}
 \caption{Peak/near-peak spectra of GRB-SNe. The spectra have been arbitrarily shifted in flux for comparison purposes, and to exaggerate their main features, and host emission lines have been manually removed.  The spectra of SNe 2012bz, 2013cq and 2013dx have been Kaiser smoothed \cite{Cano_etal_2014} in order to suppress noise. Most of the spectra are characterized by broad absorption features, while such features are conspicuously absent in the spectra of SN~2013ez and SN~2011kl.  }
 \label{fig:spectra_series}
\end{figure}

\subsection{Spectroscopic Properties}
\label{sec:Obs_spectra}

\begin{figure*}
 \centering
 \includegraphics[bb=0 0 411 417,scale=0.8,keepaspectratio=true]{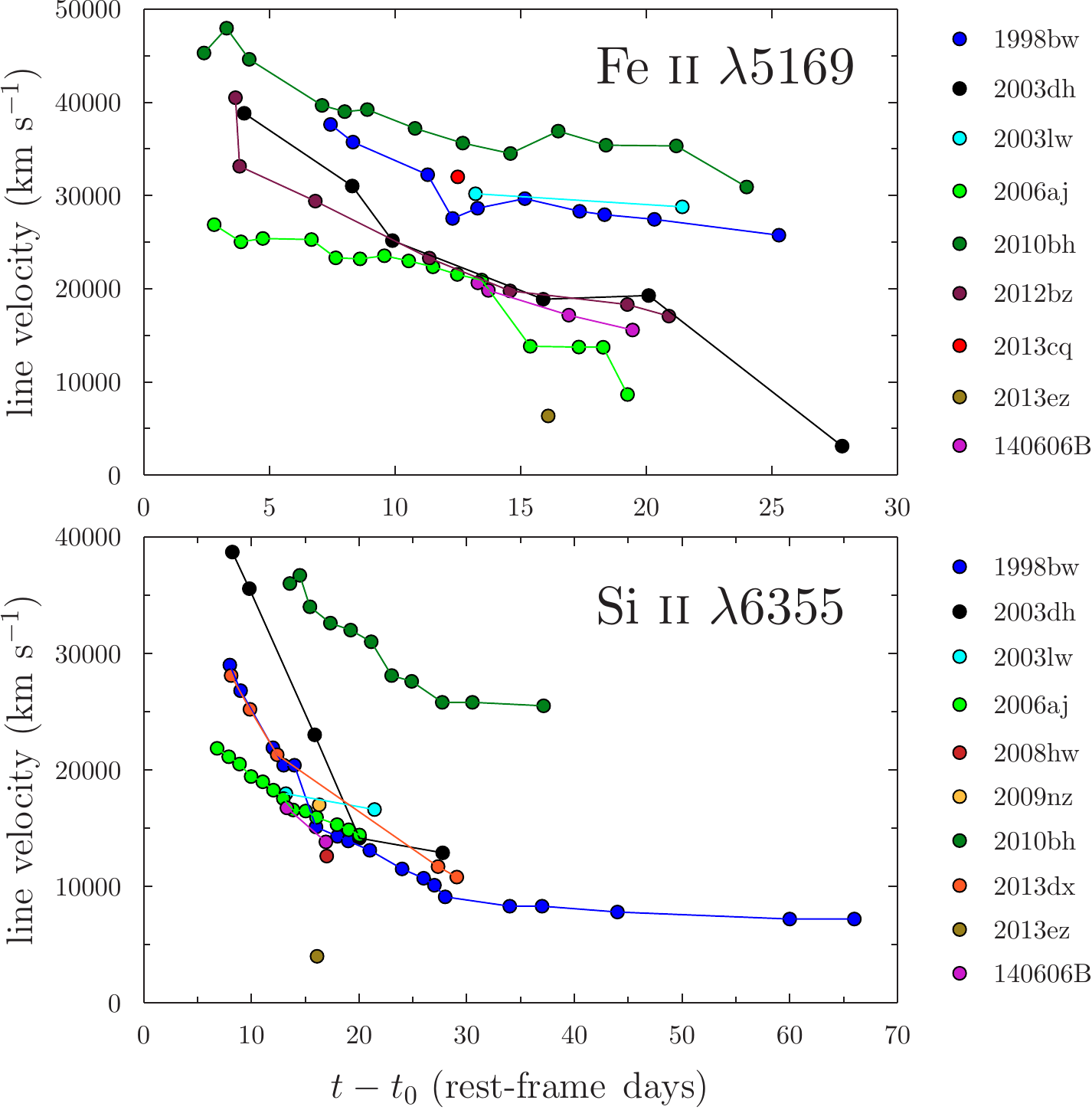}
 \caption{Measured line velocities of a sample of GRB-SNe.  See Table \ref{table:references} for their respective references.}
 \label{fig:line_vels}
\end{figure*}

Optical and NIR spectra have been obtained for more than a dozen GRB-SNe, of varying levels of quality due to their large cosmological distances.  Those of the highest quality show broad observation lines of O \textsc{i}, Ca \textsc{ii}, Si \textsc{ii} and Fe \textsc{ii} near maximum light.  The line velocities of two specific transitions (Si \textsc{ii} $\lambda$6355 and Fe \textsc{ii} $\lambda$5169; Fig. \ref{fig:line_vels}) indicate that near maximum light the ejecta that contains these elements move at velocities of order $20,000-40,000$~km~s$^{-1}$ (Fe \textsc{ii} $\lambda$5169) and about $15,000-25,000$~km~s$^{-1}$ (Si \textsc{ii} $\lambda$6355).  The weighted mean absorption velocities at peak $V$-band light of a sample of SNe IcBL that included GRB-SNe were found to be $23,800\pm9500$~km~s$^{-1}$ (Fe \textsc{ii} $\lambda$5169) by \cite{Modjaz2015} (see as well Table \ref{table:master_table_2_SN}).  SNe IcBL (including and excluding GRB-SNe) have Fe \textsc{ii} $\lambda$5169 widths that are $\sim9,000$~km~s$^{-1}$ broader than SNe Ic, while GRB-SNe appear to be, on average, about $\sim6,000$~km~s$^{-1}$ more rapid than SNe IcBL at peak light \cite{Modjaz2015}.  Si \textsc{ii} $\lambda$6355 appears to have a tighter grouping of velocities than Fe \textsc{ii} $\lambda$5169, though SN~2010bh is a notable outlier, being roughly 15,000 to 20,000~km~s$^{-1}$ more rapid than the other GRB-SNe.   SN~2013ez is also a notable outlier due to its low line velocity ($4000-6000$~km~s$^{-1}$), and inspection of its spectrum (Fig. \ref{fig:spectra_series}) reveals fewer broad features than other GRB-SNe, where it more closely resembles type Ic SNe rather than type IcBL \cite{Cano_etal_2014}.  Nevertheless, this relative grouping of line velocities may indicate similar density structure(s) in the ejecta of these SN, which in turn could indicate some general similarities in their pre-explosion progenitor configurations.  For comparison, \cite{Modjaz2015} found that the dispersion of peak SNe Ic Fe \textsc{ii} $\lambda$5169 line velocities is tighter than those measured for GRB-SNe and SNe IcBL not associated with GRBs ($\sigma = 1500, 9500, 2700$~km~s$^{-1}$, respectively).  This suggests that GRB-SNe and SNe IcBL have more diversity in their spectral velocities, and in turn their density structures, than SNe Ic.   Finally, \cite{Modjaz2015} found no differences in the spectra of $ll$GRB-SNe relative to high-luminosity GRB-SNe.

During the nebular phase of SN~1998bw (one of only a few GRB-SNe that has been spectroscopically observed during this phase due to its close proximity; see as well Section \ref{sec:geometry}), observed lines include [O \textsc{i}] $\lambda$5577, $\lambda\lambda$6300,6364; O [\textsc{ii}] $\lambda$7322; Ca \textsc{ii} $\lambda\lambda$3934,3963, $\lambda\lambda$7291,7324; Mg \textsc{i}] $\lambda$4570; Na \textsc{i} $\lambda\lambda$5890,5896; [Fe \textsc{ii}] $\lambda$4244, $\lambda$4276, $\lambda$4416, $\lambda$4458, $\lambda$4814, $\lambda$4890, $\lambda$5169, $\lambda$5261, $\lambda$5273, $\lambda$5333, $\lambda$7155, $\lambda$7172, $\lambda$7388, $\lambda$7452; [Fe \textsc{iii}] $\lambda$5270; Co \textsc{ii} $\lambda$7541; C \textsc{i}] $\lambda$8727 \cite{Mazzali2001}.  Nebular [O \textsc{i}] $\lambda\lambda$6300,6364 was also observed for nearby GRB-SNe 2006aj \cite{Maeda2007} and 100316D \cite{Bufano2012}, though in the latter case strong lines from the underlying H\textsc{ii} are considerably more dominant.  For SN~2006aj, [Ni \textsc{ii} $\lambda$7380] was tentatively detected \cite{Maeda2007}, which given the short half-life of $^{56}$Ni, implies the existence of roughly 0.05~M$_{\odot}$ of $^{58}$Ni.  Such a large amount of stable neutron-rich Ni strongly indicates the formation of a neutron star \cite{Maeda2007}.  Moreover, the absence of [Ca \textsc{ii}] lines for SN~2006aj also supported the lower kinetic energy of this event relative to other GRB-SNe, which is likely less than that attributed to a hypernova.  

\section{Phenomenological Classifications of GRB-SNe}
\label{sec:Subclasses}
\vspace*{0.5cm}

Replicating previous works \cite{Hjorth2013,Schulze2014}, in this review we divided GRB-SNe into the following sub-classes based primarily on their isotropic $\gamma$-ray luminosity $L_{\rm \gamma,iso}$:

\begin{itemize}
 \item $ll$GRB-SNe: GRB-SNe associated with low-luminosity GRBs ($L_{\rm \gamma,iso} < 10^{48.5}$~erg~s$^{-1}$).
 \item INT-GRB-SNe: GRB-SNe associated with intermediate-luminosity GRBs ($10^{48.5} < L_{\rm \gamma,iso} <10^{49.5}$~erg~s$^{-1}$).\footnote{Not to be confused with intermediate-duration GRBs, i.e. those with durations of $2-5$~s \cite{Mukherjee1998,Horvath2008,AdUP2011_INTGRB}.}
 \item GRB-SNe: GRB-SNe associated with high-luminosity GRBs ($L_{\rm \gamma,iso} > 10^{49.5}$~erg~s$^{-1}$).
 \item ULGRB-SNe: ultra-long-duration GRB-SNe, which are classified according to the exceptionally long duration of their $\gamma$-ray emission ($\sim10^4$ seconds \cite{Levan2014_ULGRB,Levan2015}) rather than on their $\gamma$-ray luminosities.
\end{itemize}

\begin{figure*}
 \centering
 \includegraphics[bb=0 0 275 182,scale=1.4,keepaspectratio=true]{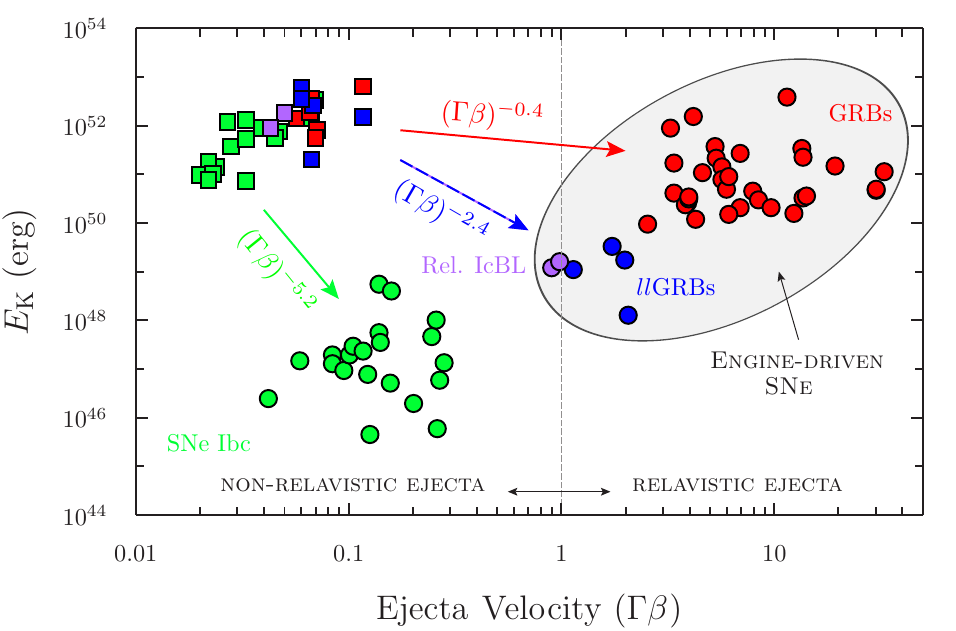}
 \caption{The positions of GRBs, SNe Ibc and GRB-SNe in the $E_{\rm K}-\Gamma\beta$ plane \cite{Xu2008,Soderberg2010,Margutti2013,Margutti2014,Cano2015}.  Ordinary SNe Ibc are shown in green, $ll$GRBs are shown in blue, relativistic SNe IcBL in purple, and jetted-GRBs in red.  Squares are used for the slow-moving SN ejecta, while circles represent the kinetic energy and velocity of the non-thermal radio-emitting ejecta associated with these events (e.g. the GRB jet). The velocities were computed for $t-t_{0}= 1$ day (rest-frame), where the value $\Gamma\beta = 1$ denotes the division between relativistic and non-relativistic ejecta.   The solid lines correspond to: (green) ejecta kinetic energy profiles of a purely hydrodynamical explosion $E_{\rm K} \propto (\Gamma\beta)^{-5.2}$ \cite{Sakurai1960,MatzMcKee1999,TanMatzMcKee2001}; (blue/purple dashed) explosions powered by a short-lived central engine (SBO-GRBs and relativistic Ic-BL SNe 2009bb and 2012ap: $E_{\rm K} \propto (\Gamma\beta)^{-2.4}$); (red) those arising from a long-lived central engine (i.e. jetted-GRBs; $E_{\rm K} \propto (\Gamma\beta)^{-0.4}$ \cite{Lazzati2012}).  Modified, with permission, from Margutti et al. \cite{Margutti2013,Margutti2014}.}
 \label{fig:ek_gammabeta}
\end{figure*}

Historically, the term X-ray flash (XRF) was used throughout the literature, which has slowly been replaced with the idiom of ``low-luminosity''.  Strictly speaking, the definition of an XRF \cite{Heise2003} arises from the detection of soft, X-ray rich events detected by the Wide Field Camera on \emph{BeppoSax} in the energy range $2-25$~keV.  Here we make no distinction based on the detection of a given satellite and instrumentation, where the ``$ll$'' nomenclature refers only to the magnitude of a given GRB's $L_{\rm \gamma,iso}$.

The luminosity, energetics and shape of the $\gamma$-ray pulse of a given GRB can reveal clues to the origin of its high-energy emission, and thus its emission process.  Of particular importance is whether the $\gamma$-rays emitted by $ll$GRBs arise from the same mechanism as high-luminosity GRBs (i.e. from a jet), or whether from a relativistic shock breakout (SBO) \cite{Kulkarni1998,Tan2001,Campana2006,Waxman2007,Cano2011a,NakarSari2012} (see as well Section \ref{sec:Theoretical_models}).  \cite{Kaneko2007,Bromberg2011} demonstrated that a key observable of $ll$GRBs are their single-peaked, smooth, non-variable $\gamma$-ray LCs compared to the more erratic $\gamma$-ray LCs of jetted-GRBs, which become softer over time.  \cite{NakarSari2012} showed that an SBO is likely present in all LGRB events, but for any realistic configuration the energy in the SBO pulse is lower by many orders of magnitude compared to those observed in the GRB prompt emission ($E_{\rm SBO} = 10^{44}-10^{47}$~erg, for reasonable estimates of the ejecta mass and progenitor radii).  These low energies (compared with $E_{\rm \gamma,iso}$) suggests that relativistic SBOs are not likely to be detected at redshifts exceeding $z\approx0.1$. In cases where they are detectable, the SBO may be in the form of a short pulse of photons with energies $>1$~MeV.  Inspection of the $E_{p}$ values in Table \ref{table:master_table_I_highE} show that only a few events have photons with peak $\gamma$-ray energies close to this value: GRB~140606B has $E_{p}\approx 800$~keV \cite{Cano2015}, however suspected $ll$GRBs 060218 and 100316D only have $E_{p}=5,30$~keV, respectively. It should be noted that while the SBO model of \cite{NakarSari2012} successfully explains the observed properties (namely the energetics, temperature and duration of the prompt emission) of GRBs 980425, 031203, 060218 and 100316D, their SBO origins are still widely debated \cite{Ghisellini2007,IrwinChev2016}, with no firm consensus yet achieved.

Thermal, blackbody (BB) components in UV and X-ray spectra have been detected for several events, including: GRB~060218 (X-ray: $kT = 0.17$~keV, time averaged from first 10,000~s, \cite{Campana2006}); GRB~100316D (X-ray: $kT = 0.14$~keV, time averaged from $144-737$~s, \cite{Starling2011}); GRB~090618 (X-ray: $kT = 0.3-1$~keV up to first 2500~s, \cite{Page2011}); GRB~101219B (X-ray: $kT = 0.2$~keV, \cite{SparreStarling2012}); and GRB~120422A (UV: $kT = 16$~eV at observer-frame $t-t_{0} = 0.054$~d, \cite{Schulze2014}).  A sample of LGRBs with associated SNe was analysed by \cite{Starling2012} who found that thermal components were present in many events, which could possibly be attributed to thermal emission arising from a cocoon that surrounds the jet \cite{Pe'er2006}, or perhaps associated with SBO emission.  \cite{SparreStarling2012} analysed a larger sample of LGRBs, and found that for several events, a model that included a BB contribution provided better fits than absorbed powerlaws.   \cite{FriisWatson2013} found that in their sample of 28 LGRBs, eight had evidence of thermal emission in their X-ray spectra, indicating such emission may be somewhat prevalent.  However, the large inferred BB temperatures ($kT$ ranging from 0.16~keV for 060218 to 3.2~keV for 061007, with an average of $\approx 1$~keV) indicates that the origin of the thermal emission may not be a SBO.  Moreover, the large superluminal expansions inferred for the thermal components instead hints at a connection with late photospheric emission.   In comparison, some studies indicate a SBO temperature of $\sim1$~keV \cite{Li2007_06aj}, while \cite{Weaver1976,Katz2010,NakarSari2012,Ohtani2013} showed that for a short while the region behind the shock is out of thermal equilibrium, and temperatures can reach as high as $\sim50$~keV.

The radius of the fitted BB component offers additional clues.  \cite{Campana2006,Waxman2007} derived a BB radius of $5-8\times10^{12}$~cm for GRB~060218; \cite{Starling2011} found $\approx 8\times10^{11}$~cm for GRB~100316D; \cite{Schulze2014} found $\approx 7\times10^{13}$~cm for GRB~120422A; and \cite{Singer2015} derived a radius of $\approx9\times10^{13}$~cm for GRB~140606B.  The radii inferred for GRBs 060218, 120422A and 140606B are commensurate with the radii of red supergiants ($200-1500$~R$_{\odot}$), while that measured for GRB~100316D is similar to that of the radius of a blue supergiant ($\le25$~R$_{\odot}$).  These radii, which are much larger than those expected for Wolf-Rayet (WR) stars (of order a few solar radius to a few tens of solar radii) were explained by these authors by the presence of a massive, dense stellar wind surrounding the progenitor star, where the thermal radiation is observed once the shock, which is driven into the wind, reaches a radius where the wind becomes optically thin.  An alternative explanation for the large BB radii was presented by \cite{Margutti2015}, (see as well \cite{Nakar2015}), where the breakout occurs in an extended ($R=100$~R$_{\odot}$) low-mass (0.01~M$_{\odot}$) envelope surrounding the pre-explosion progenitor star. The origin of envelope is likely material stripped just prior to explosion, and such an envelope is missing for high-luminosity GRB-SNe \cite{Nakar2015}.

For a given GRB-SN event there are both relativistic and non-relativistic ejecta, where the former is responsible for producing the prompt emission, and the latter is associated with the SN itself.  The average mass between the two components is large: the ejecta mass of a GRB-SN is of order $2-8$~M$_{\odot}$, while that in the jet that produces the $\gamma$-rays is of order $10^{-6}$~M$_{\odot}$, based on arguments for very low baryon loading \cite{ShemiPiran1990}.  A GRB jet decelerates very rapidly, within a few days, because the very low-mass ejecta is rapidly swept up into the comparatively larger mass of the surrounding CSM.  Conversely, SNe have much heavier ejecta and can be in free-expansion for many years or even centuries.  Measuring the amount of kinetic energy associated with each ejecta component can offer additional clues to the explosion mechanisms operating in these events. Fig. \ref{fig:ek_gammabeta} shows the position of SNe Ibc (green), GRBs (red), $ll$GRBs (blue) and relativistic SNe IcBL (purple) in the $E_{\rm K}-\Gamma \beta$ plane \cite{Xu2008,Soderberg2010,Margutti2013,Margutti2014,Cano2015}, where $\beta=v/c$ (not to be confused with the spectral PL index of synchrotron radiation) and $\Gamma$ is the bulk Lorentz factor.  Squares indicate slow-moving SN ejecta, while circles represent the kinetic energy and velocity of the non-thermal radio-emitting ejecta associated with these events (e.g. the jet in GRBs). The velocities were computed for $t-t_{0}= 1$ day (rest-frame), where the value $\Gamma\beta = 1$ denotes the division between relativistic and non-relativistic ejecta.   The solid lines show the ejecta kinetic energy profiles of a purely hydrodynamical explosion (green) $E_{\rm K} \propto (\Gamma\beta)^{-5.2}$ \cite{Sakurai1960,MatzMcKee1999,TanMatzMcKee2001}; explosions powered by a short-lived central engine (blue/purple dashed), SBO-GRBs and relativistic Ic-BL SNe 2009bb and 2012ap: $E_{\rm K} \propto (\Gamma\beta)^{-2.4}$); and those arising from a long-lived central engine (red), i.e. jetted-GRBs; $E_{\rm K} \propto (\Gamma\beta)^{-0.4}$ \cite{Lazzati2012}.

\begin{figure}
 \centering
 \includegraphics[bb=0 0 268 175,scale=1.43,keepaspectratio=true]{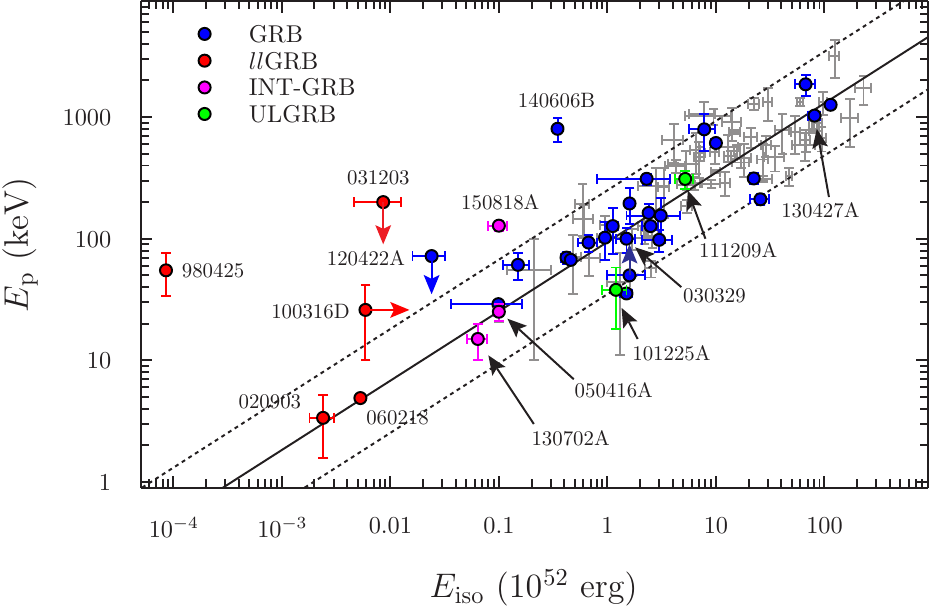}
 \caption{Properties of the prompt emission for different classes of GRBs in the $E_{\rm \gamma,iso}-E_{\rm p}$ plane \cite{Amati2002}.  Data from \cite{Amati2002,Amati2007,Amati2008} are shown in grey along with their best fit to a single powerlaw (index of $\alpha=0.57$) and the 2$\sigma$ uncertainty in their fit. Notable events that do not appear to follow the Amati relation include $ll$GRBs (980425 and 031203), INT-GRBs (150818A), and high-luminosity GRB 140606B. Both ULGRBs are consistent with the Amati relation, as are GRBs 030329 and 130427A, while GRB~120422A and $ll$GRB~100316D are marginally consistent.}
 \label{fig:Amati}
\end{figure}

It is seen that $ll$GRBs and high-luminosity GRBs span a wide range of engine energetics, as indicated by the range of PL indices seen in Fig. \ref{fig:ek_gammabeta}. The two relativistic SNe IcBL considered in this review occur at the lower-end of central engine energetics.  Modelling of GRB~060218 \cite{Soderberg2006} showed that $\sim10^{48}$~erg of energy was coupled to the mildly relativistic ejecta ($\Gamma \sim 2$).  \cite{Margutti2013} showed the presence of a very weak central engine for GRB~100316D, where $\sim10^{49}$~erg of energy was coupled to mildly relativistic ($\Gamma =1.5-2$), quasi-spherical ejecta.  \cite{Soderberg2010} showed that $\ge10^{49}$~erg was associated with the relativistic ($v=0.9c$), radio-emitting ejecta of SN~2009bb.  These authors also showed that, unlike GRB jets, the ejecta was in free-expansion, which implied it was baryon loaded.  For SN~2012ap, \cite{Chakraborti2015} estimated there was $\sim1.6\times10^{49}$~erg of energy associated with the mildly relativistic ($0.7c$) radio-emitting ejecta.  The weak X-ray emission of SN~2012ap \cite{Margutti2014} implied no late-time activity of its central engine, which led these authors to suggest that relativistic SNe IcBL represent the weakest engine-driven explosions, where the jet is unable to successfully breakout of the progenitor.  $ll$GRBs then represent events where the jet does not, or just barely escapes into space.  Note that \cite{Zhang2012_2012bz} calculated an estimate to the dividing line between SBO-GRBs and jet-GRBs, finding that for $\gamma$-ray luminosities above $10^{48}$~erg~s$^{-1}$ a jet-GRB may be possible.

Next, the distribution of $T_{90}$ (the time over which a burst emits from 5\% of its total measured counts to 95\%) as measured by the various GRB satellites can be used to infer additional physical properties of the GRB jet duration and progenitor radii.  A basic assertion of the collapsar model is that the duration of the GRB prompt phase (where $T_{90}$ is used as a proxy) is the difference between the time that the central engine operates minus the time it takes for the jet to breakout of the star: $T_{90} \sim t_{\rm engine} - t_{\rm breakout}$.  A direct consequence of this premise is that there should be a plateau in the distribution of $T_{90}$ for GRBs produced by collapsars when $T_{90}$ $<$ $t_{\rm breakout}$ \cite{Bromberg2012}.  Moreover, the value of $T_{90}$ found at the upper-limit of the plateaus seen for three satellites (\emph{BATSE}, \emph{Swift} and \emph{Fermi}) was approximately the same ($T_{90}\sim 20-30$~s), which is interpreted as the typical breakout time of the jet.  This short breakout time suggests that the progenitor star at the time of explosion is quite compact ($\sim 5$ $R_{\odot}$ \cite{Piran2013}).  \cite{Bromberg2013} then used these distributions to calculate the probability that a given GRB arises from a collapsar or not based on its $T_{90}$ and hardness ratio.  Note however, that $T_{90}$ might not always be the best indicator of the engine on-time.  For example, \cite{Zhang2012_2012bz} showed that while GRB~120422A had $T_{90} = 5$~s, the actual duration of the jet was actually 86~s, as constrained by modelling of the curvature effect\footnote{Though see \cite{KumarPana2000} who state that curvature radiation is not from a central engine that is still on but from electrons that were off-axis, and hence had a lower Lorentz factor, and which are received over a time interval that is long compared to the duration of the burst.}.

Finally, Fig. \ref{fig:Amati} shows the properties of the prompt emission for the various GRB-SN sub-classes in the $E_{\rm \gamma,iso}-E_{\rm p}$ plane, i.e. the Amati relation \cite{Amati2002}.  Data from \cite{Amati2002,Amati2007,Amati2008} are shown in grey along with their best fit to a single powerlaw (index of $\alpha=0.57$) and the 2$\sigma$ uncertainty in their fit. Several events do not appear to follow the Amati relation, including $ll$GRBs 980425 and 031203, INT-GRB~150818A, and high-luminosity GRB 140606B. Both ULGRBs are consistent with the Amati relation, as are GRBs 030329 and 130427A, while GRB~120422A and $ll$GRB~100316D are marginally consistent.  It was once supposed that the placement of a GRB in the $E_{\rm \gamma,iso}-E_{\rm K}$ plane could be a discriminant of a GRB's origins, where it is seen that SGRBs also do not follow the Amati relation.  However, over the years many authors have closely scrutinized the Amati relation, with opinions swinging back and forth as to whether it reflects a physical origin or is simply due to selection effects \cite{NakarPiran2005,Firmani2009,ShahNem2010,ShahNem2011,Ghirlanda2010,Ghirlanda2011,Collazzi2012,Preece2016}.  To date, no consensus has yet been reached.

\section{Physical Properties -- Observational Constraints}
\label{sec:Physical_Properties}
\vspace*{0.5cm}

\subsection{Bolometrics}

\begin{figure}
 \centering
 \includegraphics[bb=0 0 333 174,scale=1.4,keepaspectratio=true]{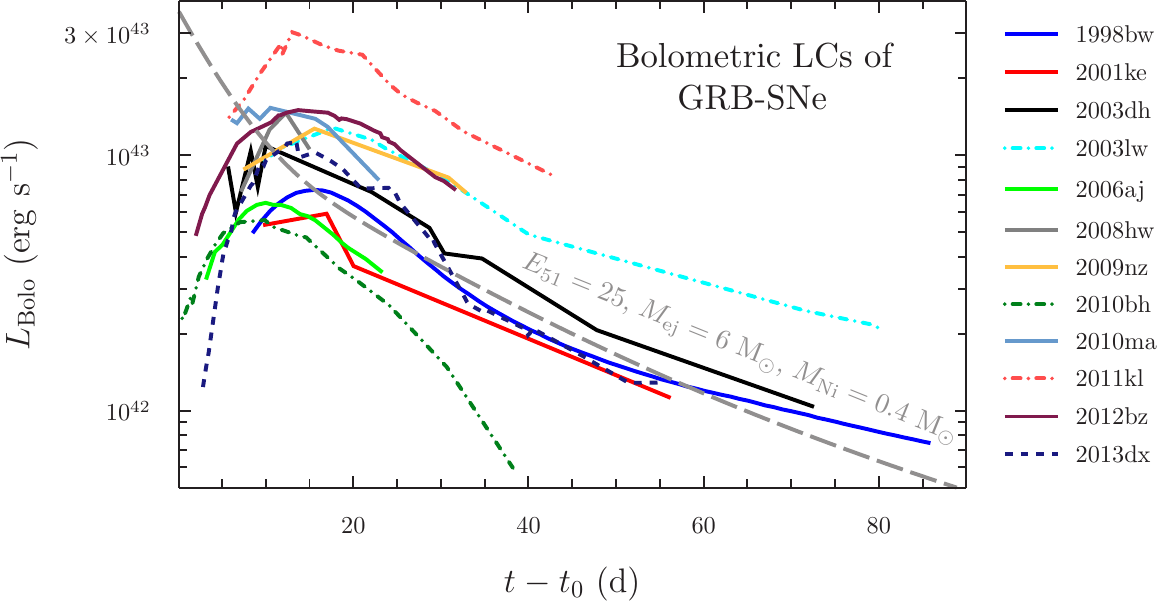}
 \caption{Bolometric LCs of a sample of GRB-SNe. Times are given in the rest-frame.  The average peak luminosity of all GRB-SNe except SN~2011kl is $\bar L_{\rm p} = 1.0\times10^{43}$~erg~s$^{-1}$, with a standard deviation of $0.36\times10^{43}$~erg~s$^{-1}$.  The peak luminosity of SN~2011kl is $L_{\rm p} = 2.9\times10^{43}$~erg~s$^{-1}$, which makes it more than 5$\sigma$ more luminous than the average GRB-SN.  The average peak time of the entire sample is $t_{\rm p} = 13.2$~d, with a standard deviation of 2.6~d.  If SN~2011kl is excluded from the sample, this changes to 13.0~d.   Plotted for reference is an analytical model that considers the luminosity produced by the average GRB-SN ($E_{\rm K} = 25\times10^{51}$~erg, $M_{\rm ej}=6$~M$_{\odot}$, and $M_{\rm Ni}=0.4$~M$_{\odot}$).}
 \label{fig:bolo_LCs}
\end{figure}

The bolometric LCs of a sample of 12 GRB-SNe, which includes $ll$GRB-SNe and ULGRB-SN~2011kl, are shown in Fig. \ref{fig:bolo_LCs}.  The Bazin function was fit to the GRB-SN bolometric LCs\footnote{Note that SNe 2001ke, 2008hw and 2009nz were excluded from the fitting, and the subsequent calculated averages, as their bolometric LCs contained too few points to be fit with the Bazin function, which has four free parameters.  As such, their luminosities and peak times were approximated by eye and are not included in the average GRB-SN properties presented here.}  in order to determine their peak luminosity ($L_{\rm p}$), the time of peak luminosity ($t_{\rm p}$) and the amount the bolometric LC fades from peak to 15 days later ($\Delta m_{15}$).  These values are presented in Table \ref{table:master_table_2_SN}.


The average peak luminosity of the GRB-SN sample, excluding SN~2011kl, is $\bar L_{\rm p} = 1.0\times10^{43}$~erg~s$^{-1}$, with a standard deviation of $\sigma_{L_{\rm p}} = 0.4\times10^{43}$~erg~s$^{-1}$.  The peak luminosities of SNe 2003dh and 2013dx are $\approx 1\times10^{43}$~erg~s$^{-1}$, meaning that they are perhaps better representatives of a typical GRB-SN than the archetype SN~1998bw ($L_{\rm p} = 7\times10^{42}$~erg~s$^{-1}$).  The peak luminosity of SN~2011kl is $L_{\rm p} = 2.9\times10^{43}$~erg~s$^{-1}$, which makes it more than 5$\sigma$ more luminous than the average GRB-SN.  This is not, however, as bright as superluminous supernovae (SLSNe), whose luminosities exceed $>7\times10^{43}$~erg~s$^{-1}$ \cite{GalYam2012}.  This makes SN~2011kl an intermediate SN event between GRB-SNe and SLSNe, and perhaps warrants a classification of a ``superluminous GRB-SNe'' (SLGRB-SN); however, in this chapter we will stick with the nomenclature ULGRB-SN.  When SN~2011kl is included in the sample, $\bar L_{\rm p} = 1.2\times10^{43}$~erg~s$^{-1}$, with $\sigma_{\bar L_{\rm p}} = 0.7\times10^{43}$~erg~s$^{-1}$.  Even using this average value, SN~2011kl is still 2.5$\sigma$ more luminous than the average GRB-SN.  

The average peak time, when SN~2011kl is and is not included in the sample, is $t_{\rm p} = 13.2$~d ($\sigma_{t_{\rm p}} = 2.6$~d) and $t_{\rm p} = 13.0$~d ($\sigma_{t_{\rm p}} = 2.7$~d), respectively.  Similarly, $\Delta m_{15} = 0.7$~mag ($\sigma_{\Delta m_{15}} = 0.1$~mag), and 0.8~mag ($\sigma_{\Delta m_{15}} = 0.1$~mag), respectively.  As such, the inclusion/exclusion of SN~2011kl has little effect on these derived values.  The fact that SN~2011kl peaks at a similar time as the average GRB-SN, but does so at a much larger luminosity, strongly suggests that ULGRB-SNe do not belong to the same class of standardizable candles as GRB-SNe.  This can be readily explained in that SN~2011kl is powered by emission from a magnetar central engine \cite{Greiner2015,Metzger2015,CJM2016,Bersten2016}, whereas GRB-SNe, including $ll$GRB-SNe, are powered by radioactive heating \cite{CJM2016}.  Whether ULGRB-SNe represent the same set of standardizable candles as type I SLSNe \cite{InserraSmartt2014,Wei2015} which are also thought to be powered by a magnetar central engine, their own sub-set, or perhaps none at all, requires additional well-monitored events.

Over the years, and since the discovery of SN~1998bw, the bolometric properties (kinetic energy, $E_{\rm K}$, ejecta mass, $M_{\rm ej}$, and nickel mass, $M_{\rm Ni}$) of the best-observed GRB-SNe have been determine by sophisticated numerical simulations (hydrodynamical models coupled with radiative transfer codes) \cite{Iwamoto1998,Woosley1999,Hoflich1999,Iwamoto1999,Mazzali2001,Nakamura2001,Maeda2002,Mazzali2003,WoosleyHeger2003,Deng2005,Maeda2006,Mazzali2006,Nagataki2006,Maeda2007,Greiner2015,Bersten2016}, and analytical modelling \cite{Maeda2003_2comp_model,Cano2011b,Olivares2012,Cano2013,Xu2013,Cano_etal_2014,Schulze2014,Cano2015,Lyman2016,Toy2016,CJM2016}.  A summary of the derived bolometric properties for individual GRB-SNe are presented in Table \ref{table:master_table_2_SN}, while a summary of the average bolometric properties, broken down by GRB-SN sub-type and compared against other subtypes of SNe Ibc, is shown in Table \ref{table:Bolo_summary}.  It should be noted that the values presented have been derived over different wavelength ranges: some are observer-frame $BVRI$, while others include UV, $U$-band and NIR contributions.  Further discussion on the effects of including additional filters when constructing a bolometric LC of a given SN can be found in \cite{Modjaz2009,Cano2011a,Bufano2012,Schulze2014,Lyman2016}, who show that including NIR flux leads to brighter bolometric LCs that decay slower at later times, and including UV flux leads to an increase in luminosity at earlier times (during the first couple of weeks, rest-frame) when the UV contribution is non-negligible.


\begin{table}
\scriptsize
\centering
\setlength{\tabcolsep}{2.1pt}
\setlength{\extrarowheight}{3pt}
\caption{Average Bolometric Properties of GRB-SNe}
\label{table:Bolo_summary}
\begin{tabular}{|r|ccc|ccc|ccc||ccc|ccc|ccc|ccc|}
\hline
	&		&		&			&		&	(M$_{\odot}$)	&			&		&	(M$_{\odot}$)	&			&		&			&		&		&	(d, rest)		&		&		&	(mag)	&		&		&	(km s$^{-1}$)	&		\\
type$^{*}$	&	$N$	&	$E^{\dagger}_{\rm K}$	&	$\sigma$	&	$N$	&	$M_{\rm ej}$	&	$\sigma$	&	$N$	&	$M_{\rm Ni}$	&	$\sigma$	&	$N$	&	$L^{\ddagger}_{\rm p}$	&	$\sigma$	&	$N$	&	$t_{\rm p}$	&	$\sigma$	&	$N$	&	$\Delta m_{15}$	&	$\sigma$	&	$N$	&	$v_{\rm ph}$	&	$\sigma$	\\
\hline																																											
GRB	&	19	&	26.0	&	18.3	&	19	&	5.8	&	4.0	&	20	&	0.38	&	0.13	&	2	&	1.26	&	0.35	&	2	&	12.28	&	0.67	&	2	&	0.85	&	0.21	&	6	&	18400	&	9700	\\
INT	&	1	&	8.2	&	-	&	1	&	3.1	&	-	&	1	&	0.37	&	-	&	1	&	1.08	&	-	&	1	&	12.94	&	-	&	1	&	0.85	&	-	&	1	&	21300	&	-	\\
LL	&	6	&	27.8	&	19.6	&	6	&	6.5	&	4.0	&	6	&	0.35	&	0.19	&	5	&	0.94	&	0.41	&	5	&	13.22	&	3.53	&	5	&	0.75	&	0.12	&	4	&	22800	&	8200	\\
ULGRB	&	2	&	18.8	&	18.7	&	2	&	6.1	&	2.9	&	1	&	0.41	&	-	&	1	&	2.91	&	-	&	1	&	14.80	&	-	&	1	&	0.78	&	-	&	1	&	21000	&	-	\\
Rel IcBL	&	2	&	13.5	&	6.4	&	2	&	3.4	&	1.0	&	2	&	0.16	&	0.05	&	2	&	0.35	&	0.50	&	2	&	12.78	&	0.84	&	2	&	0.90	&	0.21	&	2	&	14000	&	1400	\\
GRB ALL	&	28	&	25.2	&	17.9	&	28	&	5.9	&	3.8	&	28	&	0.37	&	0.20	&	9	&	1.24	&	0.71	&	9	&	13.16	&	2.61	&	9	&	0.79	&	0.12	&	12	&	20300	&	8100	\\
\hline																																											
GRB ALL**	&	27	&	25.9	&	17.9	&	27	&	5.9	&	3.9	&	28	&	0.37	&	0.20	&	8	&	1.03	&	0.36	&	8	&	12.95	&	2.72	&	8	&	0.79	&	0.13	&	11	&	20200	&	8500	\\
\hline																																											
Ib	&	19	&	3.3	&	2.6	&	19	&	4.7	&	2.8	&	12	&	0.21	&	0.22	&	-	&	-	&	-	&	-	&	-	&	-	&	-	&	-	&	-	&	11	&	8000	&	1700	\\
Ic	&	13	&	3.3	&	3.3	&	13	&	4.6	&	4.5	&	7	&	0.23	&	0.19	&	-	&	-	&	-	&	-	&	-	&	-	&	-	&	-	&	-	&	10	&	8500	&	1800	\\
\hline					
\end{tabular}
\begin{flushleft}
$^{*}$ Classifications (Section \ref{sec:Subclasses}): $ll$GRBs: GRB-SNe associated with low-luminosity GRBs ($L_{\rm \gamma,iso} < 10^{48.5}$~erg~s$^{-1}$); INT-GRBs: GRB-SNe associated with intermediate-luminosity GRBs ($10^{48.5} < L_{\rm \gamma,iso} <10^{49.5}$~erg~s$^{-1}$); GRBs: GRB-SNe associated with high-luminosity GRBs ($L_{\rm \gamma,iso} > 10^{49.5}$~erg~s$^{-1}$); ULGRBs: GRB-SNe associated with ultra-long-duration GRBs (see Section \ref{sec:Subclasses}).\\
$^{**}$ Excluding SN~2011kl. \\ 
$^{\dagger}$ units: $10^{51}$ erg\\
$^{\ddagger}$ units: $10^{43}$~erg~s$^{-1}$\\
NB: Average bolometric properties of SNe Ib and Ic are from \cite{Cano2013}.\\
\end{flushleft}
\end{table}

From this sample of $N=28$ GRB-SNe we can say that the average GRB-SN (grey dashed line in Fig. \ref{fig:bolo_LCs}) has a kinetic energy of $E_{\rm K} = 2.5\times10^{52}$~erg ($\sigma_{E_{\rm K}} = 1.8\times10^{52}$~erg), an ejecta mass of $M_{\rm ej} = 6$~M$_{\odot}$ ($\sigma_{M_{\rm ej}} = 4$~M$_{\odot}$), a nickel mass of $M_{\rm Ni} = 0.4$~M$_{\odot}$ ($\sigma_{M_{\rm Ni}} = 0.2$~M$_{\odot}$) and a peak photospheric velocity\footnote{Here we make the assumption that the line velocities of various transitions, namely Fe \textsc{ii} $\lambda$5169 and Si \textsc{ii} $\lambda$6355 are suitable proxies for the photospheric velocities.  An in-depth discussion of this assumption and its various caveats can be found in \cite{Modjaz2015}.} of $v_{\rm ph} = 20,000$~km~s$^{-1}$ ($\sigma_{v_{\rm ph}} = 8,000$~km~s$^{-1}$).  It has a peak luminosity of $L_{\rm p} = 1\times10^{43}$~erg~s$^{-1}$ ($\sigma_{L_{\rm p}} = 0.4\times10^{43}$~erg~s$^{-1}$), reaches peak bolometric light in $t_{\rm p} = 13$~days ($\sigma_{t_{\rm p}} = 2.7$~days), and has $\Delta m_{15} = 0.8$~mag ($\sigma_{\Delta m_{15}} = 0.1$~mag).  There are no statistical differences in the average bolometric properties, rise times and decay rates, between the different GRB-SN subtypes, and excluding ULGRB-SN~2011kl, there are no differences in their peak luminosities.  As found in previous studies \cite{Cano2013,Lyman2016}, relativistic SNe IcBL are roughly half as energetic as GRB-SNe, and contain approximately half as much ejecta mass and nickel content therein.  However, we are comparing GRB-SNe against a sample of two SNe, meaning we should not draw any firm conclusions as of yet.  

A few caveats to keep in mind when interpreting these results.  The first is the comparison of bolometric properties derived for SNe observed over different filter/wavelength ranges, as discussed above.  Secondly, for a GRB-SN to be observed there are several stringent requirements \cite{WoosleyBloom2006}, including AGs that fade at a reliably determined rate (for example, for GRB~030329 the complex AG behaviour led to a range of 1~mag in the peak brightness of accompanying SN~2003dh \cite{Hjorth2003,Stanek2003,Matheson2003}; thus in this case the reported peak brightness was strongly model-dependent), has a host galaxy that can be readily quantified, and to be at a relatively low redshift ($z\le1$ for current 10-m class ground telescopes and \emph{HST}).  Moreover, the modelling techniques used to estimate the bolometric properties contain their own caveats and limitations.  For example, the analytical Arnett model \cite{arnett1982} contains assumptions such as spherical symmetry, homogeneous ejecta distribution, homologous expansion, and a central location for the radioactive elements \cite{Cano2013}.


\subsection{What powers a GRB-SN?}
\label{sec:powering_GRBSNe}

Observations of GRB-SNe can act as a powerful discriminant of the different theoretical models proposed to produce them.  The analysis presented in the previous section made the assumption that GRB-SNe are powered by radioactive heating.  In this scenario, it is assumed that during the initial core-collapse (see Section \ref{sec:Theoretical_models} for further discussion), roughly 0.1~M$_{\odot}$ or so of nickel can be created via explosive nucleosynthesis \cite{MaedaTominaga2009} if the stellar material has nearly equal amounts of neutrons and protons (such as silicon and oxygen), and approximately $10^{52}$~erg of energy is focused into $\sim1$\% of the star, which occurs in the region between the newly formed compact object and $4\times10^9$~cm \cite{MacFadyen2001}.  In this scenario, temperatures in excess of $4\times10^9$~K can be attained.  However, the precise amount of $^{56}$Ni that is generated is quite uncertain, and depends greatly on how much the star has expanded (or collapsed), prior to energy deposition. The radioactive nickel decays into cobalt with a half-life of 6.077~d, and then cobalt into iron with a half-life of 77.236~d: $^{56}_{28}$Ni~$\rightarrow$~$^{56}_{27}$Co~$\rightarrow$ $^{56}_{26}$Fe \cite{arnett1982,WoosleyWeaver1986}.  Given its short half-life, the synthesized nickel must be generated during the explosion itself, and not long before core-collapse.  Gamma-rays that are emitted during the different radioactive decay processes are thermalised in the optically thick SN ejecta, which heat the ejecta that in turn radiates this energy at longer wavelengths (optical and NIR).  This physical process is expected to power other types of SNe, including all type I SNe (Ia, Ib, Ic and type Ic SLSNe), and the radioactive tail of type IIP SNe.  

\begin{figure*}
 \centering
 \includegraphics[bb=0 0 267 176,scale=1.43,keepaspectratio=true]{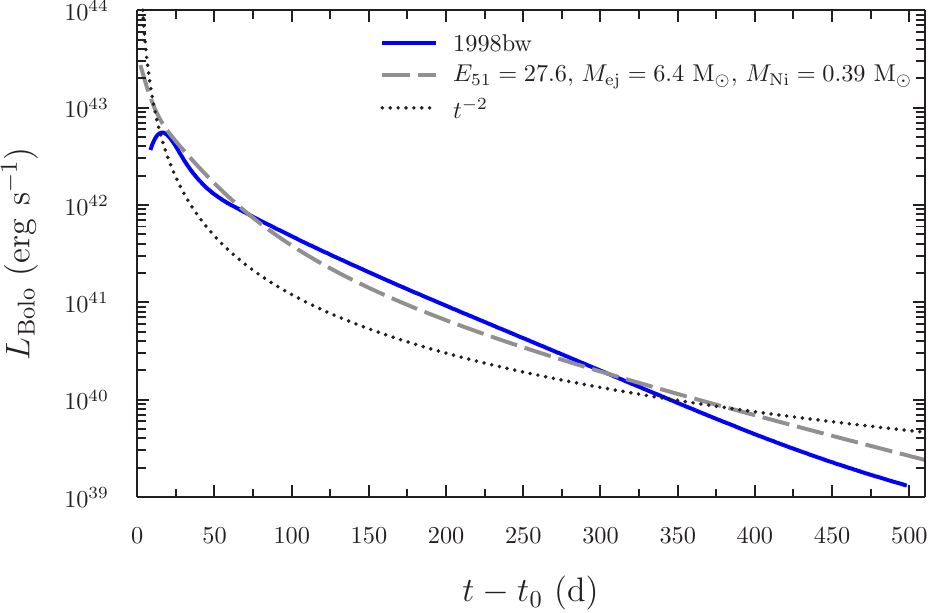}
 \caption{Late-time bolometric LC of SN~1998bw in filters $BVRI$.  Two analytical models have been plotted to match the peak luminosity: (1)  a single-zone analytical model for a fiducial SN that is powered by radioactive heating, where $E_{\rm K} = 27.6\times10^{51}$~erg, an ejecta mass of $M_{\rm ej} = 6.4$~M$_{\odot}$, a nickel mass of $M_{\rm Ni} = 0.39$~M$_{\odot}$, and (2) a $t^{-2}$ curve, which is the expected decay rate for luminosity powered by a magnetar central engine.  At late times the decay rate of model (1) provides a much better fit than the $t^{-2}$ decay, which grossly over-predicts the bolometric luminosity at times later than 400~d. This is one line of observational evidence that GRB-SNe are powered by radioactive heating, and not via dipole-extracted radiation from a magnetar central engine (i.e. a magnetar-driven SN).}
 \label{fig:1998bw_bolo_late}
\end{figure*}

Observationally, there are hints that suggest that the best-observed GRB-SNe are powered, at least in part \cite{Fryer2007}, by radioactive heating.  At late times, the decay of $^{56}$Co leads to an exponential decline in the nebular-phase bolometric LC of type I SNe.  An example of this is the grey-dashed line in Fig. \ref{fig:bolo_LCs}, which is an analytical model \cite{Maeda2003_2comp_model,Takaki2013} that considers the the luminosity produced by a fiducial SN with a kinetic energy of $E_{\rm K} = 25\times10^{51}$~erg, an ejecta mass of $M_{\rm ej}=6$~M$_{\odot}$ and a nickel mass of $M_{\rm Ni}=0.4$~M$_{\odot}$ (e.g. the ``average'' GRB-SN).  Such a model, and others of this ilk, assume full trapping of the emitted $\gamma$-rays and thermalised energy.  For comparison, the late-time LC of SN~1998bw appears to fade more rapidly than this, presumably because some of the $\gamma$-rays escape directly into space without depositing energy into the expanding ejecta.  At times later than 500~d \cite{Sollerman2002,Clocchiatti2011}, the observed flattening seen in the LC can be interpreted both in terms of more of the energy and $\gamma$-rays being retained in the ejecta, and more energy input from the radioactive decay of species in addition to cobalt.

In the collapsar model, there are additional physical processes that can lead to the creation of greater masses of radioactive nickel.  One potential source of $^{56}$Ni arises from the wind emitted by the accretion disk surrounding the newly formed black hole (BH).  According to the numerical simulations of \cite{MacFadWoosley1999}, the amount of generated nickel depends on the accretion rate as well as the viscosity of the inflow.  In theory, at least, the only upper bound on the amount of nickel that can be synthesized by the disk wind is the mass of material that is accreted.  In an analytical approach, \cite{Milosavljevic2012} demonstrated that enough $^{56}$Ni can be synthesized (in order to match observations of GRB-SNe), over the course of a few tens of seconds, in the convective accretion flow arising from the initial circularization of the infalling envelope around the BH.

In the millisecond magnetar model, it is more difficult to produce a sufficient amount of $^{56}$Ni via energy injection from a central engine.  Some simulations suggest that only a a few hundredths of a solar mass of nickel can be synthesized in the magnetar model \cite{BarkovKom2011}.  However it may be possible to generate more nickel either by tapping into the initial rotational energy of the magnetar via magnetic stresses, thus enhancing the shocks induced by the collision of the energetic wind emanated by the magnetar with material already processed by the SN shock \cite{Thompson2007,Thompson2010}.  Another route would be via a shock wave driven into the ejecta by the magnetar itself, which for certain values of $P$ and $B$ could generate the required nickel masses \cite{SuwaTominaga2015}.  However, in this scenario an isotropic-equivalent energy input rate of more than 10$^{52}$~erg is required, and the subsequent procurement of additional nickel mass via explosive nucleosynthesis will inevitably lead to a more rapid spin-down of the magnetar central engine, rendering it unable to produce energy input during the AG phase.  It is also worth considering that if a magnetar (and the subsequent GRB) is formed via the accretion-induced collapse of a white dwarf star, or perhaps the merger of two white dwarfs, there is no explosive nucleosynthesis and thus a very low $^{56}$Ni yield \cite{Metzger2007}.  

The uncertainties unpinning both models mean that neither can be ruled out at this time, though perhaps the collapsar model offers a slightly easier route for producing the necessary quantity of nickel needed to explain the observed luminosities of GRB-SNe.  But what if GRB-SNe are not powered by radioactive heating, but instead via another mechanism?  Could instead, GRB-SNe be powered by a magnetar central engine \cite{OstrikerGunn1971,ZhangMesz2001}, as has been proposed for some type I SLSNe \cite{Chatzopoulos2011,Inserra2013,Nicholl2013}?  A prediction of the magnetar-driven SN model is that at late-times the bolometric LC should decay as $t^{-2}$ \cite{OstrikerGunn1971,ZhangMesz2001,KasenBildsten2010,Woosley2010,BarkovKom2011,Chatzopoulos2011,CJM2016}).  Plotted in Fig. \ref{fig:1998bw_bolo_late} is the bolometric LC of SN~1998bw to $t-t_{0} = 500$~d.  Overplotted are two analytical models: (1) a single-zone analytical model for a fiducial SN that is powered by radioactive heating, where $E_{\rm K} = 27.6\times10^{51}$~erg, an ejecta mass of $M_{\rm ej} = 6.4$~M$_{\odot}$, a nickel mass of $M_{\rm Ni} = 0.39$~M$_{\odot}$, and (2) a $t^{-2}$ curve (i.e. the decay rate expected for luminosity powered by a magnetar central engine).  Both have been fitted to the bolometric LC of SN~1998bw to match its peak luminosity.  At late times the decay rate of the radioactive-heated analytical LC provides a much better fit than the $t^{-2}$ decay, which grossly over-predicts the bolometric luminosity at times later than 400~d.  The difference between observations and the radioactive-decay model can be attributed to incomplete trapping of $\gamma$-rays produced during the radioactive decay process.

However, it appears that not all GRB-SNe sub-types are powered by radioactive heating.  Several investigations have provided compelling evidence that ULGRB~111209A / SN~2011kl was powered instead by a magnetar central engine.  \cite{Greiner2015} showed that SN~2011kl could not be powered entirely (or at all) by radioactive heating.  Their argument was based primarily on the fact that the inferred ejecta mass ($3.2\pm0.5$~M$_{\odot}$), determined via fitting the Arnett model \cite{arnett1982} to their constructed bolometric LC, was too low for the amount of nickel needed to explain the observed bolometric luminosity ($1.0\pm0.1$~M$_{\odot}$).  The ratio of $\frac{\rm M_{Ni}}{\rm M_{ej}} = 0.3$ was much larger than that inferred for the general GRB-SN population ($\frac{\rm M_{Ni}}{\rm M_{ej}} \approx 0.07$; \cite{Cano2013}), which rules against radioactive heating powering SN~2011kl.  Secondly, the shape and relative brightness of an optical spectrum obtained of SN~2011kl just after peak SN light ($t-t_{0} = 20$~d, rest-frame) was entirely unlike the spectra observed for GRB-SNe (Fig. \ref{fig:spectra_series}), including SN~1998bw \cite{Patat2001}.  Instead, the spectrum more closely resembled those of SLSNe in its shape, including the sharp cut-off at wavelengths bluewards of 3000~\AA.  Several authors \cite{Greiner2015,Metzger2015,Bersten2016,CJM2016,Mazzali2016} modelled different phases of the entire ULGRB event to determine the ejecta mass ($M_{\rm ej}$), initial spin period ($P$), and the initial magnetic field strength ($B$), with some general consensus among the derived values: $M_{\rm ej} = 3-5$~M$_{\odot}$ (for various values of the assumed grey opacity), $P = 2-11$~ms, and $B = 0.4-2\times10^{15}$~G.  Note that some models assumed additional heating from some nucleosynthesised nickel (0.2~M$_{\odot}$ \cite{Metzger2015,Bersten2016}), while \cite{CJM2016} assumed that energy injection from the magnetar central engine was solely responsible for powering the entire event.  The general consensus of all the modelling approaches is that SN~2011kl was not powered entirely by radioactive heating, and additional energy, likely arising from a magnetar central engine, was needed to explain the observations of this enigmatic event.

\begin{figure}
 \centering
 \includegraphics[bb=0 0 677 654,scale=0.4,keepaspectratio=true]{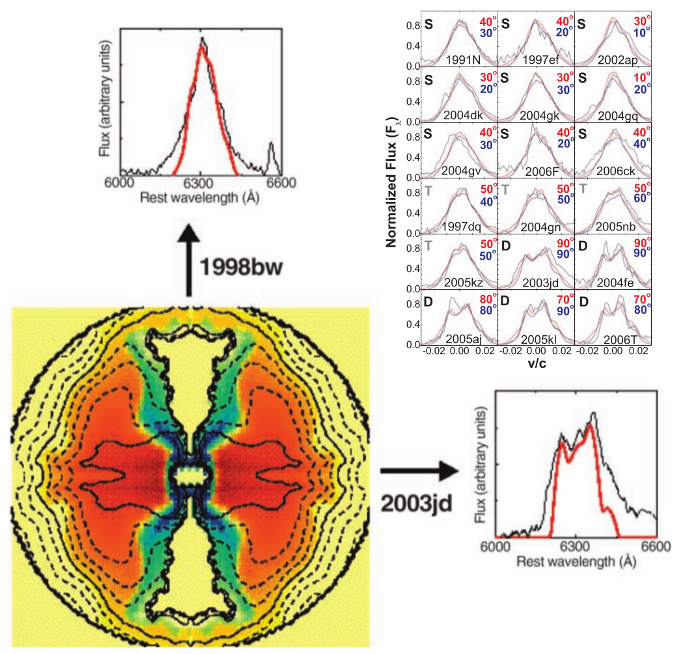}
 \caption{Observed [O \textsc{i}] $\lambda\lambda$6300,6364 emission line profiles for a sample of SNe Ibc.  \textbf{Top Right}: Emission lines classified into characteristic profiles (from \cite{Maeda2008}): single-peaked (S), transition (T), and double-peaked (D). Model predictions from a bipolar model (red curves), and a less aspherical model (blue), for different viewing directions are shown (directions denoted by the red and blue text).  \textbf{All other panels}:  Nebular line profiles observed for an aspherical explosion model for different viewing angles (from \cite{Mazzali2006}). The figure shows the properties of the explosion model: Fe (colored in green and blue) is ejected near the jet direction and oxygen (red) in a torus-like structure near the equatorial plane. Synthetic [O~\textsc{i}] $\lambda\lambda$6300,6364 emission line profiles are compared with the spectra of SN~1998bw (\textbf{top left}) and SN~2003jd (\textbf{bottom right}).    }
 \label{fig:Ibc_nebular_spectra_geometry}
\end{figure}

\section{Geometry}
\label{sec:geometry}

Measuring the geometry of GRB-SNe can lead to additional understanding of their explosion mechanism(s), and the role and degree of nickel mixing within the ejecta.  A starting point is to understand the geometry of GRB-SNe relative to other types of stripped-envelope core-collapse SNe (CCSNe), and ascertain whether any differences exist.  In this section we will recap the results of photometric, spectroscopic and polarimetric/spectropolarimetry observations of SNe Ibc.  The collective conclusion of these studies is that asphericity appears to be ubiquitous to \textit{all} SNe Ibc.

\subsection{Non-GRB SNe Ibc}

\subsubsection{Spectroscopy}

The best way to investigate the inner ejecta geometry of a given SN is through late-time spectroscopy, as done by \cite{Maeda2002,Mazzali2005,Maeda2006,Maeda2008,Modjaz2008_spectra,Tanaka2009,Taubenberger2009,Milisavljevic2010}. At $\ge200$~d after the explosion, expansion makes the density of the ejecta so low that it becomes optically thin, thus allowing optical photons produced anywhere in the ejecta to escape without interacting with the gas.  At these epochs the SN spectrum is nebular, showing emission lines mostly of forbidden transitions. Because the expansion velocity is proportional to the radius of any point in the ejecta, the Doppler shift indicates where the photon was emitted: those emitted from the near side of the ejecta are detected at a shorter (blueshifted) wavelengths, while those from the far side of the ejecta are detected at a longer (redshifted) wavelength. The late-time nebular emission profiles thus probe the geometry and the distribution of the emitting gas within the SN ejecta \cite{FranssonChevalier1987,SchlegelKirshner1989}. Importantly for SNe Ibc, nebular spectra allows the observer to look directly into the oxygen core.

One of the strongest emission lines is the [O \textsc{i}] $\lambda\lambda$6300,6364 doublet, which behaves like a single transition if the lines are sufficiently broad ($\ge0.01c$) because the red component is weaker than the blue one by a factor of three, see Fig. \ref{fig:Ibc_nebular_spectra_geometry}.  The appearance of this line can then be used to infer the approximate ejecta geometry: (1) a radially expanding spherical shell of gas produces a square-topped profile; (2) and a filled uniform sphere, where $^{56}$Ni is confined in a central high-density region with an inner hole that is surrounded by a low-density O-rich region \cite{MaedaNomoto2003}, produces a parabolic profile.  These authors also considered a third scenario: (3) a bipolar model \cite{Khokhlov1999,MacFadWoosley1999,MaedaNomoto2003} characterized by a low-density $^{56}$Ni-rich region located near the jet axis, where the jets convert stellar material (mostly O) into Fe-peak elements. The [O \textsc{i}] profile in the bipolar model depends on both the degree of asphericity and the viewing angle.  If a bi-polar SN explosion is viewed from a direction close to the jet axis, the O-rich material in the equatorial region expands in a direction perpendicular to the line of sight, and the [O \textsc{i}] emission profile is observed to be sharp and single-peaked. On the other hand, for a near-equatorial view, the profile is broader and double-peaked. It is important to note that a double-peaked profile\footnote{Furthermore, the separation of the blueshifted and redshifted peaks, which represent the forward and rear portions of an expanding torus of O-rich material, suggests that the two peaks actually originate from the two lines of the doublet from a single emitting source on the front of the SN moving toward the observer. Double emission peaks seen in asymmetric profiles with separations larger or smaller than the doublet spacing do not share this problem.  The high incidence of $\approx$64~\AA~separation between emission peaks of symmetric profiles plus the lack of redshifted emission peaks in asymmetric profiles suggests that emission from the rear of the SN may be suppressed. This implies that the double-peaked [O \textsc{i}] $\lambda\lambda$6300,6364 line profiles of SNe Ibc are not necessarily signatures of emission from a torus.  The underlying cause of the observed predominance of blueshifted emission peaks is unclear, but may be due to internal scattering or dust obscuration of emission from far side ejecta \cite{Milisavljevic2010}.} cannot be accounted for in the spherical model.  These models are for single-star progenitors, and they do not consider the effects that binary interactions or merger might impart to the observed geometry of the SN ejecta \cite{MorrisPodsiad2007}.

\cite{Maeda2008,Modjaz2008_spectra,Taubenberger2009} found that all SNe Ibc and IIb are aspherical explosions.  The degree of asphericity varies in severity, but all studies concluded that most SNe Ibc are not as extremely aspherical as GRB-SNe (specifically SN~1998bw).  Interestingly \cite{Taubenberger2009} found that for some SNe Ibc, the [O~\textsc{i}] line exhibits a variety of shifted secondary peaks or shoulders, interpreted as blobs of matter ejected at high velocity and possibly accompanied by neutron-star kicks to assure momentum conservation.  The interpretation of massive blobs in the SN ejecta is expected to be the signature of very one-sided explosions.

Some notable and relevant nebular spectra analyses include SNe IcBL 2003jd \cite{Mazzali2005}, 2009bb \cite{Pignata2011} and 2012ap \cite{Milisavljevic2015}.  \cite{Mazzali2005} interpret their double-peaked [O \textsc{i}] $\lambda\lambda$6300,6364 nebular lines of SN~2003jd as an indication of an aspherical axisymmetric explosion viewed from near the equatorial plane, and directly perpendicular to the jet axis, and suggested that this asphericity could be caused by an off-axis GRB jet.  \cite{Pignata2011} obtained moderately noisy nebular spectra of SN~2009bb, which nevertheless displayed strong nebular lines of [O \textsc{i}] $\lambda\lambda$6300,6364 and [Ca \textsc{ii}] $\lambda\lambda$7291,7324 that had all single-peaked profiles.  In their derived synthetic spectra, a single velocity provided a good fit to these lines, thus implying that the ejecta is not overly aspherical.   The nebular spectra ($>200$ d) of SN~2012ap \cite{Milisavljevic2015} had an asymmetric double-peaked [O \textsc{i}] $\lambda\lambda$6300,6364 emission profile that was attributed to either absorption in the supernova interior or a toroidal ejecta geometry.



\begin{figure}
 \centering
 \includegraphics[bb=0 0 759 529,scale=0.43,keepaspectratio=true]{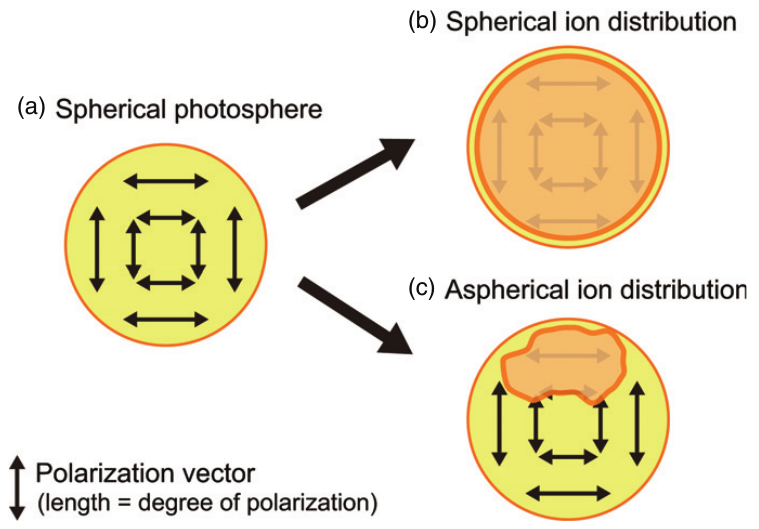}
 \caption{Schematic illustration of polarization in the SN ejecta.  (a) When the photosphere is spherical, polarization is canceled out, and no polarization is expected. At the wavelength of a line, polarization produced by the electron scattering is depolarized by the line transition. (b) When the ion distribution is spherical, the remaining polarization is canceled, and no polarization is expected. (c) When the ion distribution is not spherical, the cancellation becomes incomplete, and line polarization could be detected.  (Figure and caption taken from \cite{Tanaka2012}).}
 \label{fig:polarimetry}
\end{figure}

\subsubsection{Polarimetry}

Further enlightening clues to the geometry of SNe Ibc has arisen via polarimetric and spectropolarimetric observations (see \cite{WangWheeler2008} for an extensive review, and Fig. \ref{fig:polarimetry}).  When light scatters through the expanding debris of a SN, it retains information about the orientation of the scattering layers. Since it is not possible to spatially resolve  extragalactic SNe through direct imaging, polarization is a powerful tool to determine the morphology of the ejecta. Spectropolarimetry measures both the overall shape of the emitting region and the shape of regions composed of particular chemical elements.  Collectively, numerous polarimetric data have provided overwhelming evidence that all CCSNe are intrinsically three-dimensional phenomena with significant departures from spherical symmetry, and they routinely show evidence for strong alignment of the ejecta in single well-defined directions, suggestive of a jet-like flow.  As discussed in \cite{WangWheeler2008}, many of these CCSNe often show a rotation of the position angle with time of $30-40^{\rm o}$ that is indicative of a jet of material emerging at an angle with respect to the rotational axis of the inner layers.   Another recent investigation by \cite{Tanaka2012} showed that all SNe Ibc show non-zero polarization at the wavelength of strong lines.  More importantly, they demonstrated that five of the six SNe Ibc they investigated had a ``loop'' in their Stokes $Q-U$ diagram\footnote{Where $Q$ is the radiance linearly polarized in the direction parallel or perpendicular to the reference plane, and $U$ is the radiance linearly polarized in the directions 45$^{\circ }$ to the reference plane.}, which indicates that a non-axisymmetric, three-dimensional ion distribution is ubiquitous for SNe Ibc ejecta.

The results of \cite{WangWheeler2008} suggest that the mechanism that drives CCSNe must produce energy and momentum aspherically from the start, either induced from the pre-explosion progenitor star (i.e. rotation and/or magnetic fields) or perhaps arising from the newly-formed neutron star (NS) \cite{Akiyama2003,Thompson2004,Masada2006,Uzdensky2007}.  In any case, it appears that the asphericity is permanently frozen into the expanding matter. Collimated outflows might be caused by magnetohydrodynamic jets, as is perhaps the case for GRB-SNe \cite{Woosley1993,MacFadWoosley1999,MacFadyen2001,Mosta2015}, from accretion flows around the central neutron star, via asymmetric neutrino emission, from magnetoacoustic flux, jittering jets (jets that have their launching direction rapidly change \cite{PapishSoker2011}), or by some combination of those mechanisms. Another alternative idea, perhaps intimately related, is that material could be ejected in clumps that block the photosphere in different ways in different lines. It may be that jet-like flows induce clumping so that these effects occur simultaneously.  Alternatively, the results of \cite{Tanaka2012} suggest that the global asymmetry of SNe Ibc ejecta may rather arise from convection and pre-existing asymmetries in the stellar before and during the time of core-collapse (e.g. \cite{Couch2014,Couch2015}), rather then induced by two-dimensional jet-like asphericity.

In addition to the above analyses, there are more clues which show that asphericity is quite ubiquitous in CCSN ejecta.  A jet-model was proposed for type Ic SN~2002ap \cite{Wang2003}, where the jet was buried in the ejecta and did not bore through the oxygen mantle.  The lack of Fe polarization suggests that a nickel jet had not penetrated all the way to the surface.  For CCSNe, we know that pulsars are somehow kicked at birth in a manner that requires a departure from both spherical and up/down symmetry \cite{LyneLorimer2004}.  The spatial distribution of various elements, including $^{44}$Ti in supernova remnants \cite{Grefenstette2014} is also consistent with an aspherical explosion, arising from the development of low-mode convective instabilities (e.g. standing accretion shock instabilities \cite{Blondin2003}) that can produce aspherical or bi-polar explosions in CCSNe.  The anisotropies inferred by the oxygen distribution instead suggests that large-scale (plume-like) mixing is present, rather than small-scale (Rayleigh-Taylor) mixing, in supernova remnants.  Additionally, the Cassiopeia A supernova remnant shows signs of a jet and counterjet that have punched holes in the expanding shell of debris \cite{Orlando2016}, and there are examples other asymmetric supernova remnants \cite{Fesen2001,Wheeler2008}, and remnants with indications of being jet-driven explosions or possessing jet-like features \cite{Lopez2013,FesenMili2016}.

\subsubsection{Role of Mixing in the ejecta}

The analytical modelling of late-time ($>50-100$~d) bolometric LCs of SNe Ibc also implies a departure from spherical symmetry (or perhaps a range grey optical opacities \cite{Wheeler2015}).  Modelling performed by \cite{Maeda2003_2comp_model} showed that the late-time bolometric LC behaviour of a sample of three SNe Ic and IcBL (SN~1998bw, 1997ef and 2002ap) was better described by a two-component model (two concentric shells that approximated the behaviour of a high-velocity jet and a dense inner core/torus) than spherical models.  Their modelling also showed that there was a large degree of nickel mixing throughout the ejecta.  A similar result was inferred by \cite{CMS2014} for a sample of SNe Ibc, who showed that the outflow of SNe Ib is thoroughly mixed.  Helium lines arise via non-thermal excitation and non-local thermodynamic equilibrium \cite{Harkness1987,Lucy1991,LiMcCray1995,LiHillierDessart2012}. High-energy $\gamma$-rays produced during the radioactive decay of nickel, cobalt and iron Compton scatter with free and bound electrons, ultimately producing high-energy electrons that deposit their energy in the ejecta through heating, excitation and ionization.  

To address the question of whether the lack of helium absorption lines for SNe Ic was due to a lack of this element in the ejecta, or that the helium was located at large distances from the decaying nickel \cite{Woosley1995,LiHillierDessart2012,Hachinger2012}, \cite{CMS2014} showed that the ejecta of type SN Ic 2007gr was also thoroughly mixed, meaning that the lack of helium lines in this event could not be attributed to poor mixing.   A similar conclusion was reached by \cite{Modjaz2015} who demonstrated that He lines cannot be ``smeared out'' in the spectra of SNe IcBL, i.e. blended so much they disappear; instead He really must be absent in the ejecta (see as well \cite{Frey2013}).  A prediction of RT models \cite{Dessart2012} is if the lack of mixing is the only discriminant between SNe Ib and Ic, then well mixed SNe Ib should have higher ejecta velocities than the less well mixed SNe Ic.  The investigation by \cite{Liu_Modjaz2015} tested this prediction with a very large sample of SNe Ibc spectra, finding the opposite to be true: SNe Ic have higher ejecta velocities than SNe Ib, implying that the lack of He lines in the former cannot be attributed entirely to poor mixing in the ejecta.  Next, \cite{Taddia2015} showed that for a sample of SNe Ibc, SNe Ib, Ic \& Ic-Bl have faster rising LCs than SNe Ib, implying that the ejecta in these events are probably well-mixed.  The collective conclusion of these observational investigations state that the lack of helium features in SNe Ic spectra cannot be attributed to poor mixing but rather the absence of this element in the ejecta, which agrees with the conclusion of \cite{Hachinger2012} that no more than $0.06-0.14$~M$_{\odot}$ of He can be ``hidden'' in the ejecta of SNe Ic.

\subsection{GRB-SNe}

The key result presented in the previous sections is that all CCSNe possess a degree of asphericity: either two-dimensional \cite{WangWheeler2008} asymmetries where most CCSNe possess a jet, or three-dimensional asymmetries \cite{Tanaka2012}.  Taken at face value, if all CCSNe possess two-dimensional axisymmetric geometry, then the observation that the $30-40^{\rm o}$ rotation of the position angle with time is suggestive of a jet of material emerging at an angle with respect to the rotational axis of the inner layers.  This observation differs to that expected for GRB-SNe, where the jet angle is expected to be along, or very near to the rotation axis of the pre-explosion progenitor star.  If jets are almost ubiquitous in CCSNe, but they are usually at an angle to the rotational axis, does this suggest that GRB-SNe are different because the jet emerges along, or very near to, the rotational axis?  If so, then something is required to maintain that collimation: i.e. more rapid rotation of GRB-SN progenitors and/or strong collimation provided by magnetic fields \cite{Mosta2015}.   Moreover, is the difference between $ll$GRBs and high-luminosity GRBs due to less collimation in the former?  In turn, perhaps more SNe Ibc arise from central engine that is currently accounted for, but for whatever reason the jets very quickly lose their collimation, perhaps to under-energetic or very short-lived central engines, and deposit their energy in the interior of the star, where perhaps a combination of jets and a neutrino-driven explosion mechanism is responsible for the observed SN.  Note that this supposition is also consistent with the study of \cite{Soderberg2006_offaxis} who looked for off-axis radio emission from GRBs pointed away from Earth, finding $<10\%$ of all SNe Ibc are associated with GRBs pointed away from our line of sight.  In this scenario, no imprint of the jet in the non-GRB SNe Ibc is imparted to the ejecta.  Nevertheless, the results of \cite{Tanaka2012} need to be kept in mind when considering this speculative scenario, where the asymmetries in SNe Ibc may not be axisymmetric, but instead may be intrinsically three-dimensional.

More observations are sorely needed of nearby GRB-SNe to help address this outstanding question.  To date only two GRB-SNe have occurred at close enough distances that reasonable quality nebular spectra have been obtained: SN~1998bw ($\sim40$~Mpc) and SN~2006aj ($\sim150$~Mpc).  Even SN~2010bh was too distant ($\sim270$~Mpc) for the nebular emission lines to be reasonably modelled \cite{Bufano2012}.  In the following section we will present a brief summary of the results of spectroscopic and polarimetric analyses of these two GRB-SNe.

\cite{Mazzali2001,Maeda2002,Maeda2006} investigated the nebular spectra of SN~1998bw, which exhibited properties that could not be explained with spherical symmetry. Instead, a model with high-velocity Fe-rich material ejected along the jet axis, and lower-velocity oxygen torus perpendicular to the jet axis, was proposed. From this geometry a strong viewing-angle dependence of nebular line profiles was obtained \cite{Maeda2002}.   \cite{Mazzali2001} noted that the [Fe \textsc{ii}] lines were unusually strong for a SN Ic, and that lines of different elements have different widths, indicating different expansion velocities, where iron appeared to expand more rapidly than oxygen (i.e. a rapid Fe/Ni-jet and a slower moving O-torus).  The [O \textsc{i}] nebular lines declined more slowly than the [Fe \textsc{ii}] ones, signalling deposition of $\gamma$-rays in a slowly moving O-dominated region. These facts suggest that the explosion was aspherical. The absence of [Fe \textsc{iii}] nebular lines can be understood if the ejecta are significantly clumped.  \cite{Maeda2006} noted that their models show an initial large degree ($\sim4$ depending on model parameters) of boosting luminosity along the polar/jet direction relative to the equatorial  plane, which decreased as the SN approached peak light. After the peak, the factor of the luminosity boost remains almost constant ($\sim1.2$) until the supernova entered the nebular phase. This behaviour was attributed to an aspherical $^{56}$Ni distribution in the earlier phase and to the disk-like inner low-velocity structure in the later phase.  

Early polarization measurements of $\approx0.5\%$, possibly decreasing with time, were detected for SN~1998bw \cite{Kay1998,Patat2001}, which imply the presence of aspherical ejecta, with an axis ratio of about 2:1 \cite{Hoflich1999}.  In contrast, radio emission of GRB~980425 / SN~1998bw showed no evidence for polarization \cite{Kulkarni1998}, which suggested that the mildly relativistic ejecta were not highly asymmetric, at least in projection. However it should be noted that internal Faraday dispersion in the ejecta can suppress radio polarization.   As mentioned in the previous section, modelling of the late-time bolometric LC of SN~1998bw \cite{McKenzie1999,Sollerman2002,Nakamura2001,Clocchiatti2011} showed that some degree of asymmetry in the explosion is required to explain its decay behaviour (see as well Fig. \ref{fig:1998bw_bolo_late}).

For SN~2006aj, the [Fe \textsc{ii}] lines were much weaker than those observed for SN 1998bw, which supports its lower luminosity relative to the archetype GRB-SN \cite{Mazzali2006}.  Most of the nebular lines had similar widths, and their profiles indicated that no major asymmetries were present in the ejecta at velocities below 8000~km~s$^{-1}$.  The modelling results of \cite{Maeda2007} implied a $1.3$~M$_{\odot}$ oxygen core that was produced by a mildly asymmetric explosion.  The mildly peaked [O \textsc{i}] $\lambda\lambda$6300,6364 profile showed an enhancement of the material density at velocities less than $<$3000~km~s$^{-1}$, which also indicated an asymmetric explosion. If SN~2006 was a jetted SN explosion, the jet was wider than in SN~1998bw (intrinsically or due to stronger lateral expansion \cite{MaedaNomoto2003}), since the signature is seen only in the innermost part.  Linear polarization was detected by \cite{Gorosabel2006} between three and 39 days post explosion, which implied the evolution of an asymmetric SN expansion.  \cite{Maund2007_06aj} concluded that their polarization measurements were not very well constrained, and considering the low polarization observed between $6000-6500$~\AA, the
global asymmetry was $\le15\%$.

\begin{figure*}
 \centering
 \includegraphics[bb=0 0 576 432,scale=0.4,keepaspectratio=true]{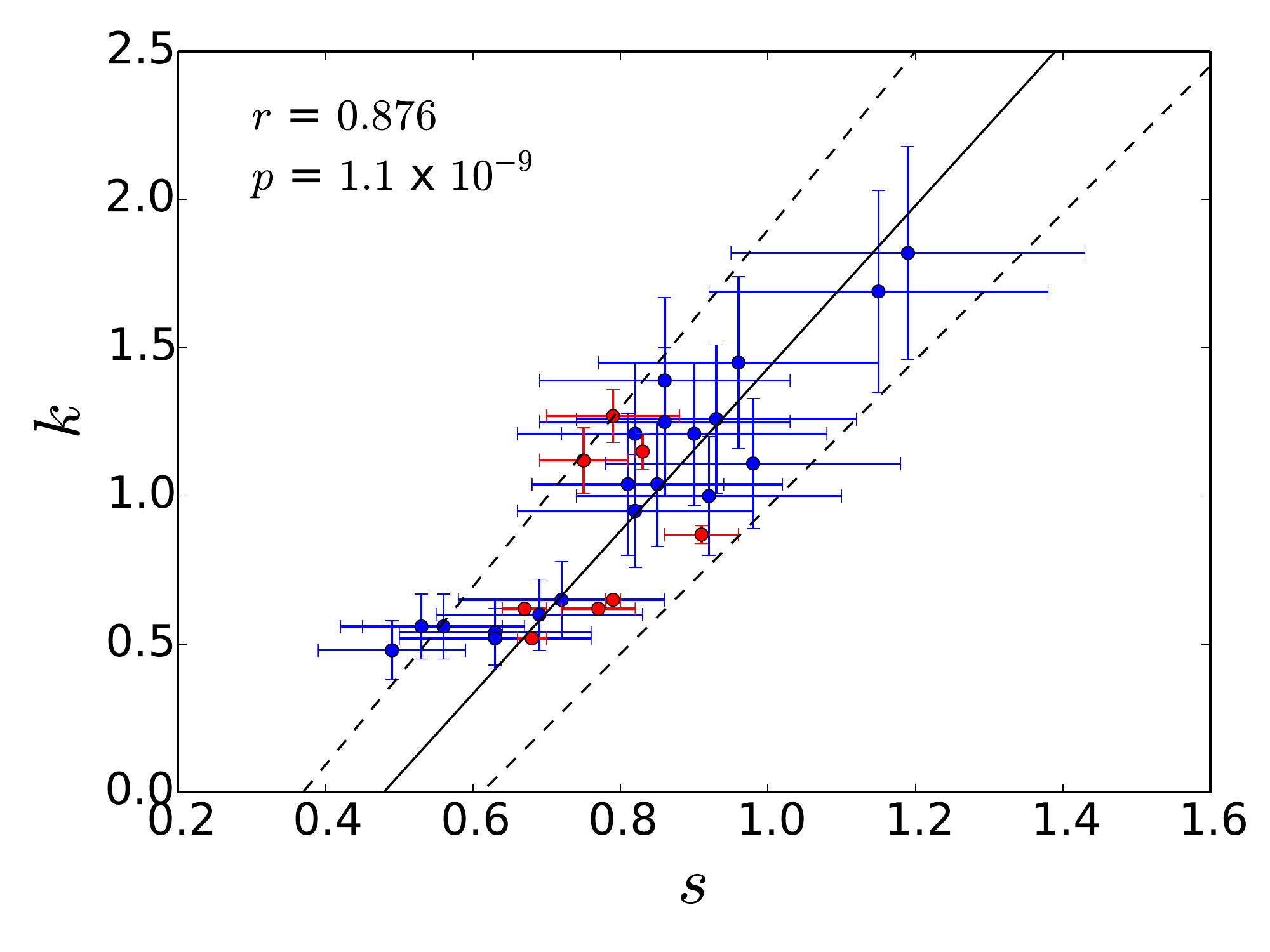}
 \caption{Luminosity ($k$) $-$ stretch ($s$) relation for relativistic type IcBL SNe \cite{Cano2014}.  For all filters from $UBVRI$, and combinations thereof, GRB-SNe are shown in blue, and the two known relativistic type IcBL SNe (2009bb and 2012ap) are shown in red.  A bootstrap analysis was performed to fit a straight-line to the dataset to find the slope ($m$) and $y$-intercept ($b$), which used Monte-Carlo sampling and $N=10,000$ simulations.  The best-fitting values are: $m=2.72\pm0.26$ and $b=-1.29\pm0.20$.  The correlation coefficient is $r=0.876$, and the two-point probability of a chance correlation is $p=1.1\times10^{-9}$.  This shows that the $k-s$ relationship is statistically significant at the 0.001 significance level.}
 \label{fig:ks}
\end{figure*}

\section{GRB-SNe as Cosmological Probes}
\label{sec:Cosmology}
\vspace*{0.5cm}

\subsection{Luminosity--Stretch/Decline Relationships}

In 2014, \cite{Cano2014,LH14,CJG14} (see as well \cite{Schulze2014}) demonstrated, using entirely different approaches, that GRB-SNe (which included $ll$GRB-SNe, INT-GRB-SNe and high-luminosity GRB-SNe) have a luminosity$-$decline relationship that is perfectly akin to that measured for type Ia SNe \cite{Phillips1993}.  All approaches investigated decomposed GRB-SN LCs (see Section \ref{sec:Obs_photo}).  In \cite{Cano2014}, a template SN LC (1998bw) was created in filters $BVRI$($1+z$) as it would appear at the redshift of the given GRB-SN.  A spline function ($g(x)$) was then fit to the template LC, and the relative brightness ($k$) and width ($s$) were determined (i.e. $f(x) = k \times g(x/s)$) for each GRB-SN in each rest-frame filter.  These were then plotted, and a straight line was fit to the data, where the slope and intercept were constrained via a bootstrap fitting analysis that used Monte-Carlo sampling.  An example of the $k-s$ relation is shown in Fig. \ref{fig:ks}, where GRB-SNe are shown in blue points, and the two relativistic SNe IcBL (2009bb and 2012ap) are shown in red.  This relation shows that GRB-SNe with larger $k$ values also have larger $s$ values $-$ i.e. brighter GRB-SNe fade slower.  The statistical significance of the fit is shown as the Pearson's correlation coefficient, where $r = 0.876$, and the two-point probability of a chance correlation is $p=1.1\times10^{-9}$, which clearly shows that the relationship is significant at more than the $p<0.001$ significance level.  This implies that not only are GRB-SNe standardizable candles, but all relativistic type IcBL SNe are.

The result in \cite{Cano2014} clearly superseded the results of \cite{Stanek2005,Ferrero2006,Cano2011a} who searched for correlations in the observer-frame $R$-band LCs of a sample of GRB-SNe, concluding that none was present.  However, the method used in \cite{Cano2014,CJG14,Cano2015} had one key difference to previous methods: they considered precise, K-corrected rest-frame filters.  Instead, previous approaches were all sampling different portions of the rest-frame spectral energy distribution (SED), which removed any trace of the $k-s$ relationship.  

\begin{figure*}
 \centering
   \includegraphics[bb=0 0 880 207,scale=0.525,keepaspectratio=true]{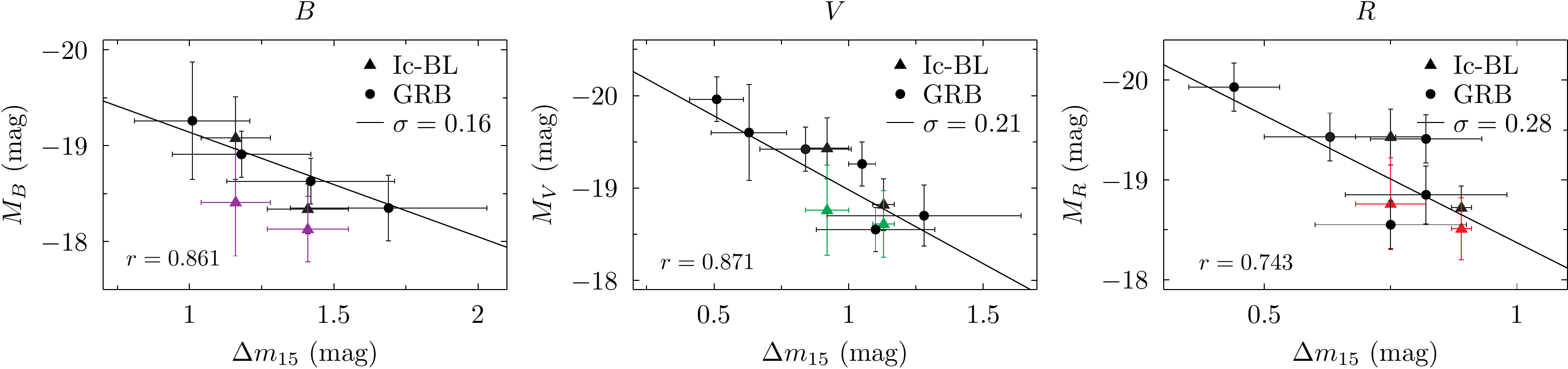}
 \caption{Luminosity$-$decline relationships of relativistic SNe IcBL (GRB-SNe: filled circles; SNe IcBL: filled triangles) in filters $B$ (purple), $V$ (green) and $R$ (red), from \cite{CJG14}.  Solid black lines and points correspond to absolute magnitudes calculated for luminosity distances, while coloured points and lines correspond to absolute magnitudes calculated for those events where independent distance measurements have been made to the SN's host galaxy.  The correlation coefficient for each dataset is shown (in black and in their respective colours) as well as the best-fitting luminosity$-$decline relationship determined using a bootstrap method, and the corresponding rms ($\sigma$) of the fitted model.  It is seen that statistically significant correlations are present for both the GRB-SNe and combined GRB-SN \& SN IcBL samples. }
 \label{fig:lumdecline}
\end{figure*}

While such a correlation implies that, like SNe Ia, there is a relationship between the brightness of a given GRB-SN and how fast it fades, where brighter GRB-SNe fade slower, this relationship is not very useful if GRB-SNe want to be used for cosmological research: the template LCs of SN~1998bw are created for a specific cosmological model, and are therefore model-dependent.  Instead, the luminosity$-$decline relationship presented by \cite{LH14,CJG14} relates the same observables as those used in SN Ia cosmology research: their peak absolute magnitude and $\Delta m_{15}$ in a given filter.  \cite{LH14} considered rest-frame $V$-band only, while \cite{CJG14} considered rest-frame $BVR$.  Fig. \ref{fig:lumdecline} shows the relationships from the latter paper, where the two relativistic SNe IcBL are included in the sample.  The amount of RMS scatter increases from blue to red filters, and is only statistically significant in $B$ and $V$ (at the $p=0.02$ level).

\subsection{Constraining Cosmological Parameters}

\begin{figure*}
 \centering
 \includegraphics[bb=0 0 910 313,scale=0.51,keepaspectratio=true]{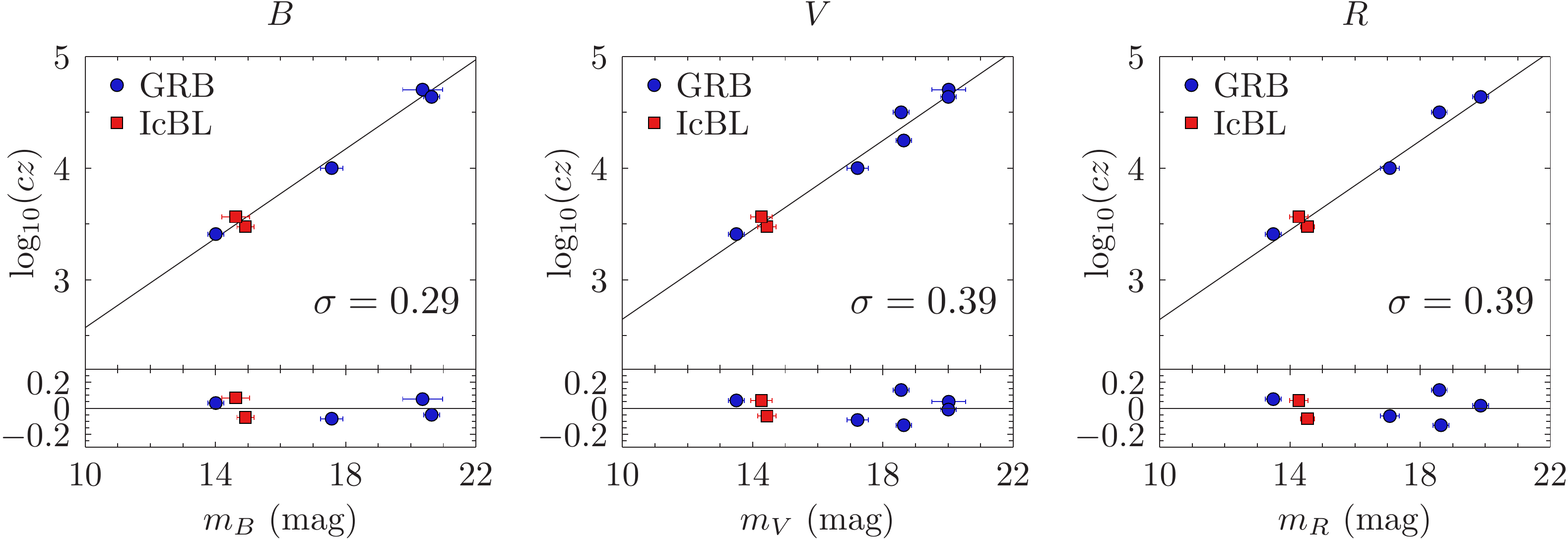}
 \caption{Hubble diagrams of relativistic SNe IcBL in filters $BVR$, from \cite{CJG14}.  GRB-SNe are shown in blue, and SNe IcBL (SNe 2009bb \& 2012ap) in red.  Plotted in each subplot are the uncorrected magnitudes of each subtype and the fitted Hubble ridge line as determined using a bootstrap method.  Also plotted are the rms values ($\sigma$) and residuals of the magnitudes about the ridge line.  In the $B$-band, the amount of scatter in the combined SNe IcBL sample is the same as that for SNe Ia up to $z=0.2$ \cite{Betoule2014,CJG14}, which is $\sigma \approx 0.3$~mag.}
  \label{fig:Hubble}
\end{figure*}

Once the luminosity$-$decline relationship was identified, the logical next step is to use GRB-SNe to constrain cosmological models, in an attempt to determine the rate of universal expansion in the local universe (the Hubble constant, $H_{0}$), and perhaps even the mass and energy budget of the cosmos.   In a textbook example of how to use any standard(izable) candle to measure $H_{0}$ in a Hubble diagram of low-redshift objects (typically $z\ll1$), \cite{CJG14}  followed the procedure outlined in numerous SNe Ia-cosmology papers  \cite{Kowal1968,BranchTamm1992,SandTamm1993,Hamuy1995,Hamuy1996a,Hamuy1996b,Riess1998,Perlmutter1999,Freedman2001}.  Fig. \ref{fig:Hubble} show Hubble diagrams of relativistic SNe IcBL in filters $BVR$ (GRB-SNe in blue, relativistic SNe IcBL in red) for redshifts less than $z=0.2$.  The amount of RMS scatter (shown as $\sigma$) is less in the $B$-filter, $\approx 0.3$~mag, and about 0.4~mag in the redder $V$ and $R$ filters.  Compared with the sample of SNe Ia in \cite{Betoule2014} over the same redshift range, it is seen that SNe Ia in the $B$-band also have a scatter in their Hubble diagram of 0.3~mag.  Moreover, when the large SNe Ia sample ($N=318$) was decreased to the same sample size of the relativistic SNe IcBL sample, the same amount of scatter was measured, meaning that GRB-SNe and SNe IcBL are as accurate as SNe Ia when used as cosmological probes.

A key observable needed to measure $H_{0}$ are independent distance measurements to one or more of the objects being used.  However, to date no independent distance has yet been determined for a GRB or GRB-SN.  However, relativistic SNe IcBL 2009bb and 2012ap were included in the same sample as the GRB-SNe, which was justified by \cite{CJG14} because both are sub-types of engine-driven SNe (Fig. \ref{fig:ek_gammabeta}), and indeed they also follow the same luminosity$-$stretch (Fig. \ref{fig:ks}) and luminosity$-$decline (Fig. \ref{fig:lumdecline}) relationship as GRB-SNe.  Thus, one can use the independent distance measurements to their host galaxies (Tully-Fisher distances), and use them as probes of the local Hubble flow to provide a model-independent estimate of $H_{0}$.  \cite{CJG14} constrained a weighted-average value of ${H}_{0,\rm w}=82.5\pm8.1$~km~s$^{-1}$~Mpc$^{-1}$.  This value is 1$\sigma$ greater than that obtained using SNe Ia, and 2$\sigma$ larger than that determined by Planck.  This difference can be attributed to large peculiar motions of the host galaxies of the two SNe IcBL, which are members of galaxy groups.  Interestingly, when the same authors used a sample of SNe Ib, Ic and IIb, they found an average value of $H_{0}$ that had a standard deviation of order $20-40$~km~s$^{-1}$~Mpc$^{-1}$, which demonstrates that these SNe are poor cosmological candles.  In a separate analysis, \cite{LHW14} used their sample of GRB-SNe, which did not include non-GRB SNe IcBL but instead covered a larger redshift range (up to $z=0.6$), to derive the mass and energy budget of the universe, finding loosely constrained values of $\Omega_{\rm M} = 0.58^{+0.22}_{-0.25}$ and  $\Omega_{\rm \Lambda} = 0.42^{+0.25}_{-0.22}$.


\subsection{Physics of the luminosity--decline relationship}

\begin{figure}
 \centering
 \includegraphics[bb=0 0 452 578,scale=0.39,keepaspectratio=true]{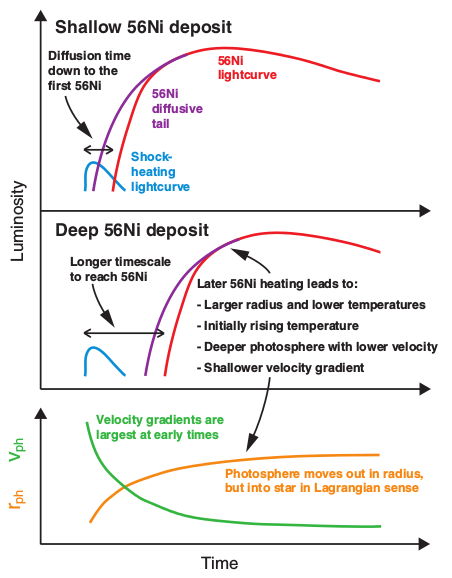}
 \caption{The effect of different degrees of nickel mixing in the ejecta of SNe Ibc on their observed LCs, from \cite{PiroNakar2013}.  \textbf{Top and middle panels}: how the relative positions of the shock-heating contribution (blue curves), $^{56}$Ni diffusive tail contribution (purple curves), and the $^{56}$Ni contribution (red curves) to the observed LC can differ depending on the depth and amount of mixing of the $^{56}$Ni. The total observed LC is the sum of these three components.  When the $^{56}$Ni is located deep in the ejecta (middle panel) and the shock-heating light curve (blue curve) is below the detection limits, there can be a significant dark phase between the time of explosion and the moment of first detection. \textbf{Bottom}: temporal evolution of the photospheric radius (orange curve) and velocity (green curve). Depending on the position of the $^{56}$Ni LC, different photospheric radii, velocities, and velocity gradients will be present during the rising LC. }
 \label{fig:SNe_Ni_mixing}
\end{figure}

A physical explanation for why GRB-SNe are standardizable candles is not immediately obvious.  If the luminosity of GRB-SNe (excluding SN~2011kl) is powered by radioactive heating (see Section \ref{sec:powering_GRBSNe}), then more nickel production leads to brighter SNe.  So far however, no correlation has been found between the bolometric properties of GRB-SNe and the properties ($E_{\rm iso}$ and $T_{90}$) of the accompanying $\gamma$-ray emission \cite{WoosleyBloom2006,Cano2015,Toy2016}.  To a first order, this is at odds with the simplest predictions of the collapsar model, which suggests that more energy input by a central engine should lead to increased nickel production and more relativistic ejecta.  However, $\gamma$-ray energetics are a poor proxy of the total energy associated with the central engine, so the absence of a correlation is perhaps not surprising.  Moreover, as pointed out by \cite{WoosleyBloom2006}, one expects large variations in the masses and rotation rates of the pre-explosion progenitor stars, especially when metallicity effects are factored in.  Different stellar rotation rates will result in different rotation rates imparted to their cores,  leading to different amounts of material being accreted, and ultimately resulting in a variation of the final BH masses.  Along with variations in the stellar density, all of these factors will result in a range of nickel masses being produced.  Moreover, even if the same amount of nickel is produced in each event, SNe that expand at a slower rate will be fainter because their LCs will peak later after which more of the nickel has decayed and suffered adiabatic degradation.  Additionally, the location of the nickel in the ejecta will also result in different looking LCs, where nickel that is located deeper in the ejecta takes longer to diffuse out of the optically thick ejecta, leading to later peak times (Fig. \ref{fig:SNe_Ni_mixing}).  If the degree of mixing in the early SN is heterogeneous for GRB-SNe, a range of rise-times is expected, along with a large variation in the velocity gradients and photospheric radii.  However, inspection of Fig. \ref{fig:line_vels} shows that, if we naively take a single transition as a proxy of the photospheric velocity, the distribution of say Si \textsc{ii} $\lambda$6355 shows that the velocity gradient of most GRB-SNe have a similar evolution, though the range of velocities of the Fe \textsc{ii} $\lambda$5169 transition imply that they still have a wide range of velocities at a given epoch.  This similar behaviour might suggest a similar degree of nickel mixing in the SN ejecta.

Nevertheless, it appears from independent studies using different approaches that GRB-SNe \textit{are} standardizable candles.  Whether this observation implies similarities in the physical properties of the central engine driving the explosion, or the SNe themselves, is uncertain.  For the most past, it is expected that most GRB-SNe are viewed close to the jet-axis \cite{Cano2014}, which also appears to apply to SN~2009bb \cite{Pignata2011}, meaning we are observing SNe more or less with the same approximate geometry.  The fact that GRB-SNe are standard\textit{izable}, and have a range of brightnesses implies that that different amounts of nickel are being generated. A naive conclusion to be drawn is that the observed luminosity$-$decline relationships suggest that a correlation exists between the strength and energetics produced by the central engine and the resultant nucleosynthetic yields of $^{56}$Ni.  Moreover, the lack a luminosity$-$decline relationship for SNe Ibc \cite{Cano2014,CJG14} implies that the explosion and nucleosynthesis mechanism(s) are not correlated.

In the context of SNe Ia, which are, of course, also standardizable candles, their LCs are also powered solely by the radioactive decay of nickel and cobalt, the amount of which determines the LC's peak brightness and width.  The width also depends on the photon diffusion time, which in turn depends on the physical distribution of the nickel in the ejecta, as well as the mean opacity of the ejecta.  In general, the opacity increases with increasing temperature and ionization \cite{WoosleyZhang2007}, thus implying that more nickel present in the ejecta leads to larger diffusion times.  This directly implies that fainter SNe Ia fade faster than brighter SNe Ia, thus satisfying the luminosity$-$decline relation \cite{Phillips1993}.  This is not the only effect however, as the distribution of nickel in the ejecta also affects how the LC evolves, where nickel located further out has a faster bolometric LC decline.  Additionally, following maximum $B$-band light, SNe Ia colours are increasingly affected by the development of Fe II and Co II lines that blanket/suppress the blue $B$-band light.  Dimmer SNe are thus cooler, and the onset of Fe III $\rightarrow$ Fe II recombination occurs quicker than in brighter SNe Ia, resulting in a more rapid evolution to redder colours \cite{KasenWoosley2007}.  Therefore the faster $B$-band decline rate of dimmer SNe Ia reflects their faster ionization evolution, and provides additional clues as to why fainter SNe Ia fade more rapidly.  Thus, as the LCs of GRB-SNe are also powered by radioactive decay, the physics that govern SNe Ia also govern those of GRB-SNe, and may go some way to explaining why GRB-SNe are also standardizable candles.

\section{Host Environments}
\label{sec:Hosts}
\vspace*{0.5cm}

Direct observations of the SNe that accompany LGRBs, and their sub-types, provide a rich range of clues as to the physical properties of their pre-explosion progenitor stars.  LGRBs represent a rare endpoint of stellar evolution, and their production and subsequent properties are likely to be a consequence of environmental factors.  As such, many in-depth investigations of their host environments, both their global/galaxy-wide properties, and where possible, host-resolved environmental conditions, have been performed.  Indeed, the information gained from this myriad of investigations warrants their own reviews, and the gathered nuances of these studies are beyond the scope of this GRB-SN review.  Instead, in this section we highlight what we regard as the most important developments in this branch of GRB phenomenology that have directly furthered our understanding of the GRB-SN connection.  For further insight, we refer the reader to excellent reviews and seminal studies by, among others, \cite{Savaglio2009,Fynbo2012,Levesque2014,Kruhler2015}, and references therein.

\subsection{Global properties}

With the advent of X-ray localizations of GRB AGs came the ability to study the type of galaxies that LGRBs occur in.  Over the years, evidence mounted that LGRBs appeared to prefer low-luminosity, low-mass, blue, star-forming galaxies, that have higher specific star-formation rates (SFRs) than the typical field galaxy \cite{MaoMo1998,HoggFruchter1999,Djorgovski2001,Djorgovski2003,LeFloch2003,Christensen2004,Tanvir2004,Conselice2005,Stanek2006,CastroCeron2006,Wainwright2007,Savaglio2009,CastroCeron2010}.  Visual inspection of optical \emph{HST} imaging of LGRB host galaxies \cite{Fruchter2006,Wainwright2007,Svensson2010} showed a high fraction of merging/interacting systems: 30\% showed clear signs of interaction, and another 30\% showed irregular and asymmetric structure, which may be the result of recent mergers.  The position of a GRB within its host also provided additional clues: both \cite{Bloom2002}, who examined the offsets of LGRBs from their host nuclei (see as well \cite{deUP2012,Blanchard2016}), and \cite{Fruchter2006} demonstrated that within their hosts, LGRBs were more likely to be localized in the brightest UV regions of the galaxy, which are associated with concentrated populations of young massive stars.

At the same time, several early studies were converging towards the idea that LGRBs favoured sub-solar, low-metallicity ($Z$) host/environments \cite{Prochaska2004,Gorosabel2005,Sollerman2005,Kewley2007}).  As the progenitors of LGRBs are massive stars with short lifetimes (of order a few million years), they are not expected to travel far from their birth in H~\textsc{ii} regions, and the measured metallicity of the associated H~\textsc{ii} region at the site of an LGRB can be used as a proxy of the natal metallicity.   \cite{Stanek2006} found that the metallicities of half a dozen low-redshift ($z<0.3$) LGRB hosts were lower than their equally luminous counterparts in the local star-forming galaxy population, and proposed that LGRB formation was limited by a strong metallicity threshold.  This was based on the observation that LGRB hosts were placed below the standard $L-Z$ relation for star-forming galaxies, where galaxies with higher masses, and therefore luminosities, generally have higher metallicities \cite{Lequeux1979,Skillman1989,Zaritsky1994,Tremonti2004}.  A metallicity cut-off for LGRB formation was also proposed by \cite{WolfPodsiad2007}.  \cite{Modjaz2008_metal} demonstrated that nearby LGRB host galaxies had systematically lower metallicities than the host galaxies of nearby ($z<0.14$) SNe IcBL.  \cite{Levesque2010} showed that most LGRB host galaxies fall below the general $L-Z$ relation for star-forming galaxies and are statistically distinct to the host galaxies of SNe Ibc and the larger star-forming galaxy population. LGRB hosts followed their own mass-metallicity relation out to $z\sim1$ that is offset from the general mass-metallicity relation for star-forming galaxies by an average of $0.4\pm0.2$~dex in metallicity. This marks LGRB hosts as distinct from the host galaxies of SNe Ibc, and reinforced the idea that that LGRB host galaxies are not representative of the general galaxy population \cite{Tanvir2004,LeFloch2006,Michalowski2008}.

For the better part of a decade, this general picture became the status-quo for the assumed host properties of LGRBs: blue, low-luminosity, low-mass, star-forming galaxies with low metal content.  However, more recently this previously quite uniform picture of GRB hosts became somewhat more diverse: several metal-rich GRB hosts were discovered \cite{Levesque2010,Elliott2013,Schady2015}, which revealed a population of red, high-mass, high-luminosity hosts that were mostly associated with dust-extinguished afterglows \cite{Kruehler2011,Hjorth2012,Rossi2012,Perley2013}.  Next, the offset of GRB-selected galaxies towards lower metallicities in the mass-metallicity relation \cite{Levesque2010} could, for example, be partially explained with the dependence of the metallicity of star-forming galaxies on SFR \cite{Mannucci2011,KocevskiWest2011}.   Moreover, it was shown that LGRBs do not exclusively formed in low-metallicity environments \cite{Berger2007,Kruehler2011,Perley2013,Perley2015}, where the results of \cite{Kruhler2015} are an excellent example of this notion.  Analysing the largest sample of LGRB-selected host spectra yet considered (up to $z=3.5$), they found that a fraction of LGRBs occur in hosts that contain super-solar ($Z> Z_{\odot}$) metal content ($<20\%$ at $z=1$).  This shows that while some LGRBs can be found in high-$Z$ galaxies, this fraction is significantly less than the fraction of star-forming regions in similar galaxies, indicating GRBs are actually quite scarce in high-metallicity hosts.  They found a range of host metallicities of 12+log(O/H) = 7.9 to 9.0, with a median of 8.5.  \cite{Kruhler2015} therefore concluded that GRB host properties at lower redshift ($z<1-2$) are driven by a given LGRB's preference to occur in lower-metallicity galaxies without fully avoiding metal-rich ones, and that one or more mechanism(s) may operate to quench GRB formation at the very highest metallicities.  This result supported similar conclusions from numerous other recent studies \cite{Kocevski2009,Levesque2010,Elliott2013,GrahamFruchter2013,Perley2013,Vergani2015,Trenti2015,Perley2015,Schulze2015} which show that LGRBs seem to prefer environments of lower metallicity, with possibly no strict cut-off in the upper limit of metal content (though see \cite{GrahamFruchter2015}).

Another revealing observation was made by \cite{Kelly2014} who showed that, low-$z$ SNe IcBL and $z<1.2$ LGRBs (i.e. core-collapse explosions in which a significant fraction of the ejecta moves at velocities exceeding $20,000-30,000$~km~s$^{-1}$) preferentially occur in host galaxies of high stellar-mass and star-formation densities when compared with SDSS galaxies of similar mass ($z<0.2$).  Moreover, these hosts are compact for their stellar masses and SFRs compared with SDSS field galaxies.  More importantly, \cite{Kelly2014} showed that the hosts of low-$z$ SNe IcBL and $z<1.2$ LGRBs have high gas velocity dispersions for their stellar masses.  It was shown that core-collapse SNe (types Ibc and II) showed no such preferences.  It appears that only SLSNe occur in more extreme environments than GRB-SNe and relativistic SNe IcBL: \cite{Leloudas2015} showed that SLSNe occur in extreme emission-line galaxies, which are on average more extreme than those of LGRBs, and that type I SLSNe may result from the the very first stars exploding in a starburst, even earlier than LGRBs.   Finally, \cite{Kelly2014} concluded that the preference for SNe IcBL and LGRBs for galaxies with high stellar mass densities and star-formation densities may be just as important as their preference for low metallicity environments.

The result of \cite{Kelly2014} is the latest in a long line of investigations that suggest that LGRBs are useful probes of high-$z$ star formation.  This results stems from a long-debated question of whether LGRBs may be good tracers of the universal star-formation rate over all of cosmic history \cite{Bloom2002,Firmani2004,PriceSchmidt2004,Natarajan2005,Chary2007,Savaglio2009,Perley2013,Kruhler2015,Perley2015,Greiner2015_hosts}.  \cite{Kruhler2015} showed that there is an increase in the (median) SFR of their sample of LGRB host galaxies at increasing redshift, where they found 0.6~M$_{\odot}$~yr$^{-1}$ at $z\approx0.6$ to 15.0~M$_{\odot}$~yr$^{-1}$ at $z\approx2.0$.  Moreover, these authors suggest that by $z\sim3$ GRBs host will probe a large fraction of the total star-formation. In absence of further secondary environmental factors, GRB hosts would then provide an extensive picture of high-redshift star-forming galaxies.  However, the connection between LGRBs and low-metallicity galaxies may hinder their utility as unbiased tracers of star formation \cite{Stanek2006,Kewley2007,Modjaz2008_metal,WangDai2014}, though if LGRBs do occur in galaxies of all types, as suggested above, then they may be only mildly biased tracers of star formation \cite{Japelj2016}.

\subsection{Immediate Environments}
\label{sec:host_immediate}

Most LGRB host galaxies are too distant for astronomers to discern their spatially resolved properties.  These limitations are important to consider when extrapolating LGRB progenitor properties from the global host properties, as it may be possible that the location of a given LGRB may differ to that of the host itself.  Where spatially resolved studies have been performed, such as for GRB~980425 \cite{Christensen2008,LeFloch2012,Michalowski2014,Arabsalmani2015}, GRB~060505 \cite{Thoene2008}, GRB~100316D \cite{Levesque2011} and GRB~120422A \cite{Levesque2012,Schulze2014}, it was found that in at least two of these cases, the metallicity and SFR of other H~\textsc{ii} regions in their hosts had comparable properties as those associated with the LGRB location (within $3\sigma$).  In these studies the host galaxies had a minimal metallicity gradient \cite{Thoene2008}, and there were multiple low-metallicity locations within the host galaxies, where in some cases the location of the LGRB was in that of the lowest metallicity \cite{Levesque2012}.  These studies suggest that in general, the host-wide metallicity measurement can be used as a first-order approximation of the LGRB site.

Next, the line ratios of [Ne \textsc{iii}] to [O \textsc{ii}] suggest that H \textsc{ii} regions associated with LGRBs are especially hot \cite{Bloom2001}, which may indicate a preference for the hosts of LGRBs to produce very massive stars.  Absorption line spectroscopy have revealed some fine-structure lines (e.g. Fe [\textsc{ii}]), which could indicate the presence of absorption occurring in fast-moving winds emanated by WR stars (i.e. stars that are highly stripped of their outer layers of hydrogen and helium).  The distances implied by variable fine-structure transitions (e.g. their large equivalent widths imply large distances to avoid photoionization) show that occur at distances of order tens to hundreds of parsecs from the GRB itself \cite{Prochaska2006,Vreeswijk2007,DElia2007,Fynbo2014}, which makes sense given that the dust and surrounding stellar material around a GRB is completely obliterated by the explosion.  Such absorption must arise from nearby WR stars whose winds dissect the line-of-sight between the GRB and Earth.

The type of environment in which a given LGRB occurs is also of interest: is it a constant interstellar medium (ISM) or a wind-like medium?  Do the progenitors of LGRB carve out large wind-blown bubbles \cite{ChevLi2000,Mirabal2003}, as has been observed for galactic WR stars \cite{TreffersChu1982,Toala2015}?  Using a statistical approach to the modelling of GRB AGs, \cite{Schulze2011} demonstrated that the majority of GRBs (L- and SGRBs) in their sample (18/27) were compatible with a constant ISM, and only six showed evidence of a wind profile at late times.  They concluded that, observationally, ISM profiles appear to dominate, and that most GRB progenitors likely have relatively small wind termination-shock radii, where a variable mass-loss history, binarity a dense ISM, a weak wind, can bring the wind-termination shock radius closer to the star \cite{vanMarle2005,vanMarle2006}.  A smaller group of progenitors, however, seem to be characterised by significantly more extended wind regions \cite{Schulze2011}.  In this study, the AG is assumed to be powered by the standard forward-shock model, which has been shown to not always be the best physical description for all LGRBs observed in nature \cite{dePasquale2016}.  

Finally, it appears that LGRBs generally occur in environments that possess strong ionization fields, which likely arise from hot, luminous massive stars in the vicinity of LGRBs.  \cite{Kruhler2015} showed that the GRB hosts in their sample occupied a different phase-space than SDSS galaxies in the Baldwin-Phillips-Terlevich (BPT) diagram \cite{Baldwin1981}: they are predominantly above the ridge line that denotes the highest density of local star-forming field galaxies. A similar offset was also observed for galaxies hosting type I SLSNe \cite{Leloudas2015}. This offset is often attributed to harder ionization fields, higher ionization parameters or changes in the ISM properties \cite{Brinchmann2008,Kewley2013,Steidel2014}.  This result is consistent with the hypothesis that the difference in the location in the BPT diagram between GRB hosts and z$\sim$0 star-forming galaxies is caused by an increase in the ionization fraction, i.e., for a given metallicity a larger percentage of the total oxygen abundance is present at higher ionization states at higher redshifts. This could be caused by a harder ionization field originating from hot O-type stars \cite{Steidel2014} that emit a large number of photons capable of ionizing oxygen into [O \textsc{iii}].

\subsection{Implications for Progenitor Stars}

Before LGRBs were conclusively associated with the core-collapse of a massive star, their massive-star origins were indirectly inferred.  If LGRBs were instead associated with the merger of binary compact objects, two ``kicks'' arising from two SN explosions would imply a long delay before coalescence, and likely lead to GRBs occurring at large distances from star-forming regions \cite{Paczynski1998,Bloom1999,Fryer1999,Belczynski2000}.  With sub-arcsecond localization came observations that showed LGRBs, on average, were offset from the apparent galactic centre by roughly 1~kpc \cite{Bloom2002}, which did not agree with a compact-object binary merger scenario.  Further statistical studies showed a strong correlation between the location of LGRBs and the regions of bluest light in their host galaxies \cite{Fruchter2006,Svensson2010}, which implied an association with massive-star formation.  Thus result was furthered by \cite{Kelly2008} that showed that LGRBs and type Ic SNe have similar locations in their host galaxies, providing additional indirect evidence of LGRBs and massive stars.

The general consensus that LGRBs occur, on average, in metal-poor galaxies (or location within more metal-rich hosts), aligned well with theoretical expectations that LGRB formation has a strong dependence on metallicity.  In theoretical models \cite{MacFadWoosley1999,KudritzkiPuls2000,MeynetMaeder2005,Hirschi2005,YoonLanger2005,WoosleyHeger2006}, the progenitors of LGRBs need to be able to lose their outer layers of hydrogen and helium (as these transitions are not observed spectroscopically), but do so in a manner that does not remove angular momentum from the core (to then power the GRB).  At high metallicities, high mass loss rates will decrease the surface rotation velocities of massive stars, and due to coupling between the outer envelopes and the core, will rob the latter of angular momentum and hence the required rapid rotation to produce a GRB.  In quasi-chemically homogeneous models \cite{YoonLanger2005,WoosleyHeger2006,Song2016}, rapid rotation creates a quasi-homogeneous internal structure, whereby the onion-like structure retained by non- or slowly 	rotating massive stars is effectively smeared out, and the recycling of material from the outer layers to the core results in the loss of hydrogen and helium in the star because it is fused in the core.  Intriguingly, quasi-chemically homogeneous stars do appear to exist in nature.  The FLAMES survey \cite{Hunter2008} observed over 100 O- and B-type stars in the Large Magellanic Cloud (LMC) and the Milky Way galaxy, and showed the presence of a group of rapidly rotating stars that were enriched with nitrogen at their surfaces.  The presence of nitrogen at the surface could only be due to rotationally triggered internal transport processes that brought nuclear processed material, in this case nitrogen, from the core to the stellar surface.  Observations of metal-poor O-type stars in the LMC by \cite{Bouret2003,Walborn2004} show the signature of CNO cycle-processed material at their surfaces, while modelling of the spectra of galactic and extragalactic oxygen-sequence WR stars shows very low surface He mass fractions, thus making them plausible single-star progenitors of SNe Ic \cite{Tramper2015}.  

However, other observations of Local Group massive star populations have revealed that the WR population actually decreases strongly at lower metallicities, particularly the carbon- and oxygen-rich subtypes \cite{Massey2003}, suggesting that these proposed progenitors may be extremely rare in LGRB host environments.  Moreover, the results of \cite{Zauderer2013}, based on the analysis of two LGRBs, suggest that some LGRBs may be associated with progenitors that suffer a great degree of mass-loss before exploding, and hence a great deal of core angular momentum.   Moreover, the association of some LGRBs with super-solar metallicity environments also contradicts the predictions of the collapsar model.  However, other recent models have considered alternative evolutionary scenarios whereby LGRB progenitors can lose a great deal of mass before exploding, but still retain enough angular momentum to power a GRB \cite{Ekstrom2012,Georgy2012,Groh2013}.  Such models consider the complex connection between surface and core angular momentum loss, and show that single stars arising from a wide range of metal content can actually produce a GRB.  Moreover, the effects of anisotropic stellar winds need to also be considered \cite{Levesque2010}.  Polar mass loss remove considerably less angular momentum than equatorial mass loss \cite{Maeder2002}, which provides the means for the progenitor to lose mass, but sustain a high rotation rate.  Alternatively, episodic mass loss, as has been observed for luminous blue variable stars may also offers another means of providing a way to lose mass but retain core angular momentum.

\section{Kilonovae associated with SGRBs}
\label{sec:SGRB_KNe}

\begin{figure}
 \centering
 \includegraphics[bb=0 0 1798 624,scale=0.26,keepaspectratio=true]{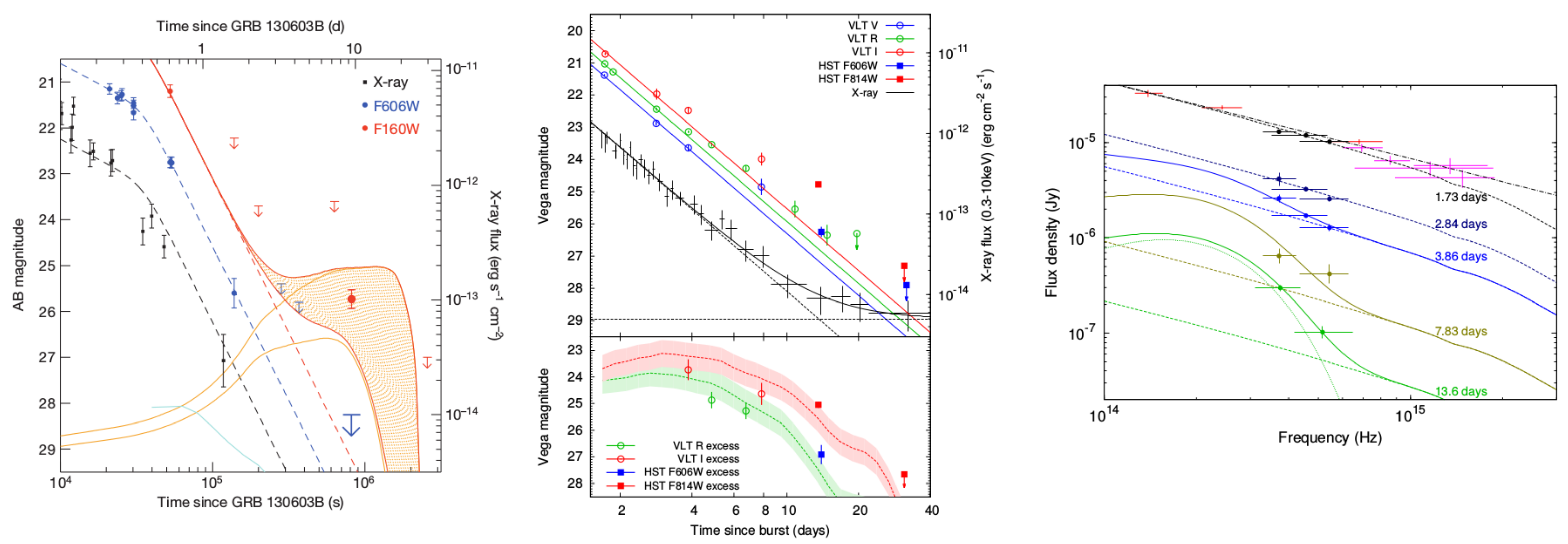}
 \caption{Observations of KNe associated with SGRBs: \textbf{Left:} GRB~130603B, from \cite{Tanvir2013}.  The decomposed optical and NIR LCs show an excess of flux in the NIR ($F160W$) filter, which is consistent with theoretical predictions of light coming from a KN.  \textbf{Middle:} GRB~060614, from \cite{Jin2015}.  Multi-band LCs show an excess in the optical LCs ($R$ and $I$), which once the AG light is removed, the resultant KN LCs match those from hydrodynamic simulations of a BH-NS merger (ejecta velocity of $\sim0.2c$ and an ejecta mass of 0.1~M$_{\odot}$ \cite{Tanaka2014}).  \textbf{Right:}  SEDs of the multi-band observations of GRB~060614, also from \cite{Jin2015}.  The early SEDs are well described by a powerlaw spectrum, which implies synchrotron radiation.  However, at later epochs the SEDs are better described by thermal, black-body spectra, with peak temperatures of $\sim2700$~K, which are in good agreement with theoretical expectations \cite{Kasen2013}. }
 \label{fig:KNe}
\end{figure}

To date, the amount of direct and indirect evidence for the massive-star origins of LGRBs is quite comprehensive, and thoroughly beyond any conceivable doubt.  The same however cannot be stated about the progenitors of SGRBs.  For many years, since the discovery that there are two general classes of GRBs \cite{Mazets1981,Kouveliotou1993}, general expectations were that they arose from different physical scenarios, where SGRBs are thought to occur via the merger of a binary compact object system containing at least one neutron star (i.e. NS-NS or NS-BH).  Circumstantial evidence for the compact object merger origins of SGRBs \cite{Nakar2007,Berger2014}, include their locations in elliptical galaxies, the lack of associated supernovae (as observed for LGRBs) \cite{Hjorth2005_050509B,DellaValle2006_060614,Fynbo2006,Gehrels2006}, the distribution of explosion-site offsets relative to their host galaxies ($0.5-75$~kpc away, median of 5~kpc \cite{Fong2010,FongBerger2013}) and a weak spatial correlation of SGRB locations with star-formation activity within their host galaxies.

The compact coalescence scenario predicts SGRB AGs at longer wavelengths \cite{Eichler1989,Narayan1992,LeeRR2007,Nakar2007,Berger2014}, which have been observed \cite{Gehrels2005}.   As well as the expected AG emission, emission from a SN-like transient was also predicted \cite{LiPacz1998,Kulkarni2005,Rosswog2005,Metzger2010,MetzBerg2012}, which have been referred to as a ``kilonova'' (KN), ``mergernova'', or ``macronova'' (see the recent review by \cite{Tanaka2016}), where we have adopted the former terminology in this review.

The KN prediction is a natural consequence of the unavoidable decompression of NS material, where a compact binary coalescence provides excellent conditions for the rapid-neutron capture process ($r$-process \cite{LattimerSchramm1974,Eichler1989,Woosley1994,Freiburghaus1999,LipRob2015}).  The neutron capture process occurs very quickly, and is completed in less than a second, and it leaves behind a broad distribution of radioactive nuclei whose decay, once the ejected material becomes transparent, powers an electromagnetic transient in a process similar to that expected to causes GRB-SNe to shine.  Hydrodynamic simulations suggest that during a merger, mass is ejected via two mechanisms: (I) during the merger, surface layers may be tidally stripped and dynamically flung out in tidal tails; (II) following the merger, material that has accumulated into a centrifugally supported disk may be blown off in a neutrino or nuclear-driven wind.   In mechanism (I), the amount of material ejected depends primarily on the mass ratio of the compact objects and the equation of state of the nuclear matter.  The material is very neutron-rich ($Y_{\rm e}\sim 0.1$), and the ejecta is expected to assemble into heavy ($Z>50$) elements (including Lanthanides, $58<Z<70$ and Actinides, $90<Z<100$) via the $r$-process.   In mechanism (II), however, neutrinos emitted by the accretion disk raise the electron fraction ($Y_{\rm e}\sim 0.5$) to values where a Lanthanide-free outflow is created \cite{MetzFern2014}.  In both cases $10^{-4}-10^{-1}$ M$_{\odot}$ of ejecta is expected to be expelled.  A direct observational consequence of mechanism (I) is a radioactively powered transient that resembles a SN, but which evolves over a rapid time-scale ($\sim$ 1 week, due to less material ejected compared with a typical SN) and whose spectrum peaks at IR wavelengths.  In contrast to other types of SNe, e.g. SNe Ia whose optical opacity is dictated by the amount of iron-group elements present in the ejecta, $r$-process ejecta that is composed of Lanthanides has a much larger expansion opacity ($\approx 100$ times greater) due to the atoms/ions having a greater degree of complexity in the number of ways in which their electrons can be arranged in their valence shells (relative to iron-group elements).


There have been a handful of observational searches for KN emission: GRB~050709 \cite{Hjorth2005_050709,Jin2016}; GRB~051221A \cite{Soderberg2006_SGRB}; GRB~060614 \cite{Yang2015,Jin2015}; GRB~070724A \cite{Berger2009,Kocevski2010}; GRB~080503 \cite{Perley2009,Gao2015}; GRB~080905A \cite{Rowlinson2010}; and GRB~130603B \cite{Tanvir2013,Berger2013}.  In almost all cases null results were obtained, with the notable exceptions being GRB~130603B, GRB~060614 (see Fig. \ref{fig:KNe}) and GRB~050709.  In these cases, the optical and NIR LCs required a careful decomposition, and once the AG components were accounted for, an excess of emission was detected.  In the case of GRB~130603B, a single NIR datapoint was found to be in excess of the extrapolated AG decay, which was interpreted by \cite{Tanvir2013} as arising from emission from a KN.  The (observer-frame) colour term $R_{F606W} - H_{F160W} < 1.7$~mag at +0.6~d, and $R_{F606W} - H_{F160W} < 2.5$~mag at +9~days, which is inconsistent with a colour change due to FS emission, and was argued to be evidence of non-synchrotron emission arising from a possible KN.  The dataset of GRB~060614 considered by \cite{Jin2015} is more extensive than that of GRB~130603B, and KN bumps were detected in two filters (observer-frame $R$ and $I$), and which peaked at $4-6$~d (rest-frame).  The decomposed KN LCs were shown to be consistent with LCs arising from hydrodynamic simulations of a BH-NS merger, which had an ejecta velocity of $\sim0.2c$ and an ejecta mass of 0.1~M$_{\odot}$ \cite{Tanaka2009}.  The larger dataset also allowed for the construction of SEDs, which showed a clear transition from a powerlaw spectrum at early epochs ($<3$~d), which appeared to transition into a thermal, blackbody spectrum over the next two weeks.  Moreover, the inferred temperature of the black-body was around 2700~K, which fitted well with theoretical expectations.  However, the precise nature of GRB~060614 is still not understood, and it is still uncertain if it is a short or a long GRB.

\section{Theoretical Overview}
\label{sec:Theoretical_models}
\vspace*{0.5cm}

While the focus of this review is geared towards what observations tell us about the GRB-SN connection, a keen understanding of the leading theoretical models is also required.  The finer intricacies of each model are presented elsewhere, and we suggest the reader to start with the comprehensive review by \cite{Piran2004} (and references therein), which is just one of many excellent reviews of the physics of the prompt emission and AGs.  As such, what is presented here is meant only as an overview of the rich and complex field of GRB phenomenology.

\subsection{Central-engine models - millisecond magnetars vs collapsars}
\label{subsec:centralengines}

The main consensus of all GRB models is that LGRBs and their associated SNe arise via the collapse of massive stars, albeit ones endowed with physical properties that must arise only seldom in nature, given the fact GRB-SNe are very rare.  In the leading theoretical paradigms, after the core-collapse of the progenitor stars, the leftover remnant is either a NS or a BH, and under the correct conditions, both can operate as a central engine to ultimately produce an LGRB.

In reality, very few solid facts are known about the true nature of the central-engine(s) operating to produce LGRBs. Nevertheless, one of the most prevailing models of the central engines of GRBs associated with SNe is the collapsar model~\cite{Woosley1993,MacFadWoosley1999,MacFadyen2001}, where the accretion of material from a centrifugally supported disk onto a BH leads to the launch a bipolar relativistic jet, and material within the jet leads to the production of $\gamma$-ray emission. The collapsar model suggests that there is enough kinetic energy ($2-5\times 10^{52}$ erg) in the accretion disk wind which can be used to explosively disrupt the star, as well as synthesizing $\sim0.5$~M$_\odot$ of $^{56}$Ni. In this model, the duration of the prompt emission is directly related to the stellar envelope infall time, and the jet structure is maintained either magnetically or via neutrino-annihilation-driven jets along the rotation axes.  The other promising mechanism that could lead to the production of an LGRB and its hypernovae is the millisecond magnetar model~\cite{Usov1992,Metzger2007,Buc2008,Metzger2011}. In this scenario, the compact remnant is a rapidly rotating ($P \sim 1-10$~ms), highly magnetized ($B~\sim~10^{14-15}$~G) NS, where the relativistic Poynting-flux jets are supported by stellar confinement \cite{Buc2008}.  

\begin{figure}
 \centering
 \includegraphics[bb=0 0 1280 545,scale=0.35,keepaspectratio=true]{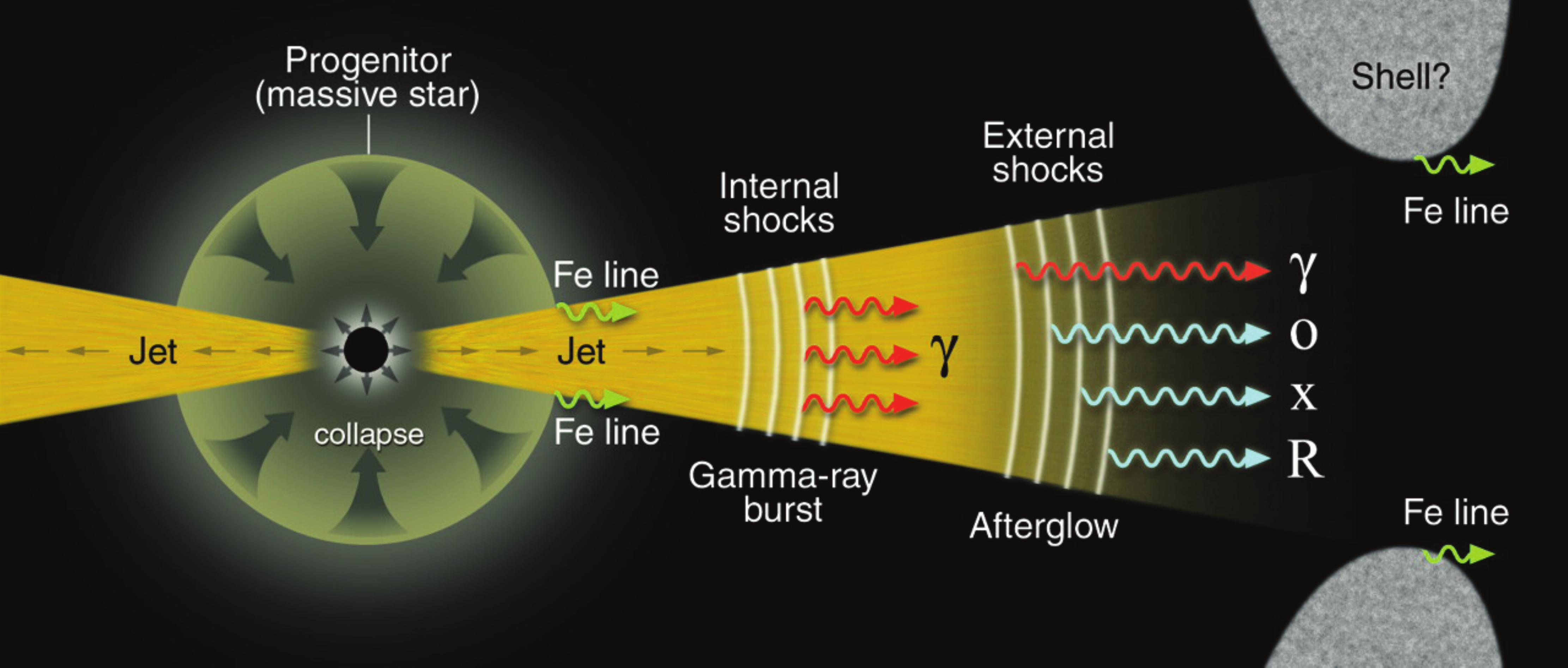}
 \caption{The death of a massive star produces a GRB (and its multi-band AG) and an energetic and bright SN (from \cite{Meszaros2001}).}
 \label{fig:meszaras2001}
\end{figure}

A cartoon visualization of the formation of an LGRB, including its AG and associated SN is shown in Fig. \ref{fig:meszaras2001}.  In the standard fireball model, shells of material within the jet interact to produce the initial burst of $\gamma$-rays, called the prompt emission, via internal shocks.  As the jet propagates away from the explosion site, it eventually collides with the surrounding medium producing external shocks that power an AG that is visible across almost the entire electromagnetic spectrum, from X-rays to radio, and which lasts for severals weeks to months.  In this leptonic model, the prompt and AG radiation is synchrotron or synchrotron-self-Compton in origin \cite{Rees1994}.  It is interesting to note that this scenario is pretty much independent of the nature of the central engine -- all that is required is the formation of an ultra-relativistic jet.  It is generally thought that luminous GRBs with bulk Lorentz factors of order $\Gamma_{\rm B} \sim 300$ must stem from ultra-relativistic collisionless jets produced by millisecond magnetars and/or collapsars.  As discussed in Section \ref{sec:Subclasses}, in order to penetrate the stellar envelope, the active timescale of the jet produced by the central engine ($t_{\rm engine}$) must be longer than the penetrating timescale, where the latter is $\sim R/v_{\rm jet}$. Some $ll$GRBs whose $\Gamma_{\rm B}\sim 2$ can also be explained by this model, but in these cases, the active timescale is likely to be slightly smaller than the penetrating timescale so that the ultra-relativistic jet from the central engine either just barely, or completely fails to completely penetrate the stellar envelope.

The first class of models for the prompt emission of GRBs was the internal-shock model, where synchrotron, or synchrotron self-Compton radiation was emitted by electrons that were accelerated by internal shocks~\cite{Rees1994,Sari1998} in the form of high-energy $\gamma$-ray photons. Inverse Compton (IC) scattering and synchrotron self-Compton (SSC) scattering can enhance the seed photons and account for the very high-energy $\gamma$-ray photons measured for some GRBs. One prediction of the internal-shock model is the production of high-energy neutrinos, which to date have not been observed by neutrino detectors such as IceCube (only upper limits have been obtained so far, see the review by \cite{Hummer2012}).  Although more detailed calculations performed by \cite{LiZ2012,He2012} have demonstrated that the internal-shock model which includes benchmark parameters (e.g., the bulk Lorentz factor $\Gamma_{\rm B}=300$) is consistent with the upper limits obtained by IceCube, these results have posed more stringent constraints on the internal shocks model because there is a possible correlation between the bulk Lorentz factor $\Gamma_{\rm B}$ and the GRB luminosity \cite{Liang2010,Ghirlanda2012,Lv2012}.  Instead, alternative scenarios in the context of the ultra-relativistic jet model are the photospheric emission model~\cite{Peer2011,Lundman2013,Ito2015,Santana2016,Peer2016}, and the Internal-Collision-induced MAgnetic Reconnection and Turbulence (ICMART) model~\cite{ZhangYan2011,Deng2016}.  Photospheric models assume that thermal energy stored in the jet is radiated as prompt emission at the Thomson photosphere \cite{Paczynski1986,Thompson1994,MesRees2000}, while ICMART models envisage that collisions between ``mini-shells'' in a Poynting-flux-dominated outflow distorts the ordered magnetic field lines in a run-away manner, which accelerates particles that then radiate synchrotron $\gamma$-ray photons at radii of $\sim 10^{15} - 10^{16}$~cm \cite{ZhangYan2011,Deng2016}.

\subsection{Shock breakout models}

It has long been believed that when the diffusion timescale of photons at the shock-wave front is comparable to the dynamical timescale, a SBO can occur (see as well Section \ref{sec:Subclasses}). The SBO of a CCSN can produce a brief and bright flash whose spectral energy distribution peaks in the near UV or X-ray regimes \cite{Colgate1968,Colgate1974,Klein1978,Falk1978,Ensman1992,MatzMcKee1999,TanMatzMcKee2001,Wang2007,Katz2010,Bromberg2011,NakarSari2012,Matzner2013,Duran2015}. When the SN progenitor is a red supergiant whose radius is larger than several hundred $R_{\odot}$, the SBO is non-relativistic (Newtonian) \cite{Klein1978,Falk1978,Ensman1992} and the emission is dominated by optical and UV radiation, which is detectable with space telescopes \cite{Schawinski2008,Gezari2015,Garnavich2016}. When the explosion is energetic enough, and the progenitor is a WR star whose radius is of order a few solar radii, the SBO emission typically peaks at X-rays or soft $\gamma$ rays, with a duration of $\sim 10-2000$~s. This class of relativistic SBOs can naturally explain some LGRBs \cite{Bromberg2011,NakarSari2012,Matzner2013,Duran2015}.  \cite{Bromberg2011} demonstrated that $ll$GRB jets either fails, or just barely pierces through the stellar envelope. This choked/stifled jet can also help accelerate the shock to a mildly relativistic velocities ($\sim 30,000-100,000$ km s$^{-1}$).  In the shock breakout model, the AG emission is produced when the stellar ejecta collides with the CSM, and \cite{Duran2015} showed that the data of the afterglows of GRBs 980425, 031203, 060218 and 100316D are in good agreement with the predictions of this model.

\section{Future Research}
\label{sec:Future_Research}
\vspace*{0.5cm}

While considerable progress has been made in the field of GRB-SNe, there are still uncertainties related to several aspects of their true nature.  Solidifying their role as standardizable candles and cosmological probes requires both more work and considerably more events.  Indeed for GRB-SNe to be used as cosmological probes, independent distance measurements to their host galaxies need to be obtained.  Sample studies of GRB-SNe are the ideal way to approach this question, and with the hopeful launch of \emph{JWST} in the next few years, their use over larger redshift ranges than SNe Ia could make them appealing cosmological candles.  Additional attention is also required to determine the physical configuration and properties of their pre-explosion progenitor stars, to help address the question of whether they arise from single vs. binary systems.  Moreover, further ULGRB-SNe are needed to address the question of whether all are ultra-luminous compared with typical GRB-SNe, as seen for SN~2011kl, or whether this event is quite anomalous.

\subsection{Role of binarity}

Throughout this review, discussions of their stellar progenitors were primarily focused on single-star candidates. However the role of binarity may prove to be one of the most important ingredients to eventually producing a GRB.   Theoretically, there are strong motivations for considering a binary evolution.  To date, the best theoretical stellar models find it hard to produce enough angular momentum in the core at the time of collapse to make a centrifugally supported disk, though some progress has been made \cite{Heger2003,YoonLanger2005,WoosleyHeger2006}.  Instead, it is possible to impart angular momentum into the core of a star through the inspiral of a companion star during a common envelope phase (CEP) \cite{Chevalier2012}: i.e. converting orbital angular momentum into core angular momentum.  The general idea is to consider a binary system comprised of, among others, a red supergiant and a NS \cite{BarkovKom2011}, a NS with the He core of a massive star \cite{ZhangFryer2001} or the merger of two helium stars \cite{FryerHeger2005}.  During the inspiral of the compact object into the secondary/companion, angular momentum is imparted to the core, which is spun up via disc accretion.  During this process, the core of the secondary will increase in mass as well as gain additional angular momentum, while the inspiralling NS will also accrete gas via the Bondi-Hoyle mechanism, which can lead to the NS reaching periods of order milliseconds before it eventually merges with the secondary's core.  If a merger of the NS with the core occurs, a collapsar can be created, where a GRB can be produced depending on the initial mass of the secondary, the spin of the newly formed BH and the amount of angular momentum imparted to the BH.  

For the binary model to be a viable route for LGRB formation, one or more mechanism is required to expel the outer envelopes out into space prior to explosion.  Generally there are different ways for this to be achieved, either through non-contact methods such as stellar winds, semi-contact processes such as roche-lobe overflow, or through contact mechanisms that operate during a CEP.  The spin-rates of a small sample of O-type star and WR binaries indicate that Roche lobe overflow mass transfer from the WR progenitor companion may play a critical role in the evolution of WR--O-star binaries, where equatorial rotational velocities of 140--500~km~s$^{-1}$ have been measured \cite{Shara2015}.  In the CE scenario, during the inspiral, the orbital separation decreases via drag forces inside the envelope which also results in a loss of kinetic energy.  Some of this energy is lost to the surrounding envelope, which heats up and expands.  Over a long-enough period the entire envelope can be lost into space.  Another mechanism to expel the CE arises via nuclear energy rather than orbital energy \cite{Podsiadlowski2010}.  For example, during the slow merger of a massive primary that has completed helium-core burning with a $1-3$~M$_{\odot}$ secondary, H-rich material from the secondary is injected into the He-burning shell of the primary.  This leads to nuclear runaway and the explosive ejection of the H and He envelopes, and produces a binary comprised of a CO star and the low-mass companion.  Should a further merger occur, this could lead to the formation of a GRB.  If GRB-SNe arise via this formation channel, then this scenario can naturally explain why GRB-SNe are all of type Ic.



A generalization of the binary-merger model is that the more massive the stars are, the more accretion will occur.  This in turn leads to more convection in the core, which results in larger magnetic fields being generated and hence more magnetic collimation for any jets that are produced.  In the case of GRB-SNe vs. SNe Ibc, if jets are ubiquitous, then the difference between them may be the mass of the merging stars, where lower masses imply lower magnetic-fields, and hence less collimation.  Moreover, the mass ratio of the secondary to the primary is also important, where higher mass ratios will result in more asymmetric explosions \cite{Chevalier2012}.

There is a growing list of observations that show that most massive stars exist in binaries, including \cite{Sana2012} who estimated that over 70\% of all massive stars will exchange mass with a companion star, which in a third of all cases will lead to a merger of the binary system.  Moreover, closely-orbiting binaries are more common at lower metallicities \cite{Linden2010}, where the progenitors of GRBs are normally found (though see \cite{NeugentMassey2014} who showed that the close binary frequency of WRs is not metallicity dependent). Additional support for the notion that the progenitors of SNe Ibc are massive stars in binary systems has come from \cite{Smith2011} who argue that for a standard initial mass function, the observed abundances of the different types of CCSNe are not consistent with expectations of single-star evolution.  Progenitor non-detections of 10 SNe Ibc strongly indicate that single massive WR stars cannot be their solitary progenitor channel \cite{Smartt2009}.  \cite{Eldridge2013} derived a 15\% probability that all SNe Ibc arise from single-star WR progenitors.  The large gas velocity dispersions measured for the host galaxies of GRBs by \cite{Kelly2014} may imply the efficient formation of tight massive binary progenitor systems in dense star-forming regions.  Rotationally supported galaxies that are more compact and have dense mass configurations are expected to have higher velocity dispersions.   Observations of extra galactic star clusters show evidence that bound-cluster formation efficiency increases with star-formation density \cite{Goddard2010,SilvaVilla2013}.  Binaries may form more frequently in bound clusters, and they evolve to become more tightly bound through dynamical interactions with other members of the cluster.  Alternatively, if the progenitors of GRBs are actually single stars, but which are more massive than those that produce SNe Ibc and II, a top-heavy initial-mass function (IMF) in dense, highly star forming regions can also explain their observations.  A similar conclusion was made by \cite{Leloudas2015}, who suggested that if the progenitors of SLSNe are single stars, the extreme emission-line galaxies in which they occur may indicate a bottom-light IMF in these systems.  However, observations of low-mass stars in elliptical galaxies that are thought to have undergone high star formation densities in their star-forming epochs, instead suggest that the IMF is bottom heavy \cite{vanDokkumConroy2010,ConroyvanDokkum2012}.

A major hurdle therefore is finding ways to provide observational evidence to distinguish between single vs. binary progenitors.  One such indication may be idea that the progenitors of GRB-SNe are ``runaway'' stars: i.e. massive stars ejected from compact massive star clusters \cite{Hammer2006,Eldridge2011}.  This notation was prompted by the observation that the very nearest GRB-SNe, which can be spatially resolved in their host galaxies, are offset from the nearest sites of star-formation by $400-800$~pc.  If GRB-SNe do arise from runaway stars, the lack of obvious wind-features in AG modelling (Sect. \ref{sec:host_immediate}) can naturally be explained:  simulations \cite{Yadav2016} suggest that a high density of OB stars are required to produce the $r^{-2}$ wind profile, in the region of $10^4 - 10^5$ OB stars within a few tens of parsecs.  This is a much larger density than has  been observed in nature, where the densest known cluster is R136 (e.g. \cite{Crowther2010,Crowther2016}) which contains many of the most massive and luminous stars known, including R136a1 (M$\sim315$~M$_{\odot}$, L$\sim8.7\times10^6$~L$_{\odot}$). Within the central five parsecs of R136 there are 32 of the hottest known type O stars (spectral type O2.0$-$3.5), 40 other O stars, and 12 Wolf-Rayet stars, mostly of the extremely luminous WNh type (which are still burning hydrogen in their cores, and have nitrogen at their surfaces).

For non-GRB related SNe, such as the very nearby peculiar type II SN~1987A (in the LMC, $D\sim50$~kpc), constraining the nature of its progenitor was made possible due to a combination of a spatially resolved SN remnant and an enormously rich photometric and spectroscopic dataset compiled over a time-span of nearly three decades.  These observations have shown that the most likely progenitor of SN~1987A was the merger of a binary system \cite{Podsiadlowski1992,MorrisPodsiad2007}, which can explain the triple-ringed structure seen in \emph{HST} images \cite{Burrows1995}, as well as explain the He-enriched outer layers of the blue supergiant progenitor \cite{Sonneborn1997}.  It was also shown that type IIb SN~1993J ($\sim3.5$~Mpc) likely originated from a binary system via analysis of its early LC \cite{Podsiadlowski1993}, hydrodynamical modelling \cite{Woosley1994_93J}, and by detection of the pre-explosion progenitor star in spatially resolved \emph{HST} images \cite{Maund2004} and a possible companion \cite{Fox2014}.  The direct imaging revealed the progenitor was a red supergiant, where excess of UV and $B$-band flux implied the presence of a hot stellar companion, or it was embedded in an unresolved young stellar cluster.  These studies are possible because of the close proximity of the SNe to our vantage point as observers on Earth.  However, the nearest GRB to date is GRB~980425, which at $\sim40$~Mpc, means the progenitor is too distant to be direct imaged.  For any progress to be made concerning single vs. binary progenitors, nearby events are required that will allow either for exceptionally detailed observations to be obtained and modelled, or even the remote chance of directly detecting the progenitor.  For lack of better ideas, what we then require is a healthy dose of patience.

\subsection{From GRB-SNe to ULGRB-SNe to SLSNe}

As discussed previously, the most luminous GRB-SNe to date is SN~2011kl, which had a peak absolute magnitude of $\approx -20$ mag \cite{Kann2016}.  This is roughly $0.5-1.0$~mag brighter than most GRB-SNe, but still one magnitude fainter than those associated with SLSNe, which peaked at $\approx -21$ mag \cite{GalYam2012}.  Moreover, it appears that SN~2011kl is not the only object that falls in this gap between ordinary SNe and SLSNe: four objects discovered by PTF and the SNLS have similar peak absolute magnitudes and LC evolution as SN~2011kl \cite{Arcavi2016}.  No accompanying $\gamma$-ray emission was detected for any of these events, which begs the question of whether they are off-axis ULGRB-SNe, or represent yet another type of explosion transient.

In contrast to the cases of GRB-SNe whose optical light curves appear to be mainly powered by heating arising from $^{56}$Ni decay, it seems that most SLSNe cannot be explained by the simple radioactive heat deposition model.  Instead the luminosity of SLSNe appears to be either driven by energy input from a magnetar \cite{Nicholl2013,Inserra2013} or powered by the interaction between SN ejecta and the CSM, which is the likely mechanism for SNe IIn.  Indeed, one could argue that the magnetar model is the most promising model to explain the luminosity of SLSNe Ic.  For the most luminous SNe Ic, such as SN~2010ay \cite{Sanders2012} and SN~2011kl, if the former event arose from radioactive heating, the ratio of the inferred nickel mass to the total ejecta mass was too large, implying that the radioactive heat deposition model was not a viable model.  Instead it is possible that events such as this could be powered by both nickel decay and a magnetar \cite{Wang2015_Ni_SLSNe}.  Then, for true SLSNe-Ic, nickel heating can be ignored, and conversely for SNe Ic, including GRB-SNe, magnetar input is negligible.  It is only for SNe Ic (all types) with peak absolute magnitudes that exceed $\approx -20$ mag that both energy sources must be considered.  Clearly more observations of luminous SNe Ic are needed to test this hypothesis.

One final point of interest is determining whether all ULGRB-SNe are superluminous compared with GRB-SNe, or whether GRB~111209A/SN~2011kl is a one-off event.  As stated previously, the number of GRB-SNe is very small, and only two are considered here: the aforementioned case and ULGRB~101225A.  Modelling of the (observer-frame) $i$-band LC of the accompanying SN in the latter event showed that its brightness was not exceptional:  we found $k=0.96\pm0.05$ and $s=1.02\pm0.03$ (Table \ref{table:master_table_2_SN}), which implies that some ULGRB-SNe have luminosities that are similar to those of other GRB-SNe.   Moreover, the definition of an ULGRB is important: here we have defined an ULGRB as an event that is still detected after several thousand seconds by a gamma-ray instrument.  This definition is inherently detector- and redshift-dependent.  Based on this definition alone, it appears that GRB~091127 is also an ULGRB $-$ an inspection of the third version of the \emph{Swift}/BAT catalog \cite{Lien2016} reveals that this event was detected by BAT at more than 5000~s.  In turn, accompanying SN~2009nz is also quite typical of the general GRB-SN population, with $k=0.89\pm0.01$ and $s=0.88\pm0.01$.  However, our definition is of course limited, and does not include additional facts of this situation: First, the BAT detection at the late times is very marginal, with a signal-to-noise ratio of just 4.36 (where a value of 7.0 is required in a typical image-trigger threshold).  Secondly, the BAT event data value of T$_{90}$ is only 7.42~s, whereas the BAT value in \cite{Lien2016} is obtained in survey mode.  Thus an alternative interpretation of the extended GRB emission seen in the survey data is that it is soft gamma-rays emitted by the very bright X-ray afterglow, and not from the promt emission.  In summary, more unambiguous ULGRB events at redshifts lower than unity are needed in order to measure the properties of their accompaning SN, and address the peculiar nature of GRB~111209A/SN~2011kl.


\section{Acknowledgements}

ZC thanks David Alexander ``Dr Data'' Kann, Steve Schulze, Maryam Modjaz, Jens Hjorth, Jochen Greiner, Elena Pian, Vicki Toy, and Antonio de Ugarte Postigo for sharing their photometric and spectroscopic data and stimulating discussions,  which without a doubt led to a much improved manuscript.  Some data was extracted using the Dexter applet \cite{Demleitner2001}, while others were downloaded from the WiseREP archive \cite{YaronGY2012}.

The work of ZC was funded by a Project Grant from the Icelandic Research Fund.  S.Q.W, Z.G.D., and X.F.W. are supported by the National Basic Research Program (973 Program) of China (grant Nos. 2014CB845800 and 2013CB834900) and the National Natural Science Foundation of China (grant Nos. 11573014 and 11322328). X.F.W. is partially supported by the Youth Innovation Promotion Association (2011231), and the Strategic Priority Research Program ``The Emergence of Cosmological Structures'' (grant No. XDB09000000) of the Chinese Academy
of Sciences.  

The authors declare that there is no conflict of interest regarding the publication of this manuscript.  This applies both to the scientific content of this work and their funding.

\begin{table}
\footnotesize
\centering
\caption{GRB-SN Master Table I: $\gamma$-ray properties}
\setlength{\tabcolsep}{4.5pt}
\setlength{\extrarowheight}{1.5pt}
\label{table:master_table_I_highE}
\begin{tabular}{cccccccc}
\hline
	&		&		&		&		&		($10^{52}$~erg)					&		(keV)					&			(erg~s$^{-1}$)				\\
GRB	&	SN	&	type	&	$z$	&	$T_{90}$ (s)	&		$E_{\gamma\rm,iso}$					&		$E_{p}$					&			$L_{\rm iso}^{\dagger}$
				\\
\hline																														
970228	&		&	GRB	&	0.695	&	56	&	$	1.6	(	0.12	)	$	&	$	195	(	64	)	$	&	$	4.84	\times 10^{	50	}	$	\\
980326	&		&	GRB	&		&		&	$	0.48	(	0.09	)	$	&	$	935	(	36	)	$	&	$					$	\\
980425	&	1998bw	&	$ll$GRB	&	0.00867	&	18	&	$	0.000086	(	0.000002	)	$	&	$	55	(	21	)	$	&	$	4.80	\times 10^{	46	}	$	\\
990712	&		&	GRB	&	0.4331	&	19	&	$	0.67	(	0.13	)	$	&	$	93	(	15	)	$	&	$	5.05	\times 10^{	50	}	$	\\
991208	&		&	GRB	&	0.7063	&	60	&	$	22.3	(	1.8	)	$	&	$	313	(	31	)	$	&	$	6.34	\times 10^{	51	}	$	\\
000911	&		&	GRB	&	1.0585	&	500	&	$	67	(	14	)	$	&	$	1859	(	371	)	$	&	$	2.75	\times 10^{	51	}	$	\\
011121	&	2001ke	&	GRB	&	0.362	&	47	&	$	7.8	(	2.1	)	$	&	$	793	(	265	)	$	&	$	2.26	\times 10^{	51	}	$	\\
020305	&		&		&		&		&							&							&							\\
020405	&		&	GRB	&	0.68986	&	40	&	$	10	(	0.9	)	$	&	$	612	(	10	)	$	&	$	4.22	\times 10^{	51	}	$	\\
020410	&		&		&		&	$>1600$	&							&							&							\\
020903	&		&	$ll$GRB	&	0.2506	&	3.3	&	$	0.0011	(	0.0006	)	$	&	$	3.37	(	1.79	)	$	&	$	4.20	\times 10^{	48	}	$	\\
021211	&	2002lt	&	GRB	&	1.004	&	2.8	&	$	1.12	(	0.13	)	$	&	$	127	(	52	)	$	&	$	8.02	\times 10^{	51	}	$	\\
030329	&	2003dh	&	GRB	&	0.16867	&	22.76	&	$	1.5	(	0.3	)	$	&	$	100	(	23	)	$	&	$	7.70	\times 10^{	50	}	$	\\
030723	&		&		&		&		&							&	$	<0.023				$	&							\\
030725	&		&		&		&		&							&							&							\\
031203	&	2003lw	&	$ll$GRB	&	0.10536	&	37	&	$	0.0086	(	0.004	)	$	&	$	<200				$	&	$	2.55	\times 10^{	48	}	$	\\
040924	&		&	GRB	&	0.858	&	2.39	&	$	0.95	(	0.09	)	$	&	$	102	(	35	)	$	&	$	7.38	\times 10^{	51	}	$	\\
041006	&		&	GRB	&	0.716	&	18	&	$	3	(	0.9	)	$	&	$	98	(	20	)	$	&	$	2.86	\times 10^{	51	}	$	\\
050416A	&		&	INT	&	0.6528	&	2.4	&	$	0.1	(	0.01	)	$	&	$	25.1	(	4.2	)	$	&	$	6.89	\times 10^{	50	}	$	\\
050525A	&	2005nc	&	GRB	&	0.606	&	8.84	&	$	2.5	(	0.43	)	$	&	$	127	(	10	)	$	&	$	4.54	\times 10^{	51	}	$	\\
050824	&		&	GRB	&	0.8281	&	25	&	$	0.041<E<0.34				$	&	$	11<E<32				$	&							\\
060218	&	2006aj	&	$ll$GRB	&	0.03342	&	2100	&	$	0.0053	(	0.0003	)	$	&	$	4.9	(	0.3	)	$	&	$	2.60	\times 10^{	46	}	$	\\
060729	&		&	GRB	&	0.5428	&	115	&	$	1.6	(	0.6	)	$	&	$	>50				$	&	$	2.14	\times 10^{	50	}	$	\\
060904B	&		&	GRB	&	0.7029	&	192	&	$	2.4	(	0.2	)	$	&	$	163	(	31	)	$	&	$	2.12	\times 10^{	50	}	$	\\
070419A	&		&	INT	&	0.9705	&	116	&	$	\approx 0.16				$	&							&	$	2.71	\times 10^{	49	}	$	\\
080319B	&		&	GRB	&	0.9371	&	124.86	&	$	114	(	9	)	$	&	$	1261	(	65	)	$	&	$	1.76	\times 10^{	52	}	$	\\
081007	&	2008hw	&	GRB	&	0.5295	&	9.01	&	$	0.15	(	0.04	)	$	&	$	61	(	15	)	$	&	$	2.54	\times 10^{	50	}	$	\\
090618	&		&	GRB	&	0.54	&	113.34	&	$	25.7	(	5	)	$	&	$	211	(	22	)	$	&	$	3.49	\times 10^{	51	}	$	\\
091127	&	2009nz	&	GRB	&	0.49044	&	7.42	&	$	1.5	(	0.2	)	$	&	$	35.5	(	1.5	)	$	&	$	3.01	\times 10^{	51	}	$	\\
100316D	&	2010bh	&	$ll$GRB	&	0.0592	&	1300	&	$	>0.0059				$	&	$	26	(	16	)	$	&	$	4.80	\times 10^{	46	}	$	\\
100418A	&		&	GRB	&	0.6239	&	8	&	$	0.0990	(	0.0630	)	$	&	$	29	(	2	)	$	&	$	2.00	\times 10^{	50	}	$	\\
101219B	&	2010ma	&	GRB	&	0.55185	&	51	&	$	0.42	(	0.05	)	$	&	$	70	(	8	)	$	&	$	1.27	\times 10^{	50	}	$	\\
101225A	&		&	ULGRB	&	0.847	&	7000	&	$	1.2	(	0.3	)	$	&	$	38	(	20	)	$	&	$	3.16	\times 10^{	48	}	$	\\
111209A	&	2011kl	&	ULGRB	&	0.67702	&	10000	&	$	58.2	(	7.3	)	$	&	$	520	(	89	)	$	&	$	9.76	\times 10^{	49	}	$	\\
111211A	&		&		&	0.478	&		&							&							&							\\
111228A	&		&	GRB	&	0.71627	&	101.2	&	$	4.2	(	0.6	)	$	&	$	58.4	(	6.9	)	$	&	$	7.12	\times 10^{	50	}	$	\\
120422A	&	2012bz	&	GRB	&	0.28253	&	5.4	&	$	0.024	(	0.008	)	$	&	$	<72				$	&	$	5.70	\times 10^{	49	}	$	\\
120714B	&	2012eb	&	INT	&	0.3984	&	159	&	$	0.0594	(	0.0195	)	$	&	$	101.4	(	155.7	)	$	&	$	5.22	\times 10^{	48	}	$	\\
120729A	&		&	GRB	&	0.8	&	71.5	&	$	2.3	(	1.5	)	$	&	$	310.6	(	31.6	)	$	&	$	5.79	\times 10^{	50	}	$	\\
130215A	&	2013ez	&	GRB	&	0.597	&	65.7	&	$	3.1	(	1.6	)	$	&	$	155	(	63	)	$	&	$	7.53	\times 10^{	50	}	$	\\
130427A	&	2013cq	&	GRB	&	0.3399	&	163	&	$	81	(	10	)	$	&	$	1028	(	50	)	$	&	$	6.65	\times 10^{	51	}	$	\\
130702A	&	2013dx	&	INT	&	0.145	&	58.881	&	$	0.064	(	0.013	)	$	&	$	15	(	5	)	$	&	$	1.24	\times 10^{	49	}	$	\\
130831A	&	2013fu	&	GRB	&	0.479	&	32.5	&	$	0.46	(	0.02	)	$	&	$	67	(	4	)	$	&	$	2.09	\times 10^{	50	}	$	\\
140606B	&		&	GRB	&	0.384	&	22.78	&	$	0.347	(	0.02	)	$	&	$	801	(	182	)	$	&	$	2.10	\times 10^{	50	}	$	\\
150518A	&		&		&	0.256	&		&							&							&							\\
150818A	&		&	INT	&	0.282	&	123.3	&	$	0.1	(	0.02	)	$	&	$	128	(	13	)	$	&	$	1.03	\times 10^{	49	}	$	\\
\hline
\end{tabular}
\begin{flushleft}
\tiny
$^{\dagger}L_{\rm iso} = E_{\rm iso}$(1+$z$)/$T_{90}$. \\
$^{\ddagger}$: $\gamma$-ray properties calculated by \cite{Amati2002} for a redshift range of $0.9 \le z \le 1.1$.\\
$ll$GRB: GRB-SN associated with a low-luminosity GRB ($L_{\rm \gamma,iso} < 10^{48.5}$~erg~s$^{-1}$); INT: GRB-SN associated with a intermediate-luminosity GRB ($10^{48.5} < L_{\rm \gamma,iso} <10^{49.5}$~erg~s$^{-1}$); GRB: GRB-SN associated with a high-luminosity GRB ($L_{\rm \gamma,iso} > 10^{49.5}$~erg~s$^{-1}$);  ULGRB: GRB-SN associated with an ultra-long-duration GRB (see Section \ref{sec:Subclasses}).\\
\end{flushleft}
\end{table}

\begin{landscape}
\begin{table}
\tiny
\centering
\caption{GRB-SN Master Table II: SN properties}
\setlength{\tabcolsep}{2pt}
\setlength{\extrarowheight}{1.2pt}
\label{table:master_table_2_SN}
\begin{tabular}{ccccccccccccccccccc}
\hline
	&		&		&		&		&		&		(mag)					&		(mag)					&		(d)					&	(erg~s$^{-1}$)	&	(d)	&	(mag)	&		($10^{51}$~erg)					&		(M$_{\odot}$)					&		(M$_{\odot}$)					&	(km~s$^{-1}$)	&							&							&		\\
GRB	&	SN	&	type	&	$z$	&	S?	&	Grade	&		$M_{V}^{*}$					&		$\Delta m_{15,V}$					&		$t_{V,\rm p}^{*}$					&	$L_{\rm p,Bol}$	&	$t_{\rm p,rest}$	&	$\Delta m_{15,\rm Bol}$	&		$E_{\rm K}$					&		$M_{\rm ej}$					&		$M_{\rm Ni}$					&	$v_{\rm ph}$	&		$\bar k$					&		$\bar s$					&	Filters	\\
\hline																																																																													
970228	&		&	GRB	&	0.695	&		&	C	&						&						&						&		&		&		&									&									&						&		&						&						&		\\
980326	&		&	GRB	&		&		&	D	&						&						&						&		&		&		&									&									&						&		&						&						&		\\
980425	&	1998bw	&	$ll$GRB	&	0.00866	&	S	&	A	&	$	-19.29	\pm	0.08	$	&	$	0.75	\pm	0.02	$	&	$	16.09	\pm	0.18	$	&	$7.33\times10^{42}$	&	15.16	&	0.80	&	$	20-30						$	&	$	6-10						$	&	$	0.3-0.6			$	&	18000	&	$	\textbf{1}			$	&	$	\textbf{1}			$	&		\\
990712	&		&	GRB	&	0.4331	&		&	C	&						&						&						&		&		&		&	$	26.1	^{	+24.6	}_{	-15.0	}	$	&	$	6.6	^{	+3.5	}_{	-2.9	}	$	&	$	0.14	\pm	0.04	$	&		&	$	0.36	\pm	0.05	$	&	$	1.10	\pm	0.20	$	&	$R$	\\
991208	&		&	GRB	&	0.7063	&		&	E	&						&						&						&		&		&		&	$	38.7	^{	+44.6	}_{	-26.0	}	$	&	$	9.7	^{	+6.8	}_{	-5.6	}	$	&	$	0.96	\pm	0.48	$	&		&	$	2.11	\pm	0.58	$	&	$	1.10	\pm	0.20	$	&	$R$	\\
000911	&		&	GRB	&	1.0585	&		&	E	&						&						&						&		&		&		&									&									&						&		&						&						&		\\
011121	&	2001ke	&	GRB	&	0.362	&	S	&	B	&						&						&						&	$\sim5.9\times10^{42}$	&	$\sim17$	&		&	$	17.7	^{	+8.8	}_{	-6.4	}	$	&	$	4.4	\pm	0.8				$	&	$	0.35	\pm	0.01	$	&		&	$	1.13	\pm	0.23	$	&	$	0.84	\pm	0.17	$	&	$BV^{*}$	\\
020305	&		&		&		&		&	E	&						&						&						&		&		&		&									&									&						&		&						&						&		\\
020405	&		&	GRB	&	0.68986	&		&	C	&						&						&						&		&		&		&	$	8.9	^{	+5.4	}_{	-3.8	}	$	&	$	2.2	^{	+0.6	}_{	-0.5	}	$	&	$	0.23	\pm	0.02	$	&		&	$	0.82	\pm	0.14	$	&	$	0.62	\pm	0.03	$	&	$R$	\\
020410	&		&		&		&		&	D	&						&						&						&		&		&		&									&									&						&		&						&						&		\\
020903	&		&	$ll$GRB	&	0.2506	&	S	&	B	&						&						&						&		&		&		&	$	28.9	^{	+32.2	}_{	-18.9	}	$	&	$	7.3	^{	+4.9	}_{	-4.0	}	$	&	$	0.25	\pm	0.13	$	&		&	$	0.61	\pm	0.19	$	&	$	0.98	\pm	0.02	$	&	$R$	\\
021211	&	2002lt	&	GRB	&	1.004	&	S	&	B	&						&						&						&		&		&		&	$	28.5	^{	+45.0	}_{	-13.0	}	$	&	$	7.2	^{	+7.4	}_{	-6.0	}	$	&	$	0.16	\pm	0.14	$	&		&	$	0.40	\pm	0.19	$	&	$	0.98	\pm	0.26	$	&	$R$	\\
030329	&	2003dh	&	GRB	&	0.16867	&	S	&	A	&	$	-19.39	\pm	0.14	$	&	$	0.90	\pm	0.50	$	&	$	10.74	\pm	2.57	$	&	$1.01\times10^{43}$	&	12.75	&	0.70	&	$	20-50						$	&	$	5-10						$	&	$	0.4-0.6			$	&	20000	&	$	1.28	\pm	0.28	$	&	$	0.87	\pm	0.18	$	&	$UBV^{*}$	\\
030723	&		&		&		&		&	D	&						&						&						&		&		&		&									&									&						&		&						&						&		\\
030725	&		&		&		&		&	E	&						&						&						&		&		&		&									&									&						&		&						&						&		\\
031203	&	2003lw	&	$ll$GRB	&	0.10536	&	S	&	A	&	$	-19.90	\pm	0.16	$	&	$	0.64	\pm	0.10	$	&	$	19.94	\pm	1.48	$	&	$1.26\times10^{43}$	&	17.33	&	0.62	&	$	60.0	\pm	15				$	&	$	13.0	\pm	4.0				$	&	$	0.55	\pm	0.20	$	&	18000	&	$	1.65	\pm	0.36	$	&	$	1.10	\pm	0.24	$	&	$VRI^{*}$	\\
040924	&		&	GRB	&	0.858	&		&	C	&						&						&						&		&		&		&									&									&						&		&						&						&		\\
041006	&		&	GRB	&	0.716	&		&	C	&						&						&						&		&		&		&	$	76.4	^{	+39.8	}_{	-28.7	}	$	&	$	19.2	^{	+3.9	}_{	-3.6	}	$	&	$	0.69	\pm	0.07	$	&		&	$	1.16	\pm	0.06	$	&	$	1.47	\pm	0.04	$	&	$R$	\\
050416A	&		&	INT	&	0.6528	&		&	D	&						&						&						&		&		&		&									&									&						&		&						&						&		\\
050525A	&	2005nc	&	GRB	&	0.606	&	S	&	B	&	$	-18.59	\pm	0.31	$	&	$	1.17	\pm	0.88	$	&	$	11.08	\pm	3.37	$	&		&		&		&	$	18.9	^{	+10.7	}_{	-7.5	}	$	&	$	4.8	^{	+1.1	}_{	-1.0	}	$	&	$	0.24	\pm	0.02	$	&		&	$	0.69	\pm	0.03	$	&	$	0.83	\pm	0.03	$	&	$R$	\\
050824	&		&	GRB	&	0.8281	&		&	E	&						&						&						&		&		&		&	$	5.7	^{	+9.3	}_{	-3.7	}	$	&	$	1.4	^{	+1.6	}_{	-0.6	}	$	&	$	0.26	\pm	0.17	$	&		&	$	1.05	\pm	0.42	$	&	$	0.52	\pm	0.14	$	&	$R$	\\
060218	&	2006aj	&	$ll$GRB	&	0.03342	&	S	&	A	&	$	-18.85	\pm	0.08	$	&	$	1.08	\pm	0.06	$	&	$	9.96	\pm	0.18	$	&	$6.47\times10^{42}$	&	10.42	&	0.83	&	$	1.0	\pm	0.5				$	&	$	2.0	\pm	0.5				$	&	$	0.20	\pm	0.10	$	&	20000	&	$	0.58	\pm	0.13	$	&	$	0.67	\pm	0.14	$	&	$UBVR^{*}$	\\
060729	&		&	GRB	&	0.5428	&		&	D	&						&						&						&		&		&		&	$	24.4	^{	+14.3	}_{	-9.9	}	$	&	$	6.1	^{	+1.6	}_{	-1.4	}	$	&	$	0.36	\pm	0.05	$	&		&	$	0.94	\pm	0.10	$	&	$	0.92	\pm	0.04	$	&	$RI$	\\
060904B	&		&	GRB	&	0.7029	&		&	C	&						&						&						&		&		&		&	$	9.9	^{	+5.1	}_{	-3.7	}	$	&	$	2.5	\pm	0.5				$	&	$	0.12	\pm	0.01	$	&		&	$	0.42	\pm	0.02	$	&	$	0.65	\pm	0.01	$	&	$R$	\\
070419A	&		&	GRB	&	0.9705	&		&	D	&						&						&						&		&		&		&									&									&						&		&						&						&		\\
080319B	&		&	GRB	&	0.9371	&		&	C	&						&						&						&		&		&		&	$	22.7	^{	+19.1	}_{	-11.9	}	$	&	$	5.7	^{	+2.6	}_{	-2.2	}	$	&	$	0.86	\pm	0.45	$	&		&	$	2.30	\pm	0.90	$	&	$	0.89	\pm	0.10	$	&	$I$	\\
081007	&	2008hw	&	GRB	&	0.5295	&	S	&	B	&						&						&						&	$\sim1.4\times10^{43}$	&	$\sim12$	&		&	$	19.0	\pm	15.0				$	&	$	2.3	\pm	1.0				$	&	$	0.39	\pm	0.08	$	&	12600	&	$	0.71	\pm	0.10	$	&	$	0.85	\pm	0.11	$	&	$riz$	\\
090618	&		&	GRB	&	0.54	&		&	C	&	$	-19.34	\pm	0.13	$	&	$	0.65	\pm	0.17	$	&	$	17.54	\pm	1.64	$	&		&		&		&	$	36.5	^{	+20.0	}_{	-14.2	}	$	&	$	9.2	^{	+2.1	}_{	-1.9	}	$	&	$	0.37	\pm	0.03	$	&		&	$	1.11	\pm	0.22	$	&	$	0.98	\pm	0.20	$	&	B*	\\
091127	&	2009nz	&	GRB	&	0.49044	&	S	&	B	&						&						&						&	$\sim1.2\times10^{43}$	&	$\sim15$	&	$\sim0.5$	&	$	13.5	\pm	0.4				$	&	$	4.7	\pm	0.1				$	&	$	0.33	\pm	0.01	$	&	17000	&	$	0.89	\pm	0.01	$	&	$	0.88	\pm	0.01	$	&	$I$	\\
100316D	&	2010bh	&	$ll$GRB	&	0.0592	&	S	&	A	&	$	-18.89	\pm	0.10	$	&	$	1.10	\pm	0.05	$	&	$	8.76	\pm	0.37	$	&	$5.67\times10^{42}$	&	8.76	&	0.89	&	$	15.4	\pm	1.4				$	&	$	2.5	\pm	0.2				$	&	$	0.12	\pm	0.02	$	&	35000	&	$	0.53	\pm	0.15	$	&	$	0.53	\pm	0.11	$	&	$VRI^{*}$	\\
100418A	&		&	INT	&	0.6239	&		&	D/E	&						&						&						&		&		&		&									&									&						&		&						&						&		\\
101219B	&	2010ma	&	GRB	&	0.55185	&	S	&	A/B	&						&						&						&	$1.5\times10^{43}$	&	11.80	&	0.99	&	$	10.0	\pm	6.0				$	&	$	1.3	\pm	0.5				$	&	$	0.43	\pm	0.03	$	&		&	$	1.16	\pm	0.63	$	&	$	0.76	\pm	0.10	$	&	$griz$	\\
101225A	&		&	ULGRB	&	0.847	&		&	D	&						&						&						&		&		&		&	$	32.0	\pm	16.0				$	&	$	8.1	\pm	1.5				$	&	$	0.41	\pm	0.03	$	&		&	$	0.96	\pm	0.05	$	&	$	1.02	\pm	0.03	$	&	$i$	\\
111209A	&	2011kl	&	ULGRB	&	0.67702	&		&	A/B	&						&						&						&	$2.91\times10^{43}$	&	14.80	&	0.78	&		$20-90$							&		$3-5$							&						&	21000	&	$	1.81	\pm	0.19	$	&	$	1.08	\pm	0.11	$	&	$iz$	\\
111211A	&		&		&	0.478	&	S	&	B/C	&						&						&						&		&		&		&									&									&						&		&						&						&		\\
111228A	&		&		&	0.71627	&		&	E	&						&						&						&		&		&		&									&									&						&		&						&						&		\\
120422A	&	2012bz	&	$ll$GRB	&	0.28253	&	S	&	A	&	$	-19.50	\pm	0.03	$	&	$	0.73	\pm	0.06	$	&	$	14.20	\pm	0.34	$	&	$1.48\times10^{43}$	&	14.45	&	0.62	&	$	25.5	\pm	2.1				$	&	$	6.1	\pm	0.5				$	&	$	0.57	\pm	0.07	$	&	20500	&	$	1.13	\pm	0.25	$	&	$	0.93	\pm	0.19	$	&	$BV^{*}$	\\
120714B	&	2012eb	&		&	0.3984	&	S	&	B	&						&						&						&		&		&		&									&									&						&		&						&						&		\\
120729A	&		&	GRB	&	0.8	&		&	D/E	&						&						&						&		&		&		&									&									&	$	0.42	\pm	0.11	$	&		&	$	1.02	\pm	0.26	$	&	$	1^{\ddagger}$				&	$ri$	\\
130215A	&	2013ez	&	GRB	&	0.597	&	S	&	B	&						&						&						&		&		&		&									&									&		$0.25-0.30$				&	6000	&		$0.6-0.75$				&	$	1^{\ddagger}$				&	$ri$	\\
130427A	&	2013cq	&	GRB	&	0.3399	&	S	&	B	&						&						&						&		&		&		&	$	64.0	\pm	7.0				$	&	$	6.3	\pm	0.7				$	&	$	0.28	\pm	0.02	$	&	35000	&	$	0.85	\pm	0.03	$	&	$	0.77	\pm	0.03	$	&	$r$	\\
130702A	&	2013dx	&	INT	&	0.145	&	S	&	A	&	$	-19.26	\pm	0.24	$	&	$	1.05	\pm	0.05	$	&	$	13.86	\pm	0.70	$	&	$1.08\times10^{43}$	&	12.94	&	0.85	&	$	8.2	\pm	0.4				$	&	$	3.1	\pm	0.1				$	&	$	0.37	\pm	0.01	$	&	21300	&	$	0.98	\pm	0.07	$	&	$	0.78	\pm	0.05	$	&	$griz$	\\
130831A	&	2013fu	&	GRB	&	0.479	&	S	&	A/B	&						&						&						&		&		&		&	$	18.7	\pm	9.0				$	&	$	4.7	\pm	0.8				$	&	$	0.30	\pm	0.07	$	&		&	$	0.95	\pm	0.19	$	&	$	0.82	\pm	0.19	$	&	$B^{*}$	\\
140606B	&		&	GRB	&	0.384	&	S	&	A/B	&						&						&						&		&		&		&	$	19.0	\pm	11.0				$	&	$	4.8	\pm	1.9				$	&	$	0.42	\pm	0.17	$	&	19800	&	$	1.04	\pm	0.24	$	&	$	0.81	\pm	0.13	$	&	$V^{*}$	\\
150518A	&		&		&	0.256	&		&	C/D	&						&						&						&		&		&		&									&									&						&		&						&						&		\\
150818A	&		&	INT	&	0.282	&	S	&	B	&						&						&						&		&		&		&									&									&						&		&						&						&		\\
\hline																																																																											
-	&	2009bb	&	Rel IcBL	&	0.009987	&	S	&	-	&	$	-18.61	\pm	0.28	$	&	$	1.13	\pm	0.04	$	&	$	13.37	\pm	0.32	$	&		&		&		&	$	18.0	\pm	8.0				$	&	$	4.1	\pm	1.9				$	&	$	0.19	\pm	0.03	$	&	15000	&	$	0.60	\pm	0.05	$	&	$	0.73	\pm	0.07	$	&	$BVRI^{*}$	\\
-	&	2012ap	&	Rel IcBL	&	0.012141	&	S	&		&	$	-18.76	\pm	0.33	$	&	$	0.92	\pm	0.08	$	&	$	14.43	\pm	0.19	$	&		&		&		&	$	9.0	\pm	3.0				$	&	$	2.7	\pm	0.5				$	&	$	0.12	\pm	0.02	$	&	13000	&	$	1.10	\pm	0.23	$	&	$	0.82	\pm	0.09	$	&	$BVRI^{*}$	\\
\hline
\end{tabular}
\begin{flushleft}
\tiny
S: Denotes one or more spectra of the SN were obtained.\\
Grades are from Hjorth \& Bloom (2012): \textbf{A}: Strong spectroscopic evidence. \textbf{B}: A clear light curve bump as well as some spectroscopic evidence resembling a GRB-SN. \textbf{C}: A clear bump consistent with other GRB-SNe at the spectroscopic redshift of the GRB. \textbf{D}: A bump, but the inferred SN properties are not fully consistent with other GRB-SNe or the bump was not well sampled or there is no spectroscopic redshift of the GRB. \textbf{E}: A bump, either of low significance or inconsistent with other GRB-SNe.\\
$^{*}$Denotes exact, $K$-corrected rest-frame filter observable. \\
$^{\ddagger}$Values fixed during fit. \\
$\bar k$ and $\bar s$ denote the filter-averaged luminosity ($k$) and stretch ($s$) factors relative to SN~1998bw.
\end{flushleft}
\end{table}
\end{landscape}

\begin{table}
\small
\centering
\setlength{\tabcolsep}{2.1pt}
\setlength{\extrarowheight}{1pt}
\caption{References}
\label{table:references}
\begin{tabular}{rl}
\hline
GRB	&	References(s)	\\
\hline			
970228	&	\cite{Reichart1999,Galama2000}	\\
980326	&	\cite{Bloom1999_260398,CTG1999}	\\
980425	&	\cite{Galama1998,Iwamoto1998,Mazzali2001,Patat2001,Maeda2002,Maeda2006,Clocchiatti2011,Cano2013,Cano2014,LH14,CJG14,Lyman2016} \\
990712	&	\cite{Bjornsson2001,Cano2013}	\\
991208	&	\cite{CT2001,Cano2013}	\\
000911	&	\cite{Lazzati2001,Masetti2005}\\
011121	&	\cite{Bloom2002_011121,Price2002,Garnavich2003,Greiner2003,Cano2013,Cano2014,CJG14}	\\
020305	&	\cite{Gorosabel2005_020305}	\\
020405	&	\cite{Price2003,Cano2013}	\\
020410	&	\cite{Nicastro2004,Levan2005}	\\
020903	&	\cite{Ricker2002,Soderberg2005,Bersier2006,Cano2013}	\\
021211	&	\cite{DellaValle2003,Pandey2003,Cano2013}\\
030329	&	\cite{Hjorth2003,Stanek2003,Matheson2003,WoosleyHeger2003,Bloom2004,Lipkin2004,Deng2005,Cano2011a,Cano2013,Cano2014,LH14,CJG14}\\
030723	&	\cite{Fynbo2004,Tominaga2004}	\\
030725	&	\cite{Pugliese2005}	\\
031203	&	\cite{Malesani2004,Soderberg2004,GalYam2004,Watson2004,Thomsen2004,Cobb2006,Mazzali2006,Cano2013,Cano2014,CJG14,LH14}\\
040924	&	\cite{Soderberg2006,Wiersema2008}	\\
041006	&	\cite{Misra2005,Stanek2005,Soderberg2006,Cano2013} \\
050416A	&	\cite{Sakamoto2006,Soderberg2007}	\\
050525A	&	\cite{DellaValle2006,Cano2013,CJG14,LH14} \\
050824	&	\cite{Krimm2005,Crew2005,Sollerman2007,Cano2013} \\
060218	&	\cite{Campana2006,Cobb2006,Ferrero2006,Mazzali2006,Mirabal2006,Modjaz2006,Pian2006,Soderberg2006,Sollerman2006,Maeda2007,Misra2011,Cano2013,Cano2014,CJG14,LH14,Lyman2016}  \\
060729	&	\cite{Grupe2007,ButlerKoc2007,Cano2011b,Cano2013}	\\
060904B	&	\cite{Markwardt2006,Cano2013} \\
070419A	&	\cite{Cenko2007,Hill2007} \\
080319B	&	\cite{Kann2008,Bloom2009,Tanvir2010} \\
081007	&	\cite{Soderberg2008,Berger2008,DellaValle2008,Jin2013,Olivares2015} \\
090618	&	\cite{Cano2011b,Cano2013,Cano2014,LH14} \\
091127	&	\cite{Cobb2010,Berger2011,Cano2013,Olivares2015} \\
100316D	&	\cite{Chornock2010,Cano2011a,Starling2011,Bufano2012,Olivares2012,Cano2013,Margutti2013,Cano2014,CJG14,LH14,Lyman2016} \\
100418A	&	\cite{Holland2010,Marshall2011,AdUP2011}  \\
101219B	&	\cite{Sparre2011,Olivares2015}  \\
101225A	&	\cite{Thone2011,Levan2014_ULGRB}, \textbf{here}	\\
111209A	&	\cite{Golenetskii2011,Levan2014_ULGRB,Greiner2015,Metzger2015,Bersten2016,CJM2016}, \textbf{here}	\\
111211A	&	\cite{AdUP2012} \\
111228A	&	\cite{DAvanzo2012} \\
120422A	&	\cite{Melandri2012,Zhang2012_2012bz,Cano2013,Schulze2014,CJG14,LH14} \\
120714B	&	\cite{Klose2012,Cummings2012} \\
120729A	&	\cite{Cano_etal_2014} \\
130215A	&	\cite{Cano_etal_2014} \\
130427A	&	\cite{Xu2013,Levan2014_27A,Melandri2014,dePasquale2016}  \\
130702A	&	\cite{CJG14,Delia2015,Toy2016} \\
130831A	&	\cite{Klose2013,Cano2014,Cano_etal_2014} \\
140606B	&	\cite{Singer2015,Cano2015} \\
150518A	&	\cite{Pozanenko2015} \\
150818A	&	\cite{AdUP2015,Golenetskii2015,Palmer2015} \\
SN~2009bb	&	\cite{Soderberg2010,Levesque2010,Pignata2011,Cano2013,Cano2014,CJG14,Lyman2016} \\
SN~2012ap	&	\cite{Chakraborti2015,Liu2015,Margutti2015,Milisavljevic2015}, \textbf{here}	\\
\hline			
\end{tabular}
\end{table}

\vspace*{0.5cm}


\bibliographystyle{article}
\bibliography{ref}


\end{document}